\documentclass[fleqn,usenatbib,twocolumn]{aastex631}
\shorttitle{Type Ia SNe in massive ellipticals}
\shortauthors{Mohapatra \& Quataert}

\usepackage{graphicx}	% Including figure files
\usepackage{amsmath}	% Advanced maths commands
\usepackage[caption=false]{subfig}
\usepackage{ragged2e}
\usepackage{placeins}
\usepackage{bigints} %To be able to write bigger integral symbols
\usepackage{physics}
\usepackage{xcolor}
\usepackage{cleveref}
\usepackage{longtable}
\usepackage{xtab}
\usepackage{bm}
\usepackage{cancel}
\allowdisplaybreaks
\newcommand{\change}[1]{{\textcolor{black}{#1}}}
\usepackage{scalerel,tikz}

\def\mean#1{\left< #1 \right>}
\def\sun{\odot}

\defcitealias{Mohapatra2022characterising}{M22b}
\defcitealias{Mohapatra2022VSF}{M22a}
\defcitealias{Mohapatra2023MNRAS}{M23}
\defcitealias{MLi2020ApJa}{MLi20a}
\defcitealias{MLi2020ApJb}{MLi20b}
\begin{document}
\title[SNIa in a stratified medium]{Multiphase gas in elliptical galaxies: the role of Type Ia supernovae}

% The list of authors, and the short list which is used in the headers.
% If you need two or more lines of authors, add an extra line using \newauthor
% \author[Mohapatra et al]{
% Rajsekhar Mohapatra$^{\orcidicon{0000-0002-1600-7552}\,1}$\thanks{E-mail: rmohapatra@princeton.edu (RM)},
% Eliot Quataert$^{\orcidicon{0000-0001-9185-5044}\,1}$\thanks{E-mail: quataert@princeton.edu (EQ)}
% \\
\correspondingauthor{Rajsekhar Mohapatra}
\email{rmohapatra@princeton.edu}

\author[0000-0002-1600-7552]{Rajsekhar Mohapatra}
\affiliation{Department of Astrophysical Sciences, Princeton University, Princeton, NJ 08544, USA}
\author[0000-0001-9185-5044]{Eliot Quataert}
\affiliation{Department of Astrophysical Sciences, Princeton University, Princeton, NJ 08544, USA}

\begin{abstract}
Massive elliptical galaxies harbor large amounts of hot gas ($T\gtrsim10^6~\mathrm{K}$) in their interstellar medium (ISM) but are typically quiescent in star formation. Active-galactic nuclei (AGNs) jets and Type Ia supernovae (SNIa) inject energy into the ISM which offsets its radiative losses and keeps it hot. SNIa deposit their energy locally within the galaxy compared to the larger few$\times10~\mathrm{kpc}$-scale AGN jets. 
% Multiphase condensation triggered by SNIa-driven fluctuations can also potentially explain the observations of cooler filaments (at $10$--$10^4~\mathrm{K}$) in the ISM. 
In this study, we perform high-resolution ($512^3$) hydrodynamic simulations of a local ($1~\mathrm{kpc}^3$) density-stratified patch of massive galaxies' ISM. We include radiative cooling and shell-averaged volume heating, as well as randomly exploding SNIa. We study the effect of different fractions of supernova heating  (with respect to the net cooling rate), different initial ISM density/entropy (which controls the thermal-instability growth time $t_\mathrm{ti}$) and different degrees of stratification (which affects the free-fall time $t_\mathrm{ff}$). 
We find that the SNIa drive predominantly compressive turbulence in the ISM with a velocity dispersion $\sigma_v$ up to $40~\mathrm{km}s^{-1}$ and logarithmic density dispersion $\sigma_s\sim0.2$--$0.4$. These fluctuations trigger multiphase condensation in regions of the ISM where \change{$\min(t_\mathrm{ti})/t_\mathrm{ff}\lesssim 0.6\exp(6 \sigma_s)$}, in agreement with theoretical expectations that large density fluctuations efficiently trigger multiphase gas formation. Since the SNIa rate is not self-adjusting, when the net cooling drops below the net heating rate the SNIa drive a hot wind which sweeps out most of the mass in our local model.  Global simulations are required to assess the ultimate fate of this gas. 
%This reduces the gas density, increases $t_\mathrm{ti}$ and prevents further condensation.
\end{abstract}
% \begin{keywords}
% methods: numerical -- hydrodynamics -- turbulence -- galaxies: elliptical galaxies
% \end{keywords}
% \keywords{\update{methods: numerical -- hydrodynamics -- turbulence -- galaxies: elliptical galaxies}}
%%%%%%%%%%%%%%%%%%%%%%%%%%%%%%%%%%%%%%%%%%%%%%%%%%

%%%%%%%%%%%%%%%%% BODY OF PAPER %%%%%%%%%%%%%%%%%%

\section{Introduction}\label{sec:introduction}
The interstellar medium (ISM) in massive galaxies is hot, with temperatures $T\sim10^6$--$10^7~\mathrm{K}$ and the dense central regions (number density $n\sim0.01$--$1~\mathrm{cm}^{-3}$) can cool in less than $1~\mathrm{Gyr}$ \citep{Werner2012MNRAS}. However, the ISM in most of these early-type galaxies is not undergoing a cooling flow \citep{Peterson2003ApJ}. They show little to no current star formation and their stellar populations are typically$\gtrsim10~\mathrm{Gyrs}$ old, barring a few exceptions \citep[e.g., see][]{Calzadilla2022ApJ}. % maybe include some 
Hence the ISM of these massive ellipticals needs to be heated constantly to keep them in rough thermal balance, see \cite{Donahue2022PhR} for a recent review. 

Heating of the ISM due to active galactic nuclei (AGNs) jets powered by the accretion of matter onto the supermassive black hole (SMBH) is one significant energy source \citep[e.g.][]{Allen2006MNRAS,Fabian2012ARA&A,Main2017MNRAS,Gaspari2017MNRAS,MGuo2023ApJ}. AGN jets are typically found to deposit their energy further out (few$\times10~\mathrm{kpc}$) from the central regions of the galaxy \citep{Birzan2004ApJ}. However, on $\sim\mathrm{kpc}$ scales, the jets are highly collimated and the cavities/bubbles associated with them are not volume-filling. While the mechanical energy input from AGN jets matches the net cooling of the halo gas in several systems \citep{McNamara2007ARA&A,Rafferty2008ApJ,Olivares2023ApJ}, the distribution of this heating throughout the ISM is still an open question.  
% Further, the bubbles inflated by the AGN jets are observed to be localized to small regions and the mechanism of distributing this energy throughout the ISM is still an open question.  

Heating due to supernovae is another attractive model. In early-type galaxies with old stellar populations, type Ia supernovae (SNIa) dominate the stellar energy output channel and are a natural outcome of stellar evolution. The spatial distribution of the SNIa is expected to follow the stellar distribution in the galaxy. Hence, these supernovae deposit their energy more uniformly and within the ISM of the galaxy compared to the AGN jets. Studies such as \cite{Voit2015ApJ803L21V} have shown that the energy injected by the SNIa exceeds the energy lost due to radiative cooling in high-entropy outer regions of the elliptical galaxies. They propose that the mass deposited into the ISM by AGB stars can be swept out by an SNIa-driven wind, helping to keep the ISM density low and its cooling time ($t_\mathrm{cool}$) long. \cite{YPLi2018ApJ} find that SN heating can suppress star formation in low mass ellipticals, while AGN feedback is more efficient for systems with stellar mass$\gtrsim10^{11}M_\sun$  \citep[also see][]{Ciotti1991ApJ,Conroy2015ApJ,Voit2020ApJ,Molero2023MNRAS}.

Since the net energy output of SNIa is lower than that of the AGN outbursts, many individual galaxy/galaxy cluster-scale simulations have ignored their effect. Cosmological simulations have generally included the mass and energy injection due to supernovae using sub-grid models \citep[for eg.][]{Crain2015MNRAS,Pillepich2018MNRAS}.    Alternatively, in simulations that do include discrete SNe \citep[eg.][]{KySu2019MNRAS} it is not clear that they have the resolution to resolve individual supernova remnants of size a few$\times10~\mathrm{pc}$ in the low-density ISM of massive ellipticals. While the energy injection due to supernovae is accounted for by such models, they cannot account for the small-scale turbulence driven by the remnants (on scales of $50$--$100~\mathrm{pc}$) and the density fluctuations they generate. These density fluctuations can trigger multiphase condensation in the halo gas if the ratio of the thermal instability growth time to the free-fall time  \citep[$t_\mathrm{ti}/t_\mathrm{ff}$,][]{sharma2012thermal,Choudhury2019} or the ratio of $t_\mathrm{ti}$ to the mixing time \citep[$t_\mathrm{ti}/t_\mathrm{mix}$,][]{Gaspari2018ApJ,Mohapatra2019} falls below a critical density dispersion amplitude ($\sigma_s$)-dependent ratio \citep{Voit2021ApJ,Mohapatra2023MNRAS}. Observational studies such as \cite{Olivares2019A&A} report multiphase filamentary structures around the brightest cluster galaxies (BCGs) in low-entropy and short cooling time regions of clusters. 

Numerical studies on small scales ($\lesssim10~\mathrm{kpc}$ regions) can resolve individual remnants, either by implementing mesh-refinement techniques around the supernovae injection regions or modeling even smaller $\lesssim\mathrm{kpc}$-scale regions of the ISM with uniform resolution. For example, \cite{Tang2009MNRASa,Tang2009MNRASb} have looked at the role played by  SNIa in the Milky Way's bulge. They find that the SNIa power a galactic bulge wind with a filamentary structure. This wind sweeps out mass from the inner regions and keeps the circumgalactic medium (CGM) hot. 

In the context of massive elliptical galaxies, \cite{MLi2020ApJa,MLi2020ApJb} \change{(hereafter \citetalias{MLi2020ApJa} and \citetalias{MLi2020ApJb}, respectively)} have modeled small patches (of size $\sim$ few$\times100~\mathrm{pc}$) of the ISM at different distances from the galactic center. They show that the  non-uniform heating by supernovae produces multiphase gas even when the net heating rate moderately exceeds the net cooling rate. They also report that the heating by SNIa is different from a uniformly injected volume-heating--the localized heating caused by SNIa drives compressive turbulence and produces large density fluctuations in the ISM. 
However, \citetalias{MLi2020ApJa,MLi2020ApJb} have not included an external gravitational field in their setup. In a stably stratified atmosphere, buoyancy effects can suppress the formation of cold gas if $t_\mathrm{ti}/t_\mathrm{ff}$ is large enough. The supernova-heated bubbles would rise against gravity and generate convective instabilities (such as the Rayleigh-Taylor instability) which can generate solenoidal turbulence and increase the mixing rate of the supernovae ejecta with the ambient ISM. Further, an overheated atmosphere would drive an out-flowing wind \citep[as reported by][in the context of the Milky Way's bulge]{Tang2009MNRASb}. Such a wind can transfer energy, mass and metals to outer regions of the galaxy. Thus an external gravitational field is expected to play a critical role in understanding the impact of Type Ia SNe on the ISM of massive galaxies.

A parallel set of calculations has studied the onset of thermal instability in stratified plasmas and the impact of ambient turbulence (and/or seed perturbations) on this process \citep[e.g.][]{Parrish2010ApJ,choudhury2016,Voit2018ApJ,Choudhury2019,Mohapatra2023MNRAS}.  This is important for understanding the origin of multiphase gas in massive galaxies, groups and clusters, and its role in fueling star formation and black hole growth.  Such calculations typically either seed initial density fluctuations of a fixed amplitude or drive turbulence on large scales. \cite{MLi2020ApJb} have shown that the properties of turbulence driven by small-scale SNIa remnants are quite different from the large-scale forcing typically employed in turbulence simulations. The remnants typically drive turbulence on sub-$100~\mathrm{pc}$ scales with a large compressive to solenoidal ratio and seed much larger density perturbations than expected from $\sigma_s$--$\mathcal{M}$ (rms Mach number) scaling relations in forced turbulence \citep{Federrath2008,federrath2010,konstandin2012}. These differences in driving can affect the mixing rate between the hot remnants and the ambient ISM, as well as the occurrence of multiphase condensation for a given set of turbulence parameters.

In this work, we aim to study the properties of SNIa energy injection in a density-stratified, $\mathrm{kpc}$-scale patch model of the ISM. We conduct hydrodynamic simulations including randomly injected supernovae, radiative cooling and shell-by-shell volume heating. In our fiducial set, we study the effect of increasing supernova injection rate on multiphase condensation, properties of turbulence in the ISM and the development of an SNIa-powered wind. We also check the dependence of our results on the ambient stratification (which control $t_\mathrm{ff}$), different initial gas density (which affect $t_\mathrm{ti}$) as well as the effect of including mass  ejected from AGB stars. %occurence of multiphase condensation, mechanisms of energy transport, %condensation criterion
We note from the outset that our study in this paper is restricted to local simulations that model a patch of the ISM/ICM.   These calculations are analogous to the extensive literature on simulations of the impact of core-collapse SNe on the ISM of disk galaxies \citep[e.g.][]{CGKim2015ApJ,Martizzi2015MNRAS,Fielding2017MNRAS}.   One important difference is that the local approximation is much less well motivated for the Type I/elliptical galaxy application because of the absence of a thin disk.   Given, however, the extensive literature on local simulations in studying the onset of thermal instability it is valuable to assess the role of Type Ia supernovae in that context before proceeding to global simulations. 

% organisation of the paper
In the next section, we introduce some theoretical background and estimate the different relevant energy sources and sinks. We state our model and numerical methods in Section \ref{sec:Methods}. In Section \ref{sec:results-fid}, we present the results from our fiducial set of simulations, where we progressively increase the SNIa heating rate keeping all other parameters constant. We study the effect of other simulation parameters and summarize the astrophysical implications of this work in Section \ref{sec:other-params-summary}. We discuss our caveats and prospects in Section \ref{sec:caveats-future} and finally conclude in Section \ref{sec:Conclusion}.

\section{Theoretical background} \label{sec:theoretical_background}
% In this section we begin by providing some estimates of the different energy and mass sources and sinks. We then provide arguments for our supernova injection model and introduce some important time-scales of the system. For an M87-like galaxy, we use the parameters from 
\subsection{Analytic estimates}\label{subsec:analytic_estimates}

\begin{subequations}
    
% \subsubsection{Cooling rate of the ISM}\label{subsec:cooling_rate}
\begin{figure}
    \centering
    \includegraphics[width=\columnwidth]{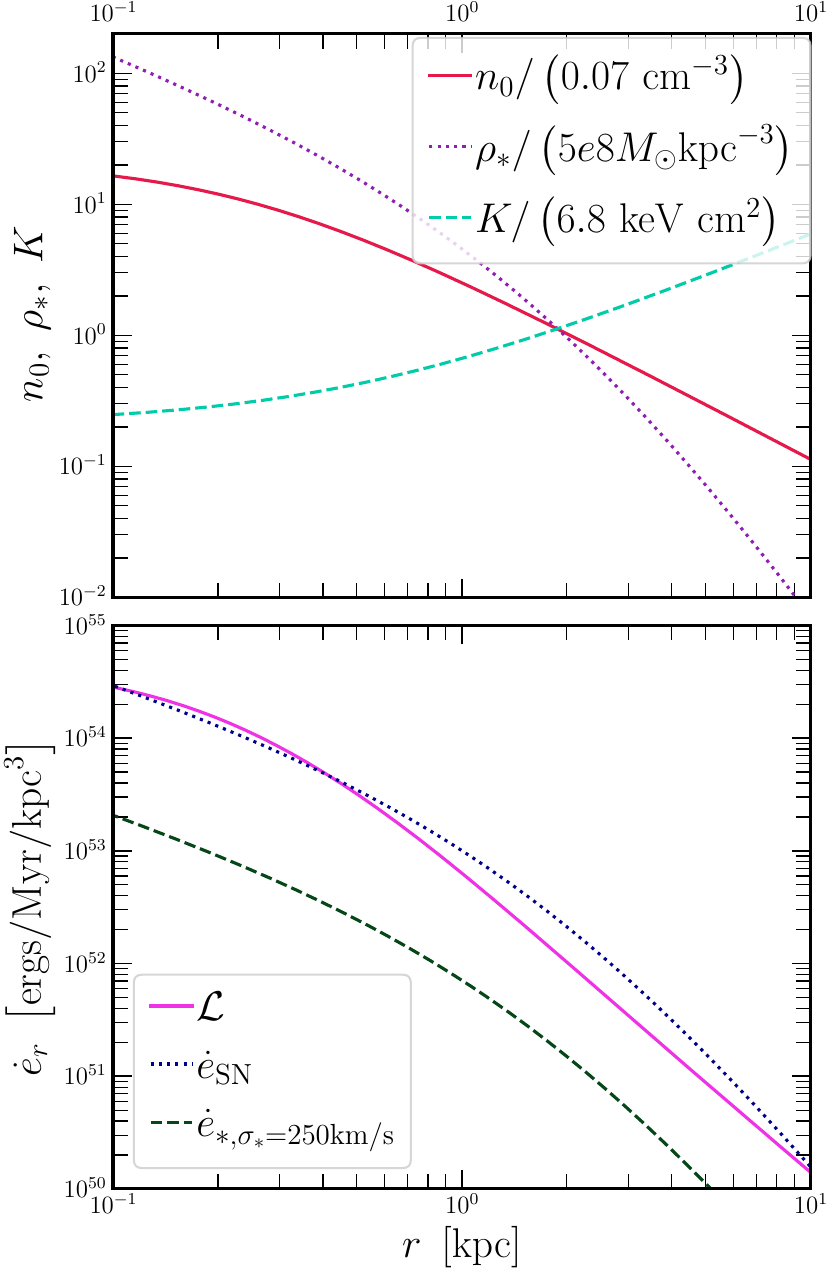}
    \caption{\emph{Upper panel:} Radial profiles of gas number density $n$, stellar mass density $\rho_*$ and entropy $S$ for a typical massive elliptical galaxy, normalized to their values at $2~\mathrm{kpc}$. \emph{Lower panel:} Radial profiles of the cooling rate density $\mathcal{L}$, supernova heating rate density $\dot{e}_\mathrm{SN}$ and stellar wind heat rate density $\dot{e}_*$, assuming the stellar velocity dispersion $\sigma_*=250\mathrm{km}s^{-1}$. }
    \label{fig:rad_profiles_analytic}
\end{figure}

In this section, we begin by providing some estimates of the different energy sources and sinks for  the typical massive elliptical galaxies studied in \cite{Werner2012MNRAS,Werner2014MNRAS}.  

\subsubsection{Radial profiles of stars, gas density and entropy}
We first obtain the radial profiles of gas number density $n$, entropy $S$ from Fig.~6 of \cite{Werner2012MNRAS}. 
% respectively, and stellar mass and effective radius $M_\mathrm{*}=3\times10^{11}~M_\sun$, $r_\mathrm{*}=2~\mathrm{kpc}$, respectively. The total enclosed mass at a radius $r$ is given by:
% \begin{align}
% &M_\mathrm{enclosed}(r) = M_\mathrm{BH}\label{eq:mass_enclosed_calc_analytic}\\
% &+ M_*\left(\ln(1+r/r_*)-\frac{r}{r+r_*}\right) \nonumber \\
% &+\frac{4\pi}{3}\rho_\mathrm{dm}r_\mathrm{dm}^3\left(\ln(1+r/r_\mathrm{dm})-\frac{r}{r+r_\mathrm{dm}}\right)   . \nonumber
% \end{align}
We normalize the gas number density $n=0.07~\mathrm{cm}^{-3}$ and entropy $S_0=6.8~\mathrm{keV}~\mathrm{cm}^2$ at $r=2~\mathrm{kpc}$, in agreement with the X-ray observations. The entropy profile flattens in the center and follows a power-law scaling $\propto r^{1}$  at large $r$.
In the upper panel of \Cref{fig:rad_profiles_analytic}, we show the radial profiles of gas number density $n$, stellar mass density $\rho_*$ and entropy $S$. 

For calculating the stellar mass distribution, we assume a Hernquist profile \citep{Hernquist1990ApJ} with $m_*=2\times10^{11}M_\sun$ \citep[see Table 1 in][for typical Bulge masses of these galaxies]{Merritt2001MNRAS} and $r_*=2~\mathrm{kpc}$ \citep[$r_*\approx r_\mathrm{eff}/1.8$, we obtain $r_\mathrm{eff}$ using mass-size relations from Fig.~1 of][]{Trujillo2011MNRAS}.

\subsubsection{Cooling rate of the hot ISM}\label{subsubsec:estimate_cooling_rate}
The ISM of an early-type galaxy is generally hot, with temperatures around $10^6$--$10^7~\mathrm{K}$. It cools through free-free Bremsstrahlung emission and far UV/X-ray line-emission. In order to calculate the cooling rate, we set the gas metallicity $Z=0.5Z_\sun$ and generate a temperature-dependent cooling table $\Lambda(T)$ using Grackle \citep{Smith2017MNRAS}. The cooling rate density $\mathcal{L}$ is given by:
\begin{equation}
    \mathcal{L}=n^2\Lambda(T).\label{eq:cooling}
\end{equation}
We show $\mathcal{L}$ as a function of $r$ in the lower panel of \Cref{fig:rad_profiles_analytic}.

\subsubsection{Heating due to SNIa}\label{subsubsec:estimate_SNIa}
%SNIa are a natural outcome of the stellar evolution channel. 
For a stellar population of age $\sim10~\mathrm{Gyr}$, the SNIa energy injection rate is given by 
\begin{equation}
    \dot{E}_\mathrm{SN}\simeq10^{51}~\mathrm{ergs} \times 300 (\mathrm{Myr})^{-1} (M_*/10^{10}~M_\sun),\label{eq:SNIa_heat_rate_estimate}
\end{equation} 
assuming that each supernova injects $E_\mathrm{SN}=10^{51}~\mathrm{ergs}$ of energy and that SNIa go off at a rate $300 (\mathrm{Myr})^{-1} (M_*/10^{10}~M_\sun)$, consistent with the delay-time distribution of SNIa in \cite{Maoz2017ApJ} \citep[also see][]{Scannapieco2005ApJ,Barkhudaryan2019MNRAS}.
We show $\dot{e}_\mathrm{SN}$ ($\dot{E}_\mathrm{SN}$ per unit volume) in the lower panel of \Cref{fig:rad_profiles_analytic}. While the exact values of the cooling rate and the SN heating rate will vary across different systems, the comparison in Figure \ref{fig:rad_profiles_analytic} motivates that the SNe heating rate is  likely within an order of magnitude of the cooling rate density within the central $10~\mathrm{kpc}$ of the ISM in massive galaxies.  This emphasizes the  potential importance of Type Ia SNe as a heating source that can compensate for  cooling losses 
%in the central regions of massive elliptical galaxies 
and affect the development of thermal instability. \emph{This is one of the main motivations for studying the effect of SNIa heating in local patches of the ISM in this paper.}

\subsubsection{Heating due to thermalisation of stellar winds}\label{subsubsec:heating_AGB_wind_rate}
Stars on the AGB eject their outer envelopes into the ISM, at a rate 
\begin{equation}
 \dot{M}_*=M_*\times 2.5\times10^{-6}\mathrm{Myr}^{-1},\label{eq:stellar_heating_norm}   
\end{equation} 
assuming a stellar age $\sim10~\mathrm{Gyrs}$ and using models of \cite{Conroy2009ApJ}. Since the stars typically have a velocity dispersion $\sigma_*\sim200$--$300~\mathrm{km}s^{-1}$ with respect to the ISM, the wind material can become thermalized with the ISM and contribute to its heating.\footnote{The AGB ejecta may not always become fully mixed with the ISM and may survive as cold gas embedded in the hot ISM, especially in high-pressure environments; see \cite{YLi2019ApJ} for a detailed study. For simplicity, we have ignored this effect in our calculations.}
The heating rate density due to the stellar winds is given by:
\begin{align}
    \dot{e}_*&=\dot{M}_*\sigma_*^2 \label{eq:AGB_wind_heat_rate_estimate}\\
    &=2.5\times10^{-6}M_*\sigma_*^2\mathrm{Myr}^{-1}.\nonumber
\end{align}
We show the radial profile of $\dot{e}_*$ in the lower panel of \Cref{fig:rad_profiles_analytic}, assuming $\sigma_*=250~\mathrm{km}s^{-1}$. The stellar heating rate density is roughly an order of magnitude smaller than the cooling rate density and the supernova heating rate density. However, it could be more important in compact galaxies with large $\sigma_*$, as discussed in \cite{Conroy2015ApJ}.

\end{subequations}
\subsection{Condition for absence of shell formation} \label{subsec:condition_shell_formation}
\begin{subequations}
In the standard picture of a supernova going off in the ISM \citep[see chapter 39 of][]{Draine2011piimbook}, the evolution of the remnant can be divided into four stages--(1) free-expansion, (2) Sedov-Taylor phase (energy conservation), (3) Snowplow phase (where a dense cooling shell forms from the swept-up material) and (4) Fadeaway (where the shock wave fades into a sound wave in pressure equilibrium with the ambient medium). In the early-type galaxies considered in this study, the ISM is hotter (temperature $T\gtrsim10^6~\mathrm{K}$) and has a lower density ($n\lesssim0.1~\mathrm{cm}^{-3}$) compared to typical conditions in a galactic disk. Under such conditions, the shock wave may reach pressure equilibrium before the snowplow phase, i.e., before the onset of significant cooling losses (\citetalias{MLi2020ApJa}).

The fade radius $R_\mathrm{fade}$, at which the shock-wave evolves into a sound wave is given by:
% Till here 24 Aug 2023
\begin{align}
    R_\mathrm{fade}&=\left(\alpha(\gamma-1)\frac{E_{\mathrm{SN}}}{4\pi P/3}\right)^{1/3} \\ \label{eq:R_fade}
    &=46.0~\mathrm{pc}\ \alpha^{1/3}(E_{\mathrm{SN}}/10^{51}~\mathrm{ergs})^{1/3} \nonumber \\
    & \times(n/0.08~\mathrm{cm}^{-3})^{-1/3}(T/3\times10^6~\mathrm{K})^{-1/3},\nonumber
\end{align}
where $P$ and $T$ are the pressure and temperature of the ambient ISM, respectively and $\alpha$ is a free parameter of order unity.  The fade radius is less than the radius at end of the Sedov-Taylor phase (when strong cooling commences) when the ambient density $n$ falls below a critical number density $n_\mathrm{crit}$, which is given by \citetalias{MLi2020ApJa}:

\begin{equation}
    n_\mathrm{crit}=7.56~\mathrm{cm}^{-3}(T/(3\times10^6~\mathrm{K})/\alpha)^{3.85}\sqrt{E_\mathrm{SN}/10^{51}~\mathrm{ergs}}.\label{eq:shell_formation} 
\end{equation}
The strongest dependence in equation \ref{eq:shell_formation} is on the temperature of the ambient medium.  Higher temperatures -- when the temperature is above the peak of the atomic cooling curve -- correspond to less efficient cooling and higher ambient pressure, both of which lead to SNR reaching pressure equilibrium prior to significant cooling losses.
The condition $n<n_\mathrm{crit}$ is satisfied for most regions of the ISM in the early-type galaxies that we study. 
%We refer the reader to Appendix B of \cite{MLi2020ApJa} for a more detailed calculation.

\end{subequations}

% \subsection{Important timescales}\label{subsec:imp_timescales}
%tti, tff, t_outflow?, t_mix?
% === Methods ===
\section{Methods}\label{sec:Methods}
\subsection{Model equations}\label{subsec:ModEq}
We use Euler equations to model the ISM as an ideal gas with an adiabatic index $\gamma=5/3$. We evolve the following equations:
\begin{subequations}
	\begin{align}
	\label{eq:continuity}
	&\frac{\partial\rho}{\partial t}+\nabla\cdot (\rho \bm{v})=\dot{n}_\mathrm{SN}M_\mathrm{SN}+\dot{\rho}_*,\\
	\label{eq:momentum}
	&\frac{\partial(\rho\bm{v})}{\partial t}+\nabla\cdot (\rho\bm{v}\otimes\bm{v})+ \nabla P=\rho \bm{g},\\% + (\cancel{ \dot{n}_\mathrm{SN}M_\mathrm{SN}}+\dot{\rho}_*)\bm{v},\\
	\label{eq:energy}
	&\frac{\partial e}{\partial t}+\nabla\cdot ((e+P)\bm{v})=-\rho (\bm{v}\cdot\nabla)\Phi \\
    &\qquad \qquad \qquad+\dot{e}_\mathrm{SN}+\dot{e}_*+\dot{e}_\mathrm{shell}-\mathcal{L}, \nonumber\\
	&e=\frac{\rho\bm{v}\cdot\bm{v}}{2} + \frac{P}{\gamma-1},\label{eq:tot_energy}
	\end{align}
where $\rho$ is the gas mass density, $\bm{v}$ is the velocity, $P=\rho k_B T/(\mu m_p)$, where $\mu\approx0.6m_p$ is the mean particle weight, $m_p$ is the proton mass and $k_B$ is the Boltzmann constant. In the continuity equation (\ref{eq:continuity}), $n_\mathrm{SN}$ denotes the rate of supernovae injection per unit volume and $M_\mathrm{SN}$ denotes the mass injected by each supernova. In addition to the mass injected by supernovae, we also include the contribution of AGB winds, denoted by $\dot{\rho}_*$ (only in a subset of our simulations).  We denote the acceleration due to gravity as $\bm{g}$ in the momentum equation (\ref{eq:momentum}). 
% The source term $(\dot{n}_\mathrm{SN}M_\mathrm{SN}+\dot{\rho}_*)\bm{v}$ ensures that the mass that we add has the same velocity as the rest of the gas in this region. 
In the energy equation (~\ref{eq:energy}), the total energy density is given by $E$ and the gravitational potential is denoted by $\Phi$, with $\bm{g}=-\nabla\Phi$. We include $\mathcal{L}$, $\dot{e}_\mathrm{SN}$ and $\dot{e}_*$ defined by \crefrange{eq:cooling}{eq:AGB_wind_heat_rate_estimate}, as well as a shell-by-shell heating rate density $\dot{e}_\mathrm{shell}$. 
% The last source term in this equation ensures that the added mass has the same specific internal and kinetic energy.
\end{subequations}

\subsection{Numerical methods}\label{subsec:numerical_methods}
We perform hydrodynamic simulations using a performance portable version of the Athena++ \citep{Stone2020ApJS} code implemented using the Kokkos library \citep{Trott2021CSE}. We use the second-order Runge-Kutta (RK2) time integrator, the Harten-Lax-van Leer-Contact (HLLC) Riemann solver and the piece-wise linear spatial reconstruction method  (PLM). For cells with unphysically large velocities or temperatures, we use a first-order flux correction algorithm described in the appendix of \cite{Lemaster2009ApJ}.

\subsubsection{Domain size and decomposition}\label{subsubsec:domain}
We simulate a cuboidal patch of the ISM with dimensions $L\times L \times 1.5L$ using a Cartesian grid, where $L=1~\mathrm{kpc}$ and the longer dimension is oriented along the $z$-direction. By default, we divide the simulation domain into a grid of size $512\times512\times768$, such that each resolution element is cubical in size (i.e. $\mathrm{d}x=\mathrm{d}y=\mathrm{d}z$).
% \footnote{We have performed all our simulations at half of the fiducial resolution (ie $256^2\times384$) to check for the convergence of our results.}. 
Our simulation box is centered at the origin $(0,0,0)$. 

\subsubsection{Boundary conditions}\label{subsubsec:boundary_conditions}
Similar to \cite{Mohapatra2023MNRAS} (hereafter \citetalias{Mohapatra2023MNRAS}), we use periodic boundary conditions along the $x$ and $y$ directions. The proper boundary conditions in the vertical direction for this problem are non-trivial in that if a SNe driven outflow develops, one would like boundary conditions that allow such an outflow (and allow the density and pressure at the boundary to change).   If an outflow does not develop, the boundary conditions should maintain hydrostatic equilibrium with densities and pressures similar to the initial condition.  After experimentation, we settled on a compromise in which we use constant boundary conditions for $\rho$ and $P$ at the $z$ boundaries. At the upper $z$ boundary, we implement diode boundary conditions for the $z$-direction velocity $v_z$. At the lower boundary, we set $v_z=0$ for gas with $T\gtrsim T_\mathrm{floor}$ and diode otherwise. Setting $v_z=0$ at the lower boundary prevents the atmosphere from undergoing a global collapse and the second condition prevents cold infalling gas from gathering at the negative $z$-direction.

To minimize the impact of the boundary conditions on our results, we treat the regions beyond $\abs{z}>0.5L$ as boundary regions and do not include them while analyzing the results of our simulations. 

\subsection{Problem setup}\label{subsec:setup}

\subsubsection{Initial density and pressure profiles} \label{subsubsec:init_dens_pres_profiles}
We set up a gravitationally stratified atmosphere with a constant $\bm{g}$  oriented along the $-\hat{\bm{z}}$ direction. At time $t=0$, pressure and density follow exponentially decreasing profiles along the $+\hat{\bm{z}}$ direction and the gas is at hydrostatic equilibrium. The initial $z$-profiles are  given by:
\begin{subequations}
\begin{align}
        &P(t=0)=P_0\exp(-\frac{z}{H}) \text{,} 
        \label{eq:init_pres}\\
        &\rho(t=0)=\frac{P(t=0)}{gH} \text{, where} \label{eq:init_dens}
\end{align}
$H$ is the scale height of pressure/density and $P_0$, $\rho_0$ ($=P_0/gH$) are the initial values of pressure and density at $z=0$, respectively. 
\end{subequations}
\subsubsection{Supernova injection}\label{subsubsec:SN_injection}
The supernovae heating rate density $\dot{e}_\mathrm{SN}$ is given by:
\begin{subequations}
\begin{align}\label{eq:sn_heating_function}
&\dot{e}_\mathrm{SN} = \dot{n}_{\mathrm{SN}}E_{\mathrm{SN}}\\
&\text{In our local ISM patch simulations, we set} \nonumber\\
&\dot{n}_{\mathrm{SN}}=\dot{n}_0\exp(-z/H_{\mathrm{SN}}),\\
&\text{such that the net heating rate due to supernovae}\nonumber\\ 
&\dot{N}_\mathrm{SN}E_\mathrm{SN}=f_\mathrm{SN}\int \mathcal{L}_0\mathrm{d}V\text{, where} \label{eq:supernova_rate}\\
&\dot{N}_\mathrm{SN}=\int \dot{n}_\mathrm{SN}\mathrm{d}V.
\end{align}
and $f_\mathrm{SN}$ is a parameter that sets the normalization of the SN heating rate.  In the above set of equations, $H_\mathrm{SN}$ denotes the scale height of the stellar/supernova distribution, $\mathcal{L}_0$ is the cooling rate at $t=0$ and $f_\mathrm{SN}$ is a parameter that we vary across simulations. We set $H_\mathrm{SN}=0.5H$ so that the heating rate by supernovae has the same variation with $z$ as $\mathcal{L}$ ($\mathcal{L}\propto \rho^2 \propto \exp(-z/0.5H)$), motivated in part by the observations in Figure \ref{fig:rad_profiles_analytic}.   As we shall show, even with an initial condition balancing heating and cooling both globally and as a function of z, the simulations with significant Type Ia heating tend to rapidly blow a significant amount of gas out of the box.   We tried other values of $H_\mathrm{SN}/H$ as well and found that in those cases the transition to rapid blowout happened even more quickly.  

We fix $f_\mathrm{SN}$ at $t=0$, which sets $\dot{N}_\mathrm{SN}$. The expected number of supernovae in a time step $\mathrm{d}t$ is given by $\dot{N}_\mathrm{SN}\mathrm{d}t$. We draw the total number of supernovae in a given time-step from a Poisson distribution with this mean. We obtain the $x$ and $y$ coordinates of the center of the remnant from a uniform distribution and use an exponential distribution for the $z$ coordinate.

The SNIa remnants in the ISM of an elliptical galaxy are not expected to undergo a snowplow phase (see sec.~\ref{subsec:condition_shell_formation}) and remain in the energy-conserving phase until they fade away into sound waves. Once the center of an individual remnant is determined, we inject $E_\mathrm{SN}=10^{51}~\mathrm{ergs}$ of energy as thermal energy and $M_\mathrm{SN}=M_\sun$ of mass inside a sphere of radius $0.6\times R_\mathrm{fade}$. Since our simulations have a spatial resolution of $\sim2~\mathrm{pc}$, each remnant is resolved by $\gtrsim10$ cells.

\subsubsection{Heating due to AGB winds}\label{subsubsec:AGB_wind_injection}
In a few of our simulations, we have included the effect of mass and energy added to the ISM due to ejecta from AGB stars. 
We assume that the AGB-wind material is thermalized with the ambient ISM. After setting $\dot{N}_\mathrm{SN}$ using \cref{eq:supernova_rate}, we obtain $\dot{M}_*$ and $\dot{e}_*$ using \crefrange{eq:SNIa_heat_rate_estimate}{eq:AGB_wind_heat_rate_estimate} and $\sigma_*=250~\mathrm{km}s^{-1}$.
We inject mass and energy into the ISM with the same $z$-dependence as the supernova injection ($\propto\exp(-z/H_\mathrm{SN})$). Although we do not include the effect of AGB winds in our fiducial set, we discuss their impact in Appendix \ref{app:effect_AGB_wind}.

\subsubsection{Modifications to the cooling function}\label{subsubsec:mod_cool_func}
% To control the code-time step set by the cooling function, we reduce $\mathcal{L}$ to zero for $n>n_\mathrm{ceiling}$ or $T<T_\mathrm{cutoff}$. The modified cooling function is given by:

% \begin{equation}
%     \mathcal{L}=n^2\Lambda(T)\exp(-10(T_\mathrm{floor}/T)^4)\mathcal{H}\left(n_\mathrm{ceiling}-n\right),\label{eq:cooling_func_modified}
% \end{equation}
% where we set $T_\mathrm{floor}=3\times10^4~\mathrm{K}$,  $n_\mathrm{ceiling}=100~\mathrm{cm}^{-3}$ and $\mathcal{H}$ is the Heaviside function. 

To control the code-time step set by the cooling function, we reduce $\mathcal{L}$ to zero for $T<T_\mathrm{cutoff}$. The modified cooling function is given by:

\begin{equation}
    \mathcal{L}=n^2\Lambda(T)\exp(-10(T_\mathrm{floor}/T)^4),\label{eq:cooling_func_modified}
\end{equation}
where we set $T_\mathrm{floor}=3\times10^4~\mathrm{K}$. 

\subsubsection{Thermal heating rate and shell-by-shell energy balance}\label{subsubsec:shell_balance}
To study the impact of different fractions of supernova heating with respect to the net cooling, as well as to prevent a runaway cooling flow, we implement an additional heating source $\dot{e}_\mathrm{shell}$ which replenishes $(1-f_\mathrm{SN})$ fraction of the energy lost due to cooling of hot gas ($T\geq T_\mathrm{hot}=10^{5.5}~\mathrm{K}$) in every $z$-shell at each time-step. This method is a useful toy model that is agnostic to the origin of other sources of heating (such as AGN, radiation, cosmic rays, conduction, etc.) and has been used in many previous studies, such as \cite{sharma2012thermal,Choudhury2019} and \citetalias{Mohapatra2023MNRAS}. Despite its simple prescription, it leads to similar results as feedback heating \citep{Gaspari2012,prasad2015}. Mathematically, this is given by:

\begin{equation}
    \dot{e}_\mathrm{shell}=\frac{\rho(x,y,z,t)\int\mathcal{L}(1-f_\mathrm{SN})\exp\left(-10(T_\mathrm{hot}/T)^{4}\right)\mathrm{d}x\mathrm{d}y}{\int\rho\mathrm{d}x\mathrm{d}y}. \label{eq:shell_heating}
\end{equation}

\end{subequations}

\begin{subequations}

\subsubsection{Important time-scales}\label{subsubsec:time_scales}
The different relevant time-scales of our setup are--
\begin{align}
    &\text{The Brunt-V\"{a}is\"{a}l\"{a} oscillation time:}\nonumber\\
    &t_\mathrm{BV}=\sqrt{\gamma H_S/g}, \text{ where } H_S=(\gamma-1)H,\label{t_BV}\\
    &\text{the turbulent mixing time:}\nonumber\\
    &t_\mathrm{mix}=\ell_\mathrm{int,\change{sol}}/v_\mathrm{int,\change{sol}}, \label{eq:t_mix}\text{ where }\\  &\ell_\mathrm{int,\change{sol}}=2\pi\frac{\int k^{-1}E_\mathrm{\change{sol}}(k)\mathrm{d}k}{\int E(k)_\mathrm{\change{sol}}\mathrm{d}k},\nonumber\\
    &\text{the gravitational free-fall time:}\nonumber\\
    &t_\mathrm{ff}=\sqrt{2H/g} \label{eq:t_ff},\\
    &\text{the cooling time:}\nonumber\\
    &t_\mathrm{cool}=\frac{nk_BT}{(\gamma-1)n^2\Lambda(T)} \label{eq:t_cool} \text{,}\\
    &\text{the thermal instability growth time:}\nonumber\\
    &t_\mathrm{ti}=\frac{\gamma t_\mathrm{cool}}{2-\mathrm{d}\ln{\Lambda}/\mathrm{d}\ln{T}-\alpha_{\mathrm{heat}}}\label{eq:t_ti}\\ 
    &\approx \gamma t_\mathrm{cool}\text{ at } T=T_0, \nonumber\\
    &\text{using }(\mathrm{d}\ln{\Lambda}/\mathrm{d}\ln{T})_{T=T_0}\approx0 \text{ and }\alpha_{\mathrm{heat}}=1.\nonumber\\
    % &\text{the Rayleigh-Taylor instability growth time:}\nonumber\\
    % &t_\mathrm{RT}= \sqrt{\ell_\mathrm{int}/Ag}\text{, where } A=\frac{n-n_\mathrm{bub}}{n+n_\mathrm{bub}}\label{eq:t_RT} \text{,}\\
    % &\text{the Kelvin-Helmholtz instability growth time:}\nonumber\\
    % &t_\mathrm{KH}= \frac{n+n_\mathrm{bub}}{\sqrt{n_\mathrm{bub}n}}\frac{\ell_\mathrm{int}}{v_\mathrm{bub}}\label{eq:t_KH} \text{ and}\\
    % &\text{the bubble disruption time due to ram-pressure:}\nonumber\\
    % &t_\mathrm{RP}= \frac{2n_\mathrm{bub}}{\mathrm{d}n/\mathrm{d}z}\frac{1}{v_\mathrm{bub}}\label{eq:t_ram_pres} \text{.}
\end{align}
In the above set of equations $H_S$ is the scale height of entropy, $k$ is the wave number, \change{$E_\mathrm{\change{sol}}(k)$ is the power spectrum of the divergence-free (or solenoidal) component of the velocity field, $v_\mathrm{int,\change{sol}}$ is the solenoidal component of velocity on the integral scale  $\ell_\mathrm{int,\mathrm{sol}}$}, $\alpha_\mathrm{heat}$ is defined such that $\dot{e}_\mathrm{shell}\propto\rho^\alpha_\mathrm{heat}$, and $n_\mathrm{bub}$ and $v_\mathrm{bub}$ are the number density and the velocity of the bubble inflated by a supernova remnant, respectively.
\end{subequations}

\subsection{Initial conditions}\label{subsec:init_conditions}
We conduct a variety of simulations to model small patches of the ISM of an elliptical galaxy at different locations away from the galactic center. For all our simulations, we initialize the gas at a constant initial temperature $T_0=3\times10^6~\mathrm{K}$, which is consistent with typical ISM temperatures in elliptical galaxies \citep[e.g., see Fig.~1 in][]{Voit2015ApJ803L21V}. For our fiducial set of simulations, we set $n_0=0.08~\mathrm{cm}^{-3}$ \change{(at roughly $2~\mathrm{kpc}$ out in Fig~\ref{fig:rad_profiles_analytic})}, motivated by observations of gas a few kpc from the center of massive galaxies\change{, see for eg. Fig.~1 in \cite{Voit2015ApJ803L21V}}.  At $t=0$, we seed density fluctuations $\delta\rho/\rho$ in the gas, with $20\%$ amplitude and a flat power spectrum (i.e. no scale dependence). These are consistent with measurements of X-ray brightness fluctuations \citep[e.g., see][albeit on somewhat larger scales compared to our setup]{zhuravleva2018}. 
Previous studies such as \cite{Choudhury2019,Voit2021ApJ}; \citetalias{Mohapatra2023MNRAS} have shown that a stratified, thermally unstable medium is expected to become multiphase if the ratio $t_\mathrm{ti}/t_\mathrm{ff}$ is becomes smaller than a $\delta\rho/\rho$-dependent threshold. We choose the density/pressure scale height $H=12~\mathrm{kpc}$ such that the initial $t_\mathrm{ti}/t_\mathrm{ff}$ is close to the condensation threshold proposed by \citetalias{Mohapatra2023MNRAS}. Since we expect the supernova-driving to generate further density fluctuations in the ISM (as seen in \citetalias{MLi2020ApJb}), our choice of initial conditions lets us directly study their impact on multiphase condensation. \change{Our choice of $H$ and $T$ are also consistent with the observed profiles of density and temperature (at a few~$\mathrm{kpc}$ from the center) in massive galaxies that host extended multiphase filaments \citep[see blue scatter points in Fig.~1 of][for reference]{Voit2015ApJ803L21V}.}

\subsection{List of simulations}\label{subsec:list_of_simulations}

\begin{table*}
	\centering
	\def\arraystretch{1.5}
	\caption{Simulation parameters and volume-averaged quantities for different runs}
	\label{tab:sim_params}
    \hspace{-1in}
	\resizebox{2.4\columnwidth}{!}{
		\begin{tabular}{lccccccccccr} 
			\hline
			Label                    & $n_0$                & $H$              & $\dot{N}_\mathrm{SN}$ & $t_{\mathrm{mp}}$ & $v$               & $\mathrm{Fr}$ & $\mathcal{M}$   & $\mathcal{M}_{\mathrm{comp}}$ & $t_{\mathrm{ti}}/t_{\mathrm{ff}}$   & $t_{\mathrm{ti}}/t_{\mathrm{mix}}$ & $\sigma_{s,\mathrm{hot}}$\\
                                         & $(\mathrm{cm}^{-3})$ & $(\mathrm{kpc})$ & $(\mathrm{Myr}^{-1})$ & $(\mathrm{Myr})$  & $(\mathrm{km}s^{-1})$ &           &                 &                               &                                     &                                    & \\
			(1)                      & (2)                  & (3)              & (4)                   & (5)               & (6)               & (7)           & (8)             & (9)                           & (10)                                & (11)                               & (12)\\
			\hline                                                                           
                                                                                    
            $f_\mathrm{SN}0.01$          & 0.08                 &   12             & $0.327$               & $>500$            & $9.4\pm0.1$       & \change{$0.40\pm0.04$}   & $0.028\pm0.001$ & \change{$0.019\pm0.001$}               & $1.06\pm0.01$                       & \change{$0.78\pm0.01$}                      & $0.055\pm0.002$\\
            $f_\mathrm{SN}0.1$           & 0.08                 &   12             & $3.271$               & $>500$            & $15.5\pm0.2$      & \change{$0.31\pm0.03$} & $0.041\pm0.001$ & \change{$0.035\pm0.001$}               & $1.38\pm0.01$                       & \change{$1.45\pm0.05$}                        & $0.114\pm0.001$\\
            $f_\mathrm{SN}0.5$           & 0.08                 &   12             & $16.35$               & $90$              & $26.6\pm0.5$      & \change{$0.48\pm0.03$} & $0.071\pm0.001$ & \change{$0.062\pm0.001$}               & $1.9\pm0.1$                         & \change{$3.6\pm0.32$}                           & $0.204\pm0.001$\\
            $f_\mathrm{SN}0.99$          & 0.08                 &   12             & $32.39$               & $55$              & $33.2\pm0.1$      & \change{$0.61\pm0.03$}   & $0.088\pm0.001$ & \change{$0.077\pm0.001$}               & $3.0\pm0.1$                         & \change{$7.2\pm0.2$}                       & $0.306\pm0.003$\\
			\hline                                                                                                                                                                                    
                                                                                                                                                                                             
            $f_\mathrm{SN}0.01$hdens     & 0.16                 &   12             & $1.212$               & $130$             & $11.9\pm0.7$      & \change{$0.496\pm0.001$} & $0.036\pm0.001$ & \change{$0.021\pm0.001$}               & $0.64\pm0.01$                       & \change{$0.608\pm0.001$}                      & $0.115\pm0.003$\\
            $f_\mathrm{SN}0.1$hdens      & 0.16                 &   12             & $12.12$               & $70$              & $18.4\pm0.3$      & \change{$0.49\pm0.01$} & $0.054\pm0.001$ & \change{$0.044\pm0.001$}               & $0.80\pm0.01$                       & \change{$1.19\pm0.01$}                        & $0.200\pm0.001$\\
            $f_\mathrm{SN}0.5$hdens      & 0.16                 &   12             & $60.59$               & $40$              & $30.2\pm0.4$      & \change{$0.7\pm0.1$}   & $0.086\pm0.003$ & \change{$0.075\pm0.003$}               & $1.6\pm0.1$                         & \change{$4.2\pm0.2$}                           & $0.36\pm0.01$\\
            $f_\mathrm{SN}0.99$hdens     & 0.16                 &   12             & $120.0$               & $25$              & $36.9\pm0.5$      & \change{$0.9\pm0.1$}   & $0.103\pm0.004$ & \change{$0.091\pm0.004$}               & $2.0\pm0.1$                         & \change{$6.9\pm0.3$}                           & $0.43\pm0.04$\\
            \hline                                                                                                                                                                           
            $f_\mathrm{SN}0.01$ldens     & 0.04                 &   12             & $0.077$               & $>500$            & $6.5\pm0.1$       & \change{$0.44\pm0.01$} & $0.018\pm0.001$ & \change{$0.014\pm0.001$}               & $1.95\pm0.01$                       & \change{$1.06\pm0.01$}                      & $0.038\pm0.001$\\
            $f_\mathrm{SN}0.99$ldens     & 0.04                 &   12             & $7.645$               & $>500$            & $31.46\pm0.01$    & \change{$0.44\pm0.1$}   & $0.055\pm0.001$ & \change{$0.052\pm0.001$}               & $4.5\pm0.5$                         & \change{$9.4\pm0.7$}                           & $0.207\pm0.05$\\
            \hline                                                                                                                                                                           
            $f_\mathrm{SN}0.01H6$hdens   & 0.16                 &   6              & $1.245$               & $>500$            & $11.8\pm0.3$      & \change{$0.445\pm0.005$} & $0.034\pm0.001$ & \change{$0.020\pm0.002$}               & $1.21\pm0.02$                       & \change{$0.67\pm0.01$}                      & $0.060\pm0.001$\\
            $f_\mathrm{SN}0.99H6$hdens   & 0.16                 &   6              & $123.3$               & $25$              & $41\pm1$          & \change{$0.73\pm0.01$} & $0.110\pm0.001$ & \change{$0.094\pm0.001$}               & $3.7\pm0.3$                         & \change{$6.7\pm0.4$}                           & $0.369\pm0.001$\\
            \hline                                                                                                                                                                           
            $f_\mathrm{SN}0.01H3$        & 0.08                 &   3              & $0.321$               & $>500$            & $11.04\pm0.02$    & \change{$0.283\pm0.001$} & $0.032\pm0.001$ & \change{$0.017\pm0.001$}               & $3.72\pm0.01$                       & \change{$0.907\pm0.002$}                      & $0.044\pm0.001$\\
            $f_\mathrm{SN}0.99H3$        & 0.08                 &   3              & $31.79$               & $>500$            & $41.2\pm0.2$      & \change{$0.34\pm0.01$} & $0.102\pm0.001$ & \change{$0.089\pm0.001$}               & $5.5\pm0.4$                         & \change{$4.4\pm0.3$}                       & $0.235\pm0.005$\\
            \hline                                                                                                                                                                           
            $f_\mathrm{SN}0.01\dot{m}_*$ & 0.08                 &   12             & $0.327$               & $>500$            & $8.45\pm0.03$     & \change{$0.54\pm0.04$}   & $0.028\pm0.001$ & \change{$0.016\pm0.001$}               & $1.15\pm0.01$                       & \change{$0.93\pm0.01$}                        & $0.062\pm0.001$\\
            $f_\mathrm{SN}0.99\dot{m}_*$ & 0.08                 &   12             & $32.39$               & $40$              & $32.4\pm0.2$      & \change{$0.7\pm0.1$}   & $0.088\pm0.002$ & \change{$0.077\pm0.001$}               & $2.69\pm0.05$                       & \change{$6.88\pm0.05$}                       & $0.315\pm0.005$\\
			
            \hline
	\end{tabular}}
    \vspace{1.0em}
	\justifying \\ \begin{footnotesize} Notes: \emph{Column 1} shows the simulation label. The number following $f_\mathrm{SN}$ denotes the ratio of the supernova energy injection rate to the net cooling rate at $t=0$. The default initial number density $n_0$ is set to $0.08~\mathrm{cm}^{-3}$ unless specified as `hdens' with $n_0=0.16~\mathrm{cm}^{-3}$ or `ldens' with $n_0=0.04~\mathrm{cm}^{-3}$, respectively and are listed in \emph{column 2}. The runs with `$\dot{m}_*$' in the label include mass and energy injection into the ISM by AGB winds. In \emph{column 3}, we show the density/pressure scale height $H$ which is set to $12$ ($=12~\mathrm{kpc}$) for most of our runs ($H$ is denoted in the simulation label otherwise). We list $\dot{N}_\mathrm{SN}$, the total number of supernovae injected in the simulation per $\mathrm{Myr}$ in \emph{column 4}. In \emph{column 5}, we show the multiphase gas formation time $t_\mathrm{mp}$ ($>500$ if the gas in the simulation remains in a single phase).  We show mass-weighted volume averages of the standard deviation of velocity $v$ in \emph{column 6} and dimensionless turbulence properties--the Froude number $\mathrm{Fr}$, the Mach number $\mathcal{M}$ and the compressive component of the Mach number $\mathcal{M}_\mathrm{comp}$ in \emph{columns 7, 8 and 9}, respectively. In \emph{columns 10 and 11}, we list the ratio of thermal instability growth time-scale $t_\mathrm{ti}$ to the free-fall time-scale $t_\mathrm{ff}$ and the turbulent mixing time-scale $t_\mathrm{mix}$, respectively. Finally, we show the amplitude of logarithmic density fluctuations in the hot phase in \emph{column 12}. All averaged quantities/statistics (\emph{columns 6-12}) are calculated for gas with $10^6~\mathrm{K}\leq T\leq3\times10^7~\mathrm{K}$ within the inner cube ($\abs{x},\abs{y},\abs{z}\leq0.5$), and averaged over a $15~\mathrm{Myr}$ duration before $t_\mathrm{mp}$ (or $t_\mathrm{end}$ if $t_\mathrm{mp}>t_\mathrm{end}=500~\mathrm{Myr}$).\end{footnotesize} 
\end{table*}

For this study, we conducted a total of 16 simulations which are listed in Table \ref{tab:sim_params}. 
We mainly study the effect of varying the fraction of supernova heating rate with respect to the net cooling rate. For our fiducial set, we conduct four simulations with $f_\mathrm{SN}=0.01$, $0.1$, $0.5$ and $0.99$, as indicated in their respective simulation labels. \change{In addition to understanding the effects of resolved SNIa heating, our parameter choices are motivated by the inferred ratio between the heating and cooling rates observations of massive elliptical galaxies \citep{Werner2012MNRAS,Werner2014MNRAS,Voit2015ApJ803L21V}. For example, we find that in massive ellipticals such as NGC5846, NGC5044 which have a flat gas entropy profile, beyond the inner $10~\mathrm{kpc}$ the stellar density (and as a result the SNIa heating rate) drops off much faster with increasing radius compared to the net cooling rate. So heating due to the SNIa could dominate in the core, but other heating sources are expected to contribute in the outskirts.}

To check the effect of a faster cooling rate, we repeat the fiducial set with double the initial density ($n_0=0.16~\mathrm{cm}^{-3}$) and smaller initial seed density fluctuations ($\delta\rho/\rho=0.05$). These runs are indicated by `hdens' in the simulation label. 

For all our other parameter choices, we conduct a pair of simulations--one with a small supernova heating fraction ($f_\mathrm{SN}=0.01$) and another with a large supernova heating fraction ($f_\mathrm{SN}=0.99$). \change{The observed galaxies are have different densities, temperatures and strengths of stratification, which also vary across the inner and outer regions of the ISM.} The `ldens' runs are useful to study the effect of a weaker cooling rate due to their smaller initial gas density. We conduct a pair of simulations with smaller density/pressure scale heights--`$H6$hdens' to look for the effect of stronger stratification on the `hdens' runs. Our `$H3$' runs have initial density profiles that are directly comparable to that of a massive elliptical galaxy (see Fig.~\ref{fig:rad_profiles_analytic}), noting that we use a slightly cooler $T_0$. Finally, the `$\dot{m}$' runs include mass and energy injection due to mass loss from AGB stars, as described in Section \ref{subsubsec:AGB_wind_injection} (all other runs do not account for AGB winds). 

We evolve all the simulations till $t_\mathrm{end}=500~\mathrm{Myr}$. We define $t_\mathrm{mp}$ as the time at which the fraction of gas mass at $T<10^{4.2}~\mathrm{K}$ exceeds $1\%$ for the first time. For all simulations that have more than $1\%$ of their total mass in the cold phase, we list $t_\mathrm{mp}$ in column 5 of Table \ref{tab:sim_params}, otherwise we denote $t_\mathrm{mp}>500~\mathrm{Myr}$. In columns 6--12 we present different statistical properties of the simulations, averaged over the last $15~\mathrm{Myr}$ of their evolution before $t_\mathrm{mp}$ (or $t_\mathrm{end}$ if they never have $M_\mathrm{cold}/M_\mathrm{tot}>0.01$ during the simulation). For simulations where $t_\mathrm{mp}\leq30~\mathrm{Myr}$, we present the statistical properties for $15~\mathrm{Myr}\leq t<t_\mathrm{mp}$, to allow enough time for the system to evolve before we analyze them. 

We rerun all 16 simulations listed in Table \ref{tab:sim_params} at half the resolution (using $256^2\times384$ resolution elements) and find all of the simulation properties listed in columns 5--12 to be within a few$~\%$ of their high-resolution counterparts.

% ==================================
\section{Results - Fiducial runs}\label{sec:results-fid}
% Start with an eye-candy slice plot - compare between f_\mathrm{SN}0.01 and f_\mathrm{SN}0.99
We start by presenting results from our fiducial runs, where we progressively increase the fraction of the SNIa heating rate with respect to the net cooling rate. We study the effect of this parameter ($f_\mathrm{SN}$) on the distributions of thermodynamic quantities such as density and temperature, time evolution of mass and energy, formation of cold gas ($T\lesssim10^{4.2}~\mathrm{K}$) and vertical shell-averaged profiles of energy fluxes.   As we shall see, an important feature of our results is that when Type Ia SNe heating is important (larger $f_\mathrm{SN}$), there is no local equilibrium between SN heating and cooling possible.   SNe eventually drive gas out of the box, decreasing the cooling rate, leading to stronger SNe driven wind, etc.   The longer-timescale outcome of this inevitable instability cannot be studied in our local approximation and will ultimately require global simulations.

% After introducing properties of the fiducial set, we then present the effects of varying other simulation parameters on our results, such as increasing/decreasing the average initial density (which controls $t_\mathrm{ti}$), increasing the strength of stratification (which controls $t_\mathrm{ff}$) and including AGB winds.

% In column 4 of Table \ref{tab:sim_params}, we show the time at which different simulations have at least $1\%$ of their total mass at $T\lesssim10^{4.2}~\mathrm{K}.$ (i.e. $M_\mathrm{cold}/M_\mathrm{tot}>0.01$). We define this time as the multiphase gas-formation time $t_\mathrm{mp}$. 

\subsection{Projection maps}\label{subsec:projection_maps}
\begin{figure*}
		\centering
	\includegraphics[width=2\columnwidth]{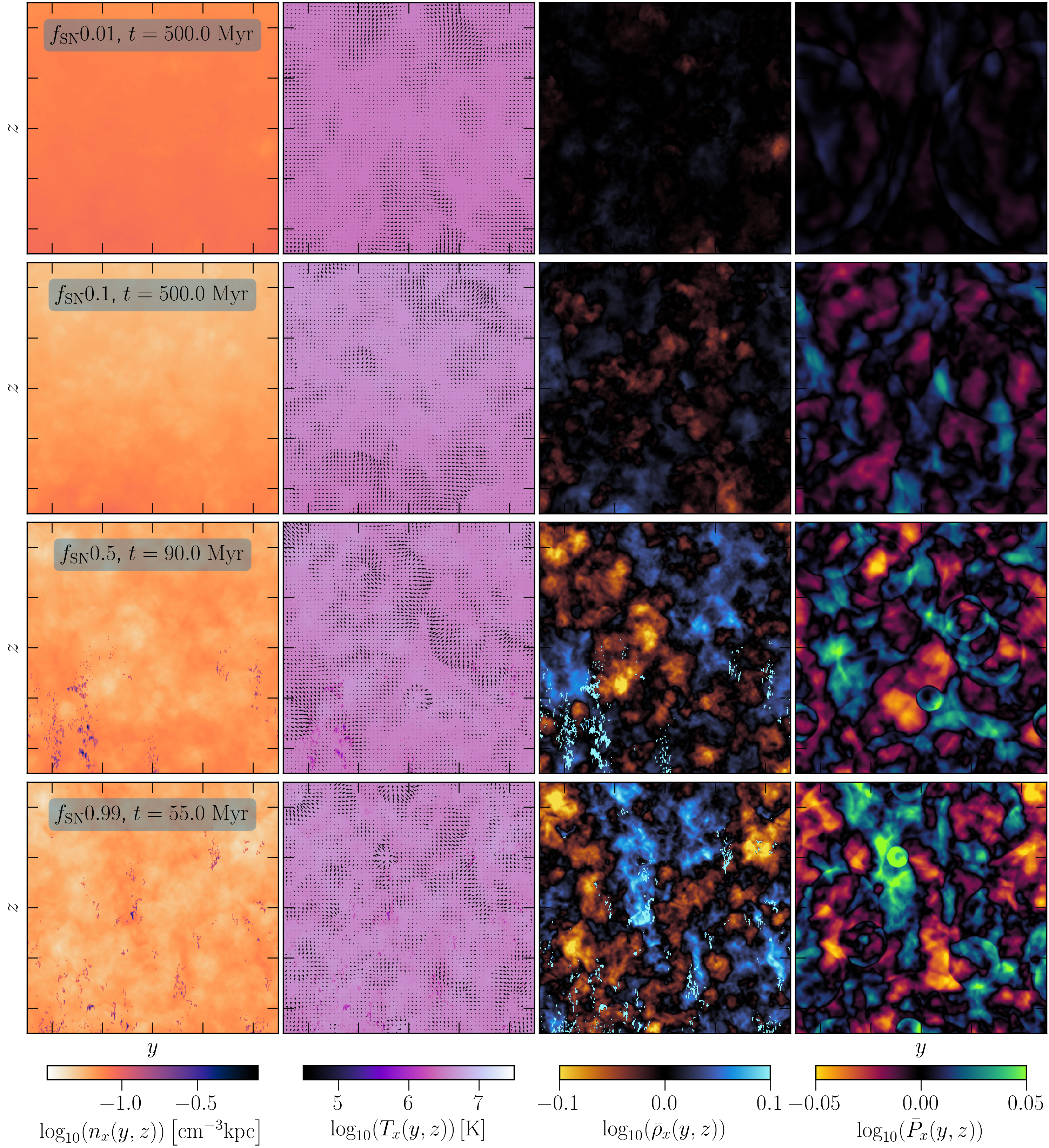}	
	\caption{Snapshots of $x$-projected quantities for our fiducial set of simulations with different fractions of heating due to supernovae at $t=t_\mathrm{mp}$ ($t=t_\mathrm{end}$ for $f_\mathrm{SN}0.01$ and $f_\mathrm{SN}0.1$ runs that do not form multiphase gas)--\emph{Column 1:} Logarithm of gas number density $n$; \emph{Column 2:} Logarithm of mass-weighted temperature $T$, \emph{Column 3:} Logarithm of $x$-projected density normalized by the $z$-profile of density; \emph{Column 4:} Logarithm of $x$-projected pressure normalized by the $z$-profile of pressure. The runs with larger $f_\mathrm{SN}$ show larger density and pressure fluctuations and are more likely to undergo multiphase condensation. The cold gas forms in small dense clumps.}
	\label{fig:2dproj_fid}
\end{figure*}

In \Cref{fig:2dproj_fid} we show snapshots of the $x$-projections of the logarithms of gas number density $\log_{10}(n_x)$ (column 1), mass-weighted temperature $\log_{10}(T_x)$ (column 2), normalized density $\log_{10}(\bar{\rho}_x)$ (column 3) and normalized pressure $\log_{10}(\bar{P}_x)$ (column 4), respectively\footnote{Here $\bar{\rho}=\rho/\mean{\rho}_z$, where $\mean{\rho}_z$ is the average $\rho$ in a $z$-shell, similarly $\bar{P}=P/\mean{P}_z$.}. For the $f_\mathrm{SN}0.01$ and $f_\mathrm{SN}0.1$ runs that do not form multiphase gas (i.e.~$M_\mathrm{cold}/M_\mathrm{tot}<0.01$ at all times), we present the snapshots at $t=t_\mathrm{end}$, whereas we show the snapshots at $t=t_\mathrm{mp}$ for the $f_\mathrm{SN}0.5$ and $f_\mathrm{SN}0.99$ runs.

Among the single phase runs ($f_\mathrm{SN}0.01$ and $f_\mathrm{SN}0.1$), we find that the density of the gas decreases with increasing $f_\mathrm{SN}$. The gas is also hotter. Since the SNIa rate is fixed throughout the course of the simulation unlike the shell-by-shell heat prescription (which self-adjusts to ($1-f_\mathrm{SN}$) fraction of the total cooling rate at each time-step), whenever the net cooling rate falls below the initial cooling rate the system overheats. The amount of overheating increases with increasing $f_\mathrm{SN}$. 

\begin{figure}
		\centering
	\includegraphics[width=\columnwidth]{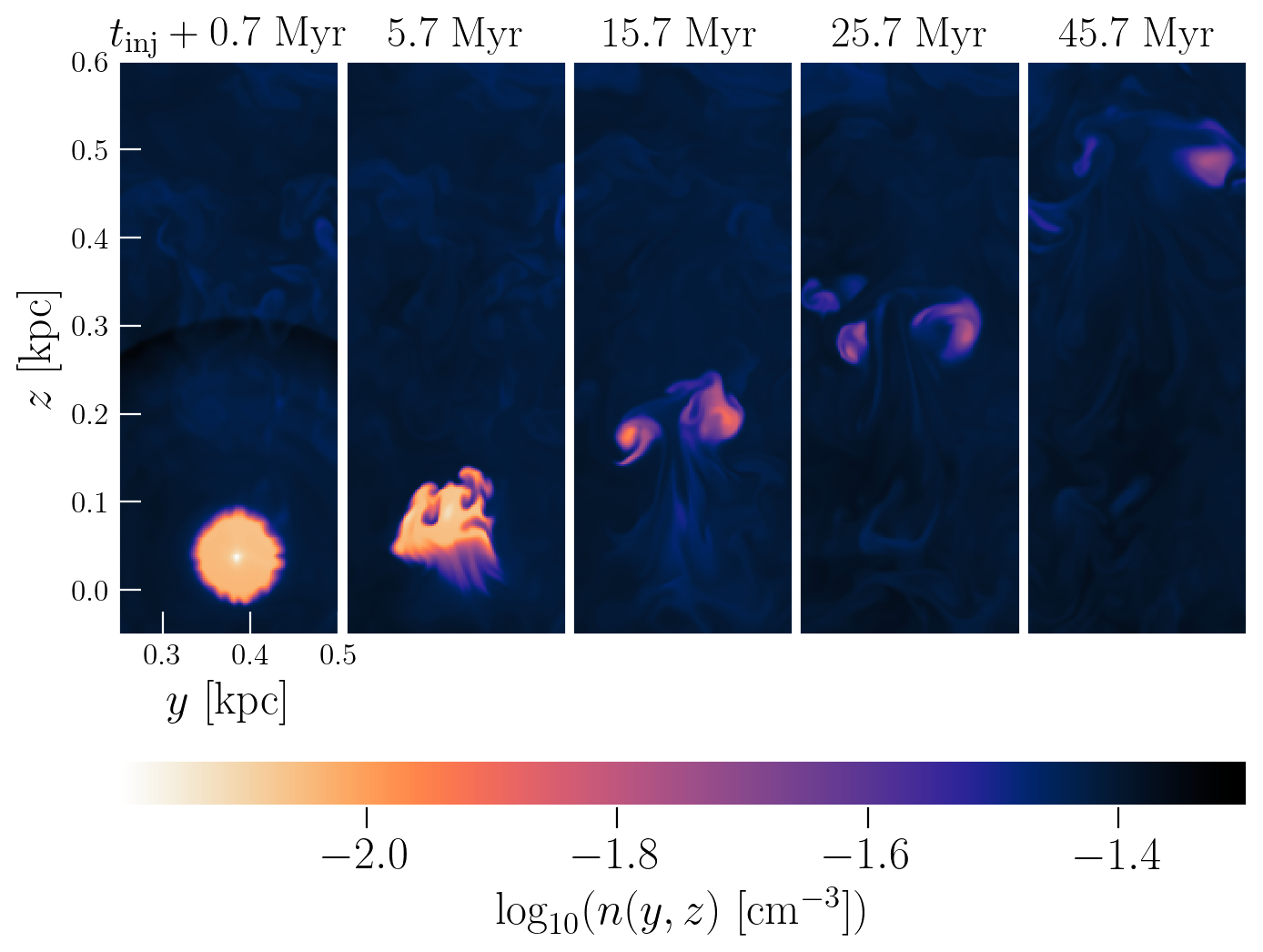}	
	\caption{Snapshots of $x$-slices of the gas number density showing the evolution (from left to right) of a supernova inflated bubble in one of our simulations. The sound wave driven by the supernova can be seen in the first column. The under-dense bubble rises buoyantly and gets disrupted due to the Rayleigh-Taylor instability. These motions drive turbulence in the ISM.}
	\label{fig:2dslice_bubble}
\end{figure}

Higher SNIa rates are also associated with larger density and pressure fluctuations. This is because the supernovae remnants are small (a few $\times10~\mathrm{pc}$-scale regions) and get overheated compared to the remainder of the ISM. The overheated regions expand and rise buoyantly (as seen in fig.~\ref{fig:2dslice_bubble}) and drive sound waves in the ISM. As these bubbles interact with the ISM, they form mushroom-shaped clouds characteristic of the Rayleigh-Taylor instability and drive turbulence in the ISM. The amplitudes of density and pressure fluctuations increase with increasing $f_\mathrm{SN}$, as seen in the right two panels of Figure \ref{fig:2dproj_fid}.

In snapshots of the multiphase runs ($f_\mathrm{SN}0.5$ and $f_\mathrm{SN}0.99$), we observe dense small-scale clouds which correspond to gas at or below the cooling cutoff temperature ($T\lesssim10^{4.2}~\mathrm{K}$). These clouds are not in hydrostatic equilibrium with the ambient ISM and rain down to the bottom of the box, forming a trail along the direction of gravity.  The structure and distribution of cold gas in supernova-driven turbulence are quite different from that of the cold clouds in turbulence driven on large box scales studied in \citetalias{Mohapatra2023MNRAS} (see their Fig.~1). In our simulations, the clouds are not volume-filling (in contrast to their natural driving runs) and do not form large, box-scale filaments (in contrast to their compressive driving runs).

\subsection{Time-evolution}\label{subsec:time_evol_plots}
\begin{subequations}
\begin{figure}
		\centering
	\includegraphics[width=\columnwidth]{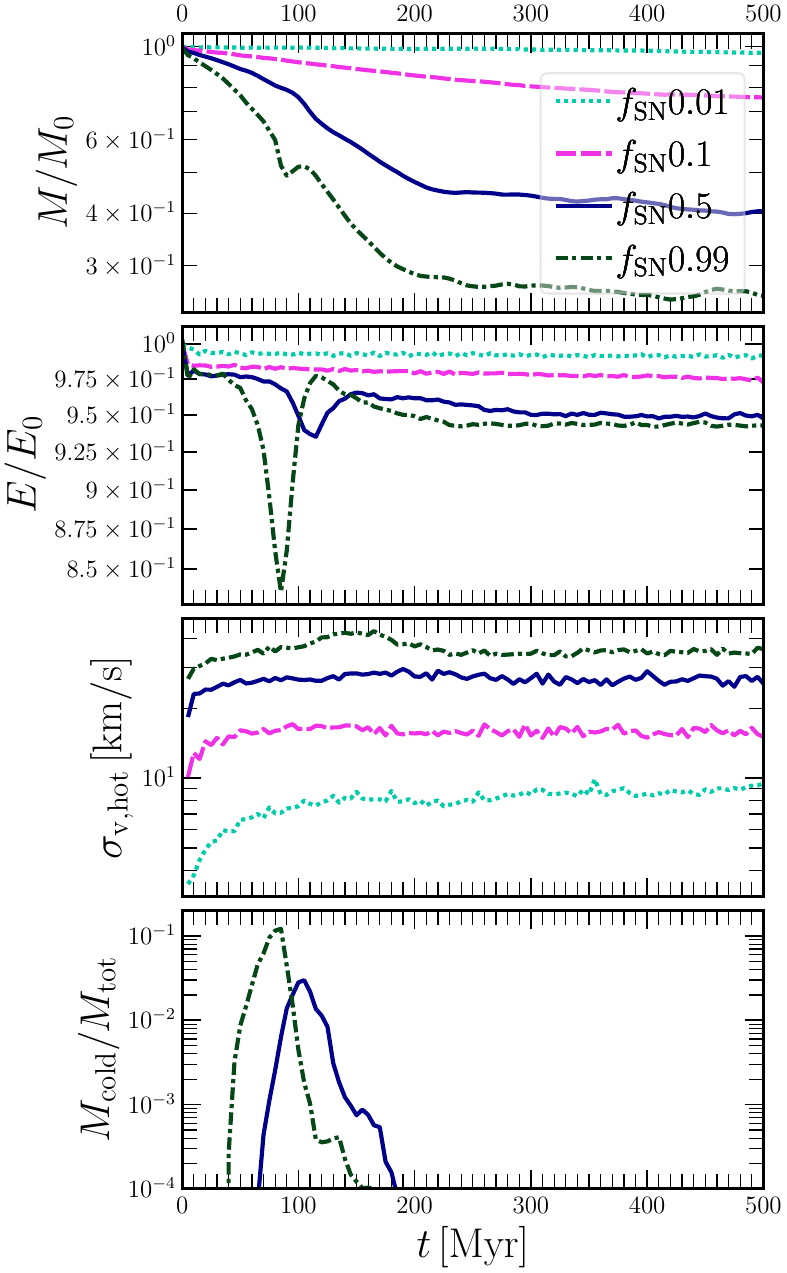}	
	\caption{Time-evolution of net mass (first row), energy (second row) normalized to their initial values, velocity dispersion ($\sigma_{v,\mathrm{hot}}$, column 3) of hot gas ($T\gtrsim10^{5.5}~\mathrm{K}$) and mass fraction of cold gas ($T\lesssim10^{4.2}~\mathrm{K}$, column 4) for our fiducial set of simulations. A higher SNIa rate (larger $f_\mathrm{SN}$) leads to a larger $\sigma_{v,\mathrm{hot}}$. Stronger driving triggers multiphase condensation in the $f_\mathrm{SN}0.5$ and $f_\mathrm{SN}0.99$ runs. Imbalanced heating due to the SNIa also gives rise to a wind. In steady state, the mass and energy within the box decrease with increasing $f_\mathrm{SN}$ due to the increasing impact of these two processes.}
	\label{fig:mass_etot_evol_fid}
\end{figure}

Here we present the time evolution of some key properties of our fiducial set of simulations. In rows 1--4 of \Cref{fig:mass_etot_evol_fid}, we show the time-evolution of net mass and energy (normalized to their values at $t=0$), the amplitude of velocity dispersion of hot gas ($\sigma_{v,\mathrm{hot}}$) and the mass fraction of cold gas ($M_\mathrm{cold}/M_\mathrm{tot}$), respectively. We first discuss the evolution of these quantities at initial times ($t<100~\mathrm{Myr}$). 

We find that the net mass decreases (almost monotonically) as a function of time. The mass loss rate increases with increasing $f_\mathrm{SN}$. The net energy also shows a similar behavior, although the energy loss rate is much smaller and the decrease is not monotonic. We observe a clear effect of different SNIa rates on $\sigma_{v,\mathrm{hot}}$, where the runs with larger SNIa rates drive stronger motions in the ISM.  \change{This is consistent with the velocity measured in observed H$\alpha$ filaments on $100~\mathrm{pc}$ scales in \cite{Li2020ApJ}. However, the cold and hot phases are not co-moving in our simulations. The condensed cold gas exits our simulation domain before it becomes entrained in the hot phase, so it does not trace the velocity structure of the hot phase.}

Only the $f_\mathrm{SN}0.5$ and $f_\mathrm{SN}0.99$ runs trigger multiphase condensation\footnote{We define a system to be multiphase if $M_\mathrm{cold}/M_\mathrm{tot}\geq0.01$ at any time, and $t_\mathrm{mp}$ as the first time that this criterion is satisfied.}. The cold gas mass fraction shows an initial increase, followed by a steep decrease. The fraction drops below $10^{-4}$ in roughly $t_\mathrm{ff}\approx80~\mathrm{Myr}$ after its peak value as the cold gas rains down through the bottom of the box. 

The $f_\mathrm{SN}0.99$ run converts a larger fraction of its mass into the cold phase at an earlier time compared to the $f_\mathrm{SN}0.5$ run. This is because the higher SNIa rate generates stronger density fluctuations where the dense regions cool faster, so the threshold $t_\mathrm{ti}/t_\mathrm{ff}$ for the simulation to remain single phase is higher (\citealt{Choudhury2019,Voit2021ApJ},\citetalias{Mohapatra2023MNRAS}). 

By $t=500~\mathrm{Myr}$, the net mass and energy reach a rough steady state and show a slow decrease with time. Their values are smaller for larger $f_\mathrm{SN}$ runs, since they lose more gas due to multiphase condensation as well as to outflows (which we will show in the next section). The motions in the hot phase have roughly the same $\sigma_{v,\mathrm{hot}}$ as initial times. The system remains in a single phase after the initial round of multiphase condensation.

% \subsection{Sources and sinks of mass and energy}\label{subsec:source_sink_mass_energy}
% The heat injected by the SNIa is highly anisotropic on small $\sim10~\mathrm{pc}$ scales, where overheated remnants are expected to rise buoyantly and drive a convective flow. Further, the SNIa rate does not depend on the net cooling rate unlike the AGN feedback loop, where the cooling and heating are expected to be coupled through a delayed feedback loop \citep{prasad2015,Tremblay2018ApJ}.
% Although the energy injection due to SNIa and the radiative cooling rate are of similar magnitude (see Fig.~\ref{fig:rad_profiles_analytic}), no local equilibrium is possible:   SNe over/underheat the ISM when the net cooling rate increases or decreases. During episodes of weaker cooling, the energy injected by the SNIa powers an outflow, at least on the scale of our local box.   
% In this subsection, we study the properties of mass and energy sources, sinks and their fluxes at the boundaries of our simulation domain.

\subsection{Mass injection by SNIa and boundary fluxes}

\begin{figure}
		\centering
	\includegraphics[width=\columnwidth]{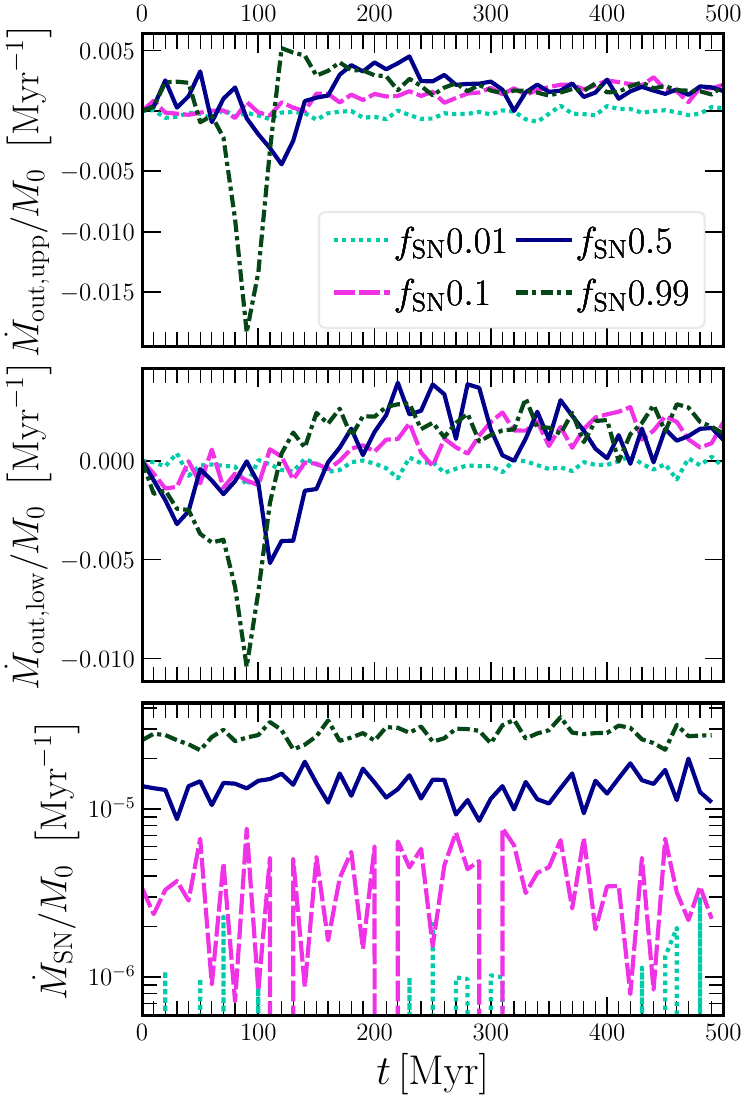}	
	\caption{Time-evolution of the mass fluxes at $z=0.5$ (upp, first row) and $z=-0.5$ (low, second row) and the supernova mass injection rate ($\dot{M}_\mathrm{SN}$, third row) for our fiducial set of simulations. The mass fluxes at the domain boundaries are an order of magnitude larger than the mass injected by the SNIa, even for the $f_\mathrm{SN}0.99$ run. Thus gas inflow/outflow at the boundaries dominates the evolution of net mass.}
	\label{fig:mass_fluxes_fid}
\end{figure}

Type Ia supernovae inject both mass and metals into the ISM. In addition to forming metals during the SNIa explosions, the wind driven by the SNIa can transport these metals to the outer regions of the galaxy/into the CGM/ICM \citep{Tang2009MNRASb}. Accounting for the contribution from SNIa is important to explain the observed radial trends of metallicity in galaxy clusters, especially for the observed amounts of \texttt{Fe} and \texttt{Ni} \citep{Gatuzz2023MNRASa,Gatuzz2023MNRASb}.\footnote{We do not evolve the metallicity of the gas in the present study and leave such a study to future work.} 
%However, we have measured the mass outflow rate at the boundaries which can provide a rough estimate if the metals are well-mixed.} 

Here we discuss the evolution of the mass fluxes and sources. The relevant sources/sinks of mass are the mass outflow rate $\dot{M}_\mathrm{out}$, and the mass injection due to the supernovae. They are defined as:
\begin{align}
    &\dot{M}_\mathrm{out} = \int\rho v_z\mathrm{d}x\mathrm{d}y     \label{eq:mass_flux} \\
    &\dot{M}_\mathrm{SN} = \int \dot{n}_\mathrm{SN}M_\mathrm{SN}\mathrm{d}V, \label{eq:net_Mdot_SN} 
\end{align}

In \Cref{fig:mass_fluxes_fid} we show the evolution of $\dot{M}_\mathrm{out}$ at the upper $z$-boundary $\dot{M}_\mathrm{out,upp}$ (row 1) and the lower $z$-boundary $\dot{M}_\mathrm{out,low}$ (row 2), respectively. In the third row, we present the mass injection rate due to the SNIa remnants $\dot{M}_\mathrm{SN}$ (row 3). All mass fluxes are divided by $M_0$, the total mass at $t=0$. 

The mass injected due to the SNIa remnants is dependent on $f_\mathrm{SN}$ as it sets the SNIa rate. However, its value is an order of magnitude (or more) lower than the mass fluxes at the boundaries and $M_0/\dot{M}_\mathrm{SN}\ll500~\mathrm{Myr}$. So $\dot{M}_\mathrm{SN}$ is too small to have a significant impact on the gas density during the course of a simulation.
On the other hand, $M_0/\dot{M}_\mathrm{out,upp}$ and $M_0/\dot{M}_\mathrm{out,low}$ range between $80$--$500~\mathrm{Myr}$ for $f_\mathrm{SN}\geq0.5$ and thus affect the overall mass distribution significantly during the course of a simulation. 

Before the formation of multiphase gas at $t=t_\mathrm{mp}$, $\dot{M}_\mathrm{out,upp}$ is positive and $\dot{M}_\mathrm{out,low}$ is negative for the $f_\mathrm{SN}0.5$ and $f_\mathrm{SN}0.99$ runs, implying that the system is losing mass due to outflows even before multiphase condensation. At $t=t_\mathrm{mp}$, we observe a sharp, correlated dip in $\dot{M}_\mathrm{out,upp}$ and $\dot{M}_\mathrm{out,low}$ for both runs, which indicates cold gas raining down and hot gas flowing in from the upper boundary to replace it. This is also associated with a sharp decrease in the net mass and energy of the box. The amplitude of this drop is larger for the larger $f_\mathrm{SN}$ run which forms more cold gas. 

The $f_\mathrm{SN}0.01$ run does not show strong outflows or inflows at any given time. The evolution of the $f_\mathrm{SN}0.1$ runs is similar to the late-time evolution of the $f_\mathrm{SN}0.5$ and $f_\mathrm{SN}0.99$ runs, where most of the material is flowing in the positive $z$-direction with a similar outflow rate. 
%The higher $f_\mathrm{SN}$ runs have a larger $v_z$ but have a smaller mean density. The product of these two is roughly constant for $f_\mathrm{SN}\geq0.1$.

\subsection{Vertical profile of density}\label{subsec:vert_prof_density}
\begin{figure}
		\centering
	\includegraphics[width=\columnwidth]{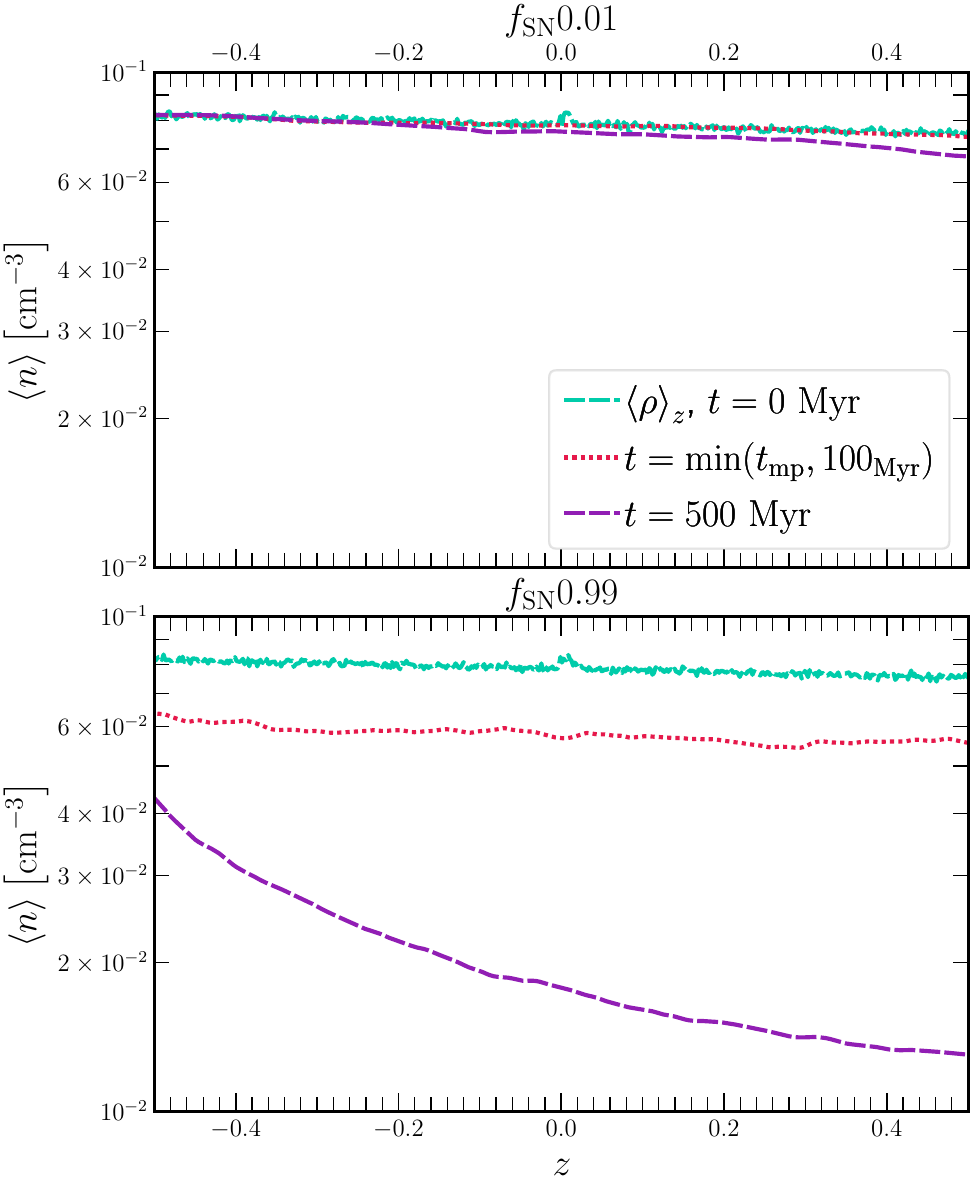}	
	\caption{The $z$-profile of number density of gas of gas with $10^6~\mathrm{K}<T<10^{7.5}~\mathrm{K}$ at $t=0~\mathrm{Myr}$, $t=\min(t_\mathrm{mp},100~\mathrm{Myr})$ and $t=500~\mathrm{Myr}$. Increasing the SNIa rate leads to multiphase condensation and a strong outflow at late times. These factors contribute to a drop in $n_0$ as well as a change in its functional form by $500~\mathrm{Myr}$.}
	\label{fig:vert_prof_dens}
\end{figure}

In \Cref{fig:vert_prof_dens}, we show the vertical profiles of the shell-averaged number density $\mean{n}$ for the $f_\mathrm{SN}0.01$ and $f_\mathrm{SN}0.99$ runs from the fiducial set at $t=0$, $\min(t_\mathrm{mp},100~\mathrm{Myr})$ and $t_\mathrm{end}$. At $t=0$, the squiggles in the profiles are due to the seed white noise of amplitude $\sigma_s=0.2$ that we add. By $t=\min(t_\mathrm{mp},100~\mathrm{Myr})$, we find that $\mean{n}$ drops while maintaining a similar dependence with $z$, as a result of the weak outflows driven at early times. 
% The fraction of decrease in $\mean{n}$ increases with increasing $f_\mathrm{SN}$ due to the increasing strength of outflows. However, $\mean{n}$ still has the functional form $\propto\exp(-z/H)$. 
We also note that most of the small-scale perturbations have disappeared, likely due to viscous dissipation. 

By $t=t_\mathrm{end}$, stronger outflows in $f_\mathrm{SN}0.99$ run lead to a further drop in $\mean{n}$. The number density profile is also steeper. This implies that the SNIa heating drives an outflow, which can affect both the total amount of gas in the ISM as well as its radial distribution. So the gas density profiles found in observations \citep[e.g.][]{Werner2012MNRAS,Werner2014MNRAS} can evolve with time due to heating by the SNIa.

\subsection{Energy sinks, sources and boundary flux}

\begin{figure}
		\centering
	\includegraphics[width=\columnwidth]{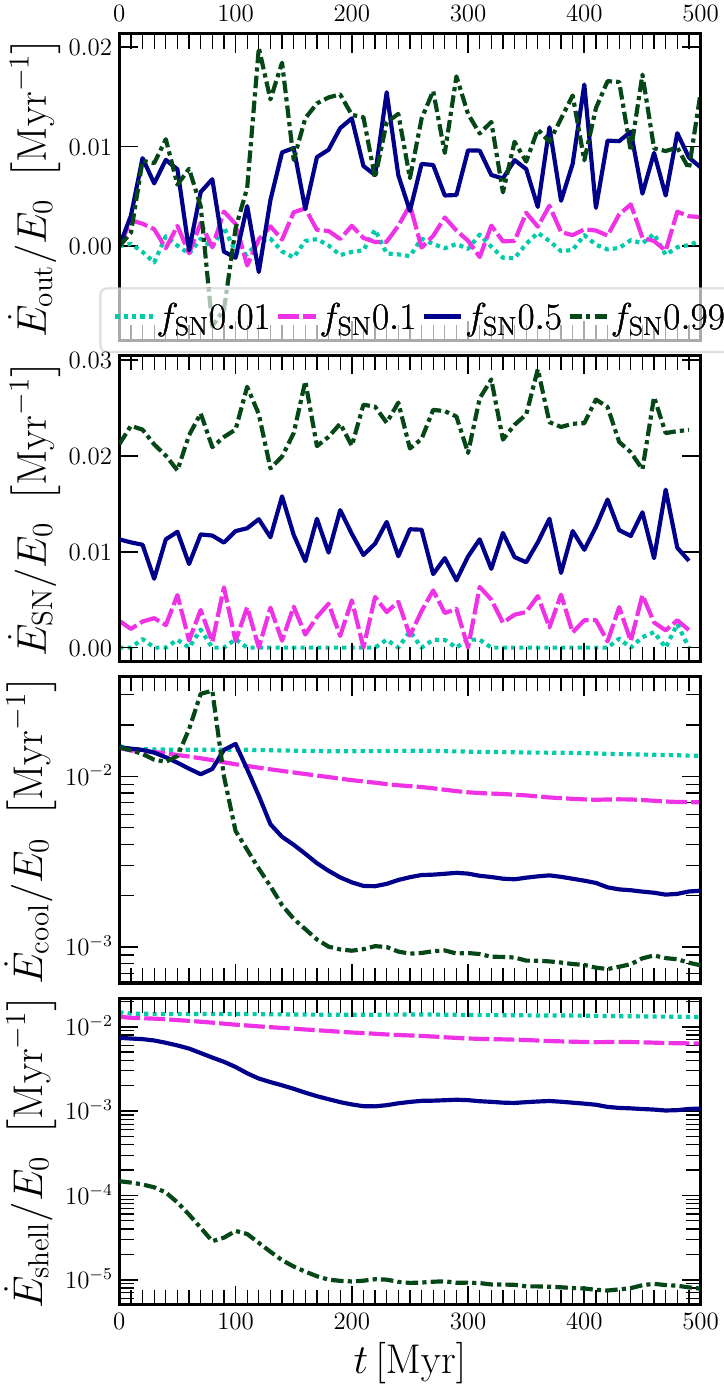}	
	\caption{Time-evolution of the energy outflow flux $\dot{E}_\mathrm{out}$ (first row), SNIa energy injection rate $\dot{E}_\mathrm{SN}$ (second row), radiative cooling rate $\dot{E}_\mathrm{cool}$ (third row) and the shell heating rate $\dot{E}_\mathrm{shell}$  sources and sinks for our fiducial set of simulations. At initial times, the supernovae energy input compensates for the net cooling and drives a weak outflow. Once the cooling rate drops (after multiphase condensation), the SNIa power a stronger outflow.}
	\label{fig:ener_evol_fid}
\end{figure}

The heat injected by the SNIa is highly anisotropic on small $\sim10~\mathrm{pc}$ scales, where overheated remnants are expected to rise buoyantly and drive a convective flow. Further, the SNIa rate does not depend on the net cooling rate unlike the AGN feedback loop, where the cooling and heating are expected to be coupled through a delayed cycle \citep{prasad2015,Tremblay2018ApJ}.
Although the energy injection due to SNIa and the radiative cooling rate are of similar magnitude (see Fig.~\ref{fig:rad_profiles_analytic}), no local equilibrium is possible:   SNe over/underheat the ISM when the net cooling rate increases or decreases. During episodes of weaker cooling, the energy injected by the SNIa powers an outflow, at least on the scale of our local box.   

We define the energy outflow rate $\dot{E}_\mathrm{out}$ and the net energy injection/loss rate densities below\footnote{Note that we have ignored $\Phi$ from our energy flux calculations because our fiducial runs have weak stratification ($H=12$) and its contribution to the total energy is small.}:
\begin{align}
    &\dot{E}_\mathrm{out} = \int\rho v_z\left(v^2/2+\frac{\gamma}{\gamma-1}P/\rho\right)\mathrm{d}x\mathrm{d}y,     \label{eq:ener_flux_tot} \\
    &\dot{E}_\mathrm{SN} = \int \dot{e}_\mathrm{SN}\mathrm{d}V = \int \dot{n}_\mathrm{SN}E_\mathrm{SN}\mathrm{d}V, \label{eq:net_Edot_SN} \\
    &\dot{E}_\mathrm{cool} = \int \mathcal{L}\mathrm{d}V = \int n^2\Lambda(T)\mathrm{d}V \text{, and} \label{eq:Edot_cool} \\
    &\dot{E}_\mathrm{shell} = \int \dot{e}_\mathrm{shell}\mathrm{d}V. \label{eq:Edot_shell} 
\end{align}

In \Cref{fig:ener_evol_fid} we present the evolution of the net outflowing energy flux ($\dot{E}_\mathrm{out,upp}-\dot{E}_\mathrm{out,low}$) (first row), $\dot{E}_\mathrm{SN}$ (second row), $\dot{E}_\mathrm{cool}$ (third row) and $\dot{E}_\mathrm{shell}$ (fourth row). All quantities are normalized by $E_0$, the net energy at $t=0$. All the fluxes, source and sink terms are significant during the course of a simulation (i.e. $E_0/\dot{E}\lesssim500~\mathrm{Myr}$). By construction, the energy injection rate due to the SNIa increases with increasing $f_\mathrm{SN}$ and does not show much variation with time.

Before $t=t_\mathrm{mp}$, we observe a net energy outflow at the boundaries. Around $t=t_\mathrm{mp}$, $\dot{E}_\mathrm{out}$ is briefly negative for the $f_\mathrm{SN}0.99$ run, possibly because cold gas exits the box from the bottom and hot gas enters from the top at this time. At late times, $\dot{E}_\mathrm{out}$ is positive and increases with increasing $f_\mathrm{SN}$, with an amplitude comparable to the SNIa heating rate. 

Since the cooling rate is proportional to the square of the gas density, its evolution follows a similar trajectory as the evolution of net mass in \Cref{fig:mass_etot_evol_fid}, barring the peak in the cooling rate at $t=t_\mathrm{mp}$. The shell-by-shell heating rate is $(1-f_\mathrm{SN})$ fraction of the radiative cooling losses at all times except for $t_\mathrm{mp}\lesssim t\lesssim t_\mathrm{mp}+t_\mathrm{ff}$ in the $f_\mathrm{SN}0.5$ and $f_\mathrm{SN}0.99$ runs. Because it only replenishes cooling losses for gas with $T\gtrsim10^{5.5}~\mathrm{K}$ (see eq.~\ref{eq:shell_heating}). Hence the absence of a peak in the net $\dot{E}_\mathrm{shell}$ corresponds to the formation of gas at intermediate and low temperatures ($T\lesssim10^{5.5}~\mathrm{K}$).%, which rains down within a free-fall time. 

For the $f_\mathrm{SN}0.01$ and $f_\mathrm{SN}0.1$ runs, $\dot{E}_\mathrm{SN}$ and $\dot{E}_\mathrm{out}$ are much smaller compared to the cooling and shell-by-shell heating rates at all times. For the two multiphase runs ($f_\mathrm{SN}0.5$ and $f_\mathrm{SN}0.99$), we find that at $t\lesssim t_\mathrm{mp}$, the cooling rate has a similar amplitude as the SNIa heating rate and is larger than the outflow rate. However, at late times the cooling rate drops significantly due to mass loss and most of the energy injected by the SNIa is accounted for by a larger $\dot{E}_\mathrm{out}$.

% \subsection{Shell-averaged profiles}\label{subsec:shell_averaged_profiles}
% In the previous subsection, we showed that the SNIa energy injection rate stays constant while the cooling rate drops after $t_\mathrm{mp}$ for the $f_\mathrm{SN}0.5$ and $f_\mathrm{SN}0.99$ runs. The unbalanced SNIa heating rate then overheats the gas, which both expands and rises buoyantly leading to an outflow in the positive $z$-direction. Here we show the effect of the SNIa heating on the vertical profiles of gas density and the channels through which the injected energy is transferred to the ambient ISM. 

\subsection{Vertical profiles of energy fluxes}\label{subsec:vert_prof_ener_fluxes}
\begin{figure}
		\centering
	\includegraphics[width=\columnwidth]{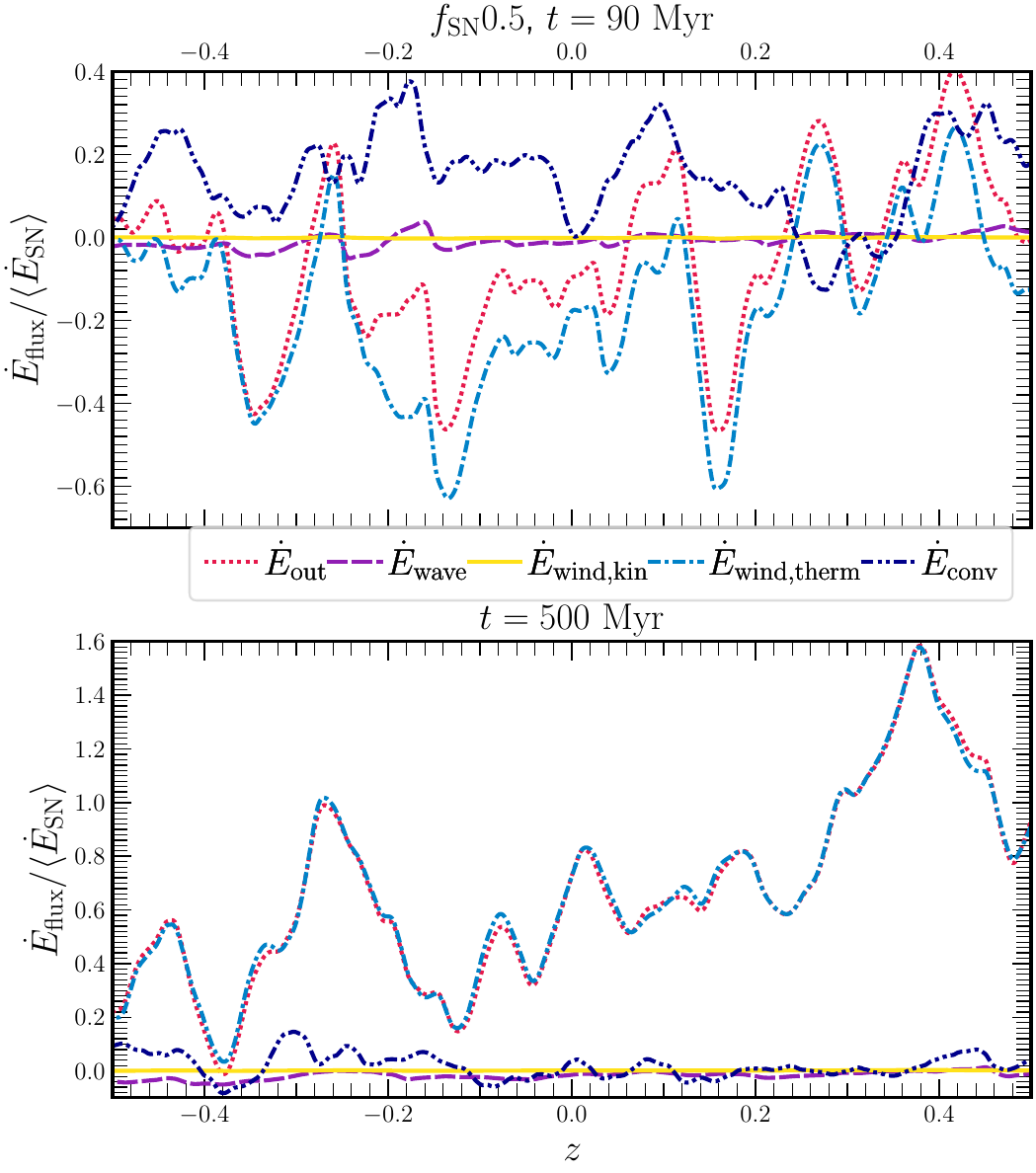}	
	\caption{Vertical profiles of energy fluxes for the $f_{\mathrm{SN}0.5}$ run at $t=t_\mathrm{mp}$ (upper panel) and $t=t_\mathrm{end}$ (lower panel), normalized by the mean SNIa rate over the entire volume. Convection and thermal wind flux are the two main contributors to the net energy flux at early times. At late times, the net energy flux is mostly positive and dominated by the thermal wind flux.}
	\label{fig:vert_prof_fsn05}
\end{figure}
% \begin{figure}
% 		\centering
% 	\includegraphics[width=\columnwidth]{edot_prof_n08_tend.pdf}	
% 	\caption{Similar to \Cref{fig:vert_prof_tmulti_fid}, plotted at $t=t_\mathrm{end}$ instead. The net energy flux is mostly positive and dominated by the thermal wind flux.}
% 	\label{fig:vert_prof_tend_fid}
% \end{figure}

Here we present the contribution of different flux components to the net energy flux at different $z$-shells. This is important to understand the mechanism through which supernovae transfer their energy to the ISM as compared to the heat transfer mechanism by AGN \citep[e.g., see][]{Tang2017MNRAS,Bambic2019ApJ,Choudhury2022MNRAS,Wang2022MNRAS}.

The total energy flux $\dot{E}_\mathrm{out}$ defined in eq.~\ref{eq:ener_flux_tot} can be decomposed into different physical components including wind (advective), convective, and wave as follows \citep[see][for a similar analysis]{Parrish2009ApJ}:

\begin{align}
    \dot{E}_\mathrm{wind} &= \int \mathrm{d}x\mathrm{d}y\mean{\rho v_z}\left(\underbrace{\frac{1}{2}\mean{v^2}}_\mathrm{kin}+\underbrace{\frac{\gamma}{\gamma-1}\mean{P/\rho}}_\mathrm{therm}\right), \label{eq:flux_wind}\\
    \dot{E}_\mathrm{conv} &= \int \mathrm{d}x\mathrm{d}y\frac{\gamma}{\gamma-1}k_B \nonumber\\
    &\times\left( \mean{n}\mean{\delta v_z\delta T} +\mean{v_z}\mean{\delta n\delta T} \right), \label{eq:flux_conv} \\
    \dot{E}_\mathrm{wave} &= \int \mathrm{d}x\mathrm{d}y\mean{\delta P\delta v_z}, \label{eq:flux_wave} 
\end{align}
\end{subequations}
where $\mean{}$ represents the average over a $z$-shell and the fluctuations such as $\delta n$ are defined as $n-\mean{n}$ for each shell. We break down this discussion into two subsections--(1) at early times ($t\lesssim\min(t_\mathrm{mp},100~\mathrm{Myr})$), shown in the upper panel of \Cref{fig:vert_prof_fsn05} and (2) at $t=t_\mathrm{end}$, shown in lower panel. Since the evolution of the $z$-profiles of the energy fluxes does not vary much across the fiducial set, we present our analysis for the $f_\mathrm{SN}0.5$ run only.

% Edited till here Sep 6 2023 11:48 pm
For $t=t_\mathrm{mp}$ (or at $t=100~\mathrm{Myr}$), we find that the net energy flux is dominated by the contribution of the thermal wind flux. The contribution of the kinetic wind flux is low, which is expected since $\dot{E}_\mathrm{wind,kin}/\dot{E}_\mathrm{wind,therm}=(\gamma-1)\mathcal{M}^2/2$ and $\mathcal{M}\lesssim0.1$ for all our simulations (see column 8 in Table \ref{tab:sim_params}).
The flux due to convection is also generally positive since the supernova remnants are overheated compared to their surroundings and rise due to buoyancy ($\delta T\times \delta v_z>0$). 

\cite{Tang2017MNRAS} show that for an outburst in a uniform medium under spherical symmetry, the fraction of energy carried away by sound waves is $\lesssim12\%$. \cite{Bambic2019ApJ} showed that this fraction could be larger if a cocoon of shocked plasma generates small-scale waves, boosting their fraction to $\sim25\%$. We do not find much energy in the form of sound waves, except small bumps in their $z$-profiles which likely correspond to recently injected SNIa. On average, the energy flux fraction in sound waves is $\lesssim1\%$ of $\dot{E}_\mathrm{SN}$.

At $t=t_\mathrm{end}$ (the lower panel of \Cref{fig:vert_prof_fsn05}), we find the total energy flux $\dot{E}_\mathrm{tot}$ is positive at all $z$-shells for $f_\mathrm{SN}\geq0.1$, implying an outflow in the positive $z$-direction. It is completely dominated by the contribution of the thermal wind flux. 
Since the system loses mass due to outflows or condensation, the gas density decreases as a function of time, hence $\dot{E}_\mathrm{conv}$ becomes weaker at late times. 

% At both early and late times, the overall amplitude of different flux components increase with increasing $f_\mathrm{SN}$, as is expected from \Cref{fig:ener_evol_fid}. However, we do not find a strong dependence of their relative importance on $f_\mathrm{SN}$.

\subsection{Distribution functions}\label{subsec:distribution_func}
\begin{figure*}
		\centering
	\includegraphics[width=1.8\columnwidth]{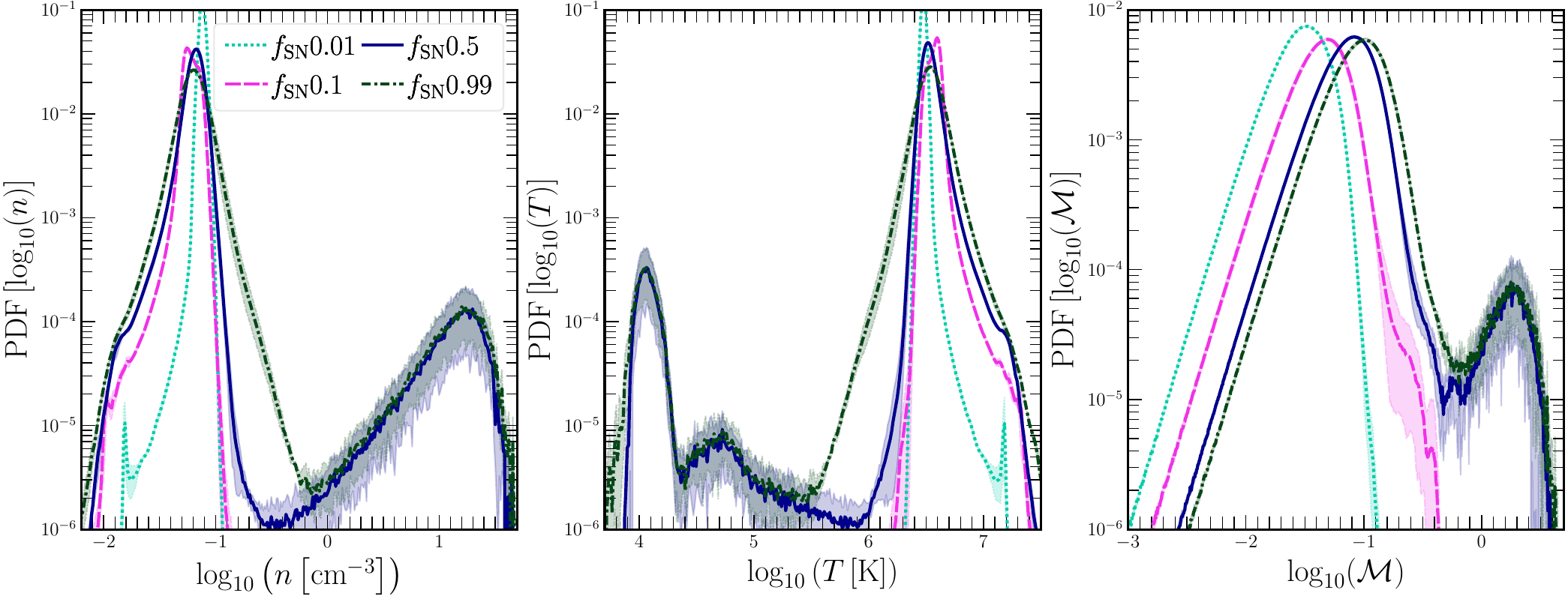}	
	\caption{Mass-weighted probability distribution functions of gas number density, temperature and Mach number for our fiducial set at $t=\min(t_\mathrm{mp},t_\mathrm{end})$. The two peaks in the distribution for the $f_\mathrm{SN}0.5$ and $f_\mathrm{SN}0.99$ runs correspond to the hot and cold phases. With increasing $f_\mathrm{SN}$, the width of the PDFs of the hot phase show an increase, consistent with increasing strength of turbulence. }
	\label{fig:dens_temp_mach_pdfs_fid}
\end{figure*}

In \Cref{fig:dens_temp_mach_pdfs_fid}, we present the mass-weighted probability distribution functions (PDFs) of gas number density (column 1), temperature (column 2) and Mach number (column 3) . These help us understand the general characteristics of turbulence driven by supernovae in a low-density ISM. All the PDFs are averaged over $10~\mathrm{Myr}$ before $\min(t_\mathrm{mp},t_\mathrm{end})$.

% In \Cref{fig:dens_temp_mach_pdfs_fid}, we show the mass-weighted distribution functions of gas number density (column 1), temperature (column 2) and Mach number (column 3), respectively. 
The two single-phase runs show a single peak, corresponding to the hot phase. The two multiphase runs show two distinct peaks, corresponding to the hot and cold phases, respectively. Unlike large-scale driven turbulence, the gas density distribution is not lognormal. For the $f_\mathrm{SN}0.01$ and $f_\mathrm{SN}0.1$runs show a low-density tail and a corresponding high-temperature tail in the temperature PDF, which are likely to be associated with gas inside the hot bubbles directly heated by the remnants. The gas motions in the hot phase are subsonic, with the peak $\mathcal{M}\lesssim0.1$ for all runs. The cold phase in the $f_\mathrm{SN}0.5$ and $f_\mathrm{SN}0.99$ runs is mildly supersonic, with $\mathcal{M}\sim2$--$3$.

With increasing $f_\mathrm{SN}$, the density and temperature PDFs become broader and the peak $\mathcal{M}$ increases--denoting stronger turbulence driven by a higher SNIa rate. 
We also find more gas at intermediate temperatures ($10^{4.2}~\mathrm{K}\lesssim T\lesssim10^{\mathrm{5.5}}~\mathrm{K}$).
%through turbulent mixing, in agreement with the results of \cite{Mohapatra2022MNRASb}. 
% Column density of the gas at these temperatures is generally measured by studying absorption features in the UV spectra of background quasars (e.g. \citealt{HWChen2023arXiv}, also see \citealt{Tumlinson2017review} for a review). Hot gas at $T\gtrsim10^6~\mathrm{K}$ is detected in X-ray emission around  early-type \citep[e.g.][]{Fabricant1980ApJ}. 
The hot, intermediate and cold phases are observed to coexist in filaments of the M87 galaxy, with temperatures ranging from $100~\mathrm{K}$ to $10^7~\mathrm{K}$ \citep{Werner2013ApJ,Anderson2018A&A}.  Since the intermediate temperature gas cools fast and has a short expected lifetime, turbulent mixing with the hot phase is one of the proposed mechanisms to continuously generate them. Heating of the cold regions (at $T\lesssim10^4~\mathrm{K}$) by the SNIa and condensation from the dense regions of the hot phase can also generate gas at these intermediate temperatures. Although the filaments in M87 are likely associated with gas uplifted by the AGN jet, the SNIa can play a role in producing gas at these temperatures through the above processes.

\subsection{Density-temperature phase diagram}
\begin{figure*}
		\centering
	\includegraphics[width=2\columnwidth]{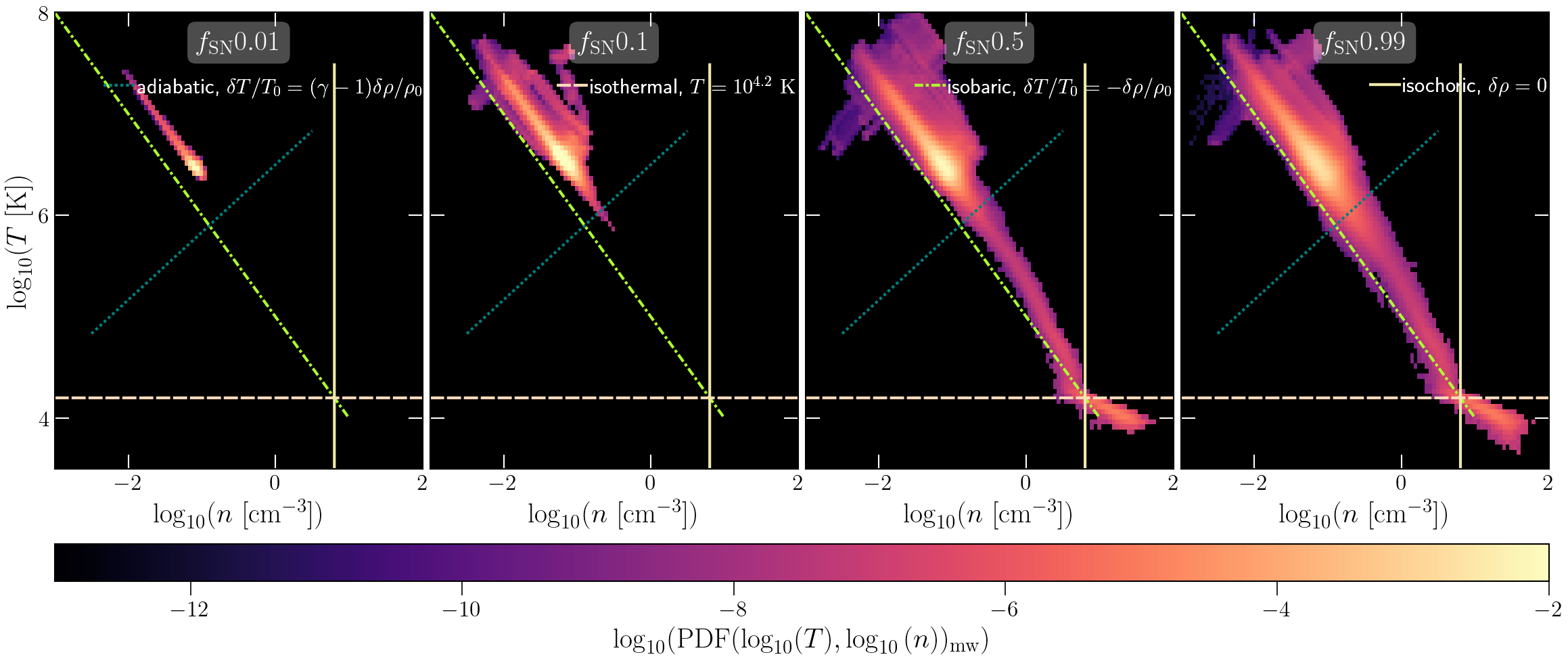}	
	\caption{Phase diagram showing the joint mass-weighted distribution of the gas temperature and density (in log-scale) for our fiducial set of simulations. The perturbations in the hot phase are mostly isobaric in nature. Recently injected SNIa appear as adiabatic fluctuations. The peak near $T\sim10^4~\mathrm{K}$ for `$f_\mathrm{SN}0.5$' and `$f_\mathrm{SN}0.99$' runs correspond to cold gas that has reached $T_\mathrm{floor}(=10^{4.2}~\mathrm{K})$.}
	\label{fig:temp_dens_phase_diagram}
\end{figure*}

Here we discuss the effects of SNIa driving on the density-temperature phase diagram. The different trend-lines shown in the plots are useful for tracking the nature of the thermodynamic perturbations in the gas. The phase diagrams let us distinguish between different physical mechanisms that are responsible for energy transfer in the ISM -- for example, adiabatic fluctuations can be due to compressive turbulence and sound waves, whereas isobaric fluctuations are caused by weak subsonic motions in a stratified medium \citep{Mohapatra2020}. Further, the growth rate of different modes of thermal instability $t_\mathrm{ti}$ (for example isobaric, isochoric) is expected to be different \citep{Das2021MNRAS}. We can determine the relevant mode for our simulations from the phase diagram and calculate the correct $t_\mathrm{ti}$. 

In \Cref{fig:temp_dens_phase_diagram}, we show the phase diagram for the fiducial set. For the $f_\mathrm{SN}0.5$ and $f_\mathrm{SN}0.99$ runs, we observe the hot phase at $T\gtrsim10^6~\mathrm{K}$ and the cold phase below the cutoff temperature at $10^{4.2}~\mathrm{K}$. At $T\gtrsim10^7~\mathrm{K}$, we find adiabatic fluctuations that are likely to be associated with recently injected supernovae. The perturbations in the intermediate temperature gas are mostly isobaric ($10^{4.2}~\mathrm{K}\lesssim T\lesssim10^{\mathrm{5.5}}~\mathrm{K}$). 
Unlike previous studies such as \cite{Mohapatra2023MNRAS}, we do not observe a sharp isochoric drop at $T\sim10^{5.5}~\mathrm{K}$, which corresponds to the peak of the cooling curve. We note a slight steepening of the PDF at $T\sim10^{4.5}~\mathrm{K}$, which could be due to the under-resolution of the coldest clumps, for which we do not resolve the cooling length \citep[$\ell_\mathrm{cool}=c_s t_\mathrm{cool}$, see][]{Fielding2020ApJ}.

With increasing $f_\mathrm{SN}$, we observe more contribution of adiabatic perturbations--which can be associated with compressive turbulence and sound waves directly driven by recently injected supernovae. \cite{Arevalo2016ApJ} have analyzed the nature of X-ray brightness fluctuations for M87 using the hardness ratio of X-ray spectra in Chandra observations \citep[also see][]{zhuravleva2018}. They find the filamentary regions to be isobaric, weakly shocked regions to be adiabatic and the remaining regions to be a combination of the two. Our findings suggest that the subsonic turbulence driven by the SNIa can contribute to either of the two observed modes. 

\subsection{Condition for multiphase gas formation--fiducial runs}\label{subsec:cond_multiphase_fid}
\begin{figure}
		\centering
	\includegraphics[width=\columnwidth]{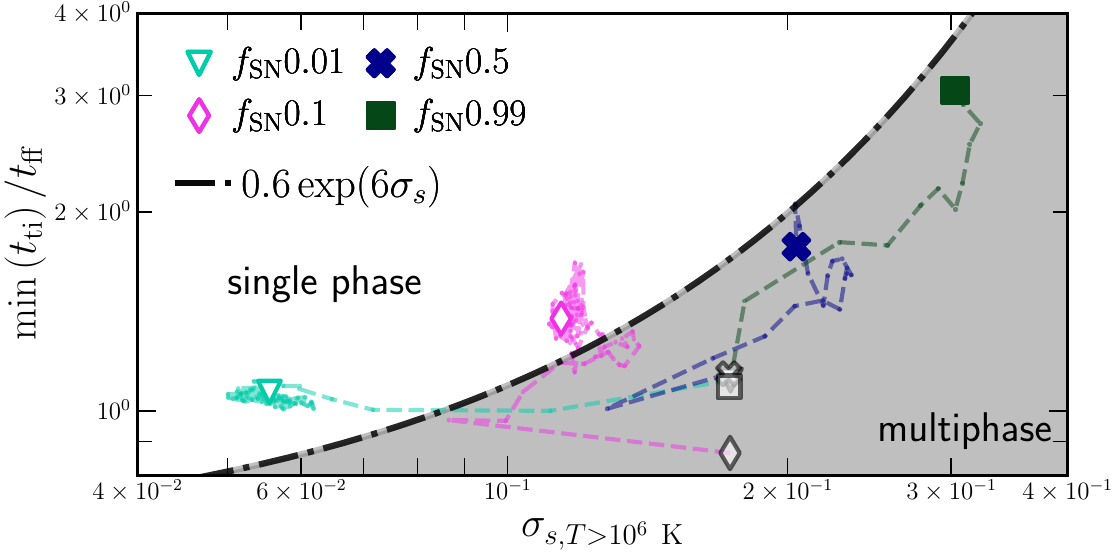}	
	\caption{The time-evolution of the minimum value of the ratio $t_\mathrm{ti}/t_\mathrm{ff}$ plotted against $\sigma_s$ for the fiducial set. The dark gray points show the value of this ratio at $t=0$. Simulations that do not form multiphase gas are shown as unfilled colored data points at $t=t_\mathrm{end}$ and simulations that form multiphase gas are shown as filled colored data points at $t=t_\mathrm{mp}$. The colored dashed lines show the evolution of this ratio from $t=0$ to $t=\min(t_\mathrm{mp},t_\mathrm{end})$. The dash-dotted trend line is a scaled version of the condensation criterion proposed in \citetalias{Mohapatra2023MNRAS}. Although all four simulations start with similar \change{$\min(t_\mathrm{ti})/t_\mathrm{ff}$}, their evolution is strongly dependent on the SNIa rate.}
	\label{fig:condensation_criterion_fid}
\end{figure}

For the gas around galaxies and galaxy clusters, the ratio $t_\mathrm{ti}/t_\mathrm{ff}$ is an important criterion to determine whether a thermally unstable system forms multiphase gas \citep{sharma2012thermal,mccourt2012}. A small $t_\mathrm{ti}/t_\mathrm{ff}$ ($\lesssim10$) has been associated with the existence of multiphase gas in both simulations \citep{prasad2018} and observations \citep{Olivares2019A&A}. 
More recent theoretical studies have indicated that the threshold value of $t_\mathrm{ti}/t_\mathrm{ff}$ for multiphase gas formation depends on the amplitude of density/entropy fluctuations (\citealt{Choudhury2019,Voit2021ApJ}; \citetalias{Mohapatra2023MNRAS}). 

In \Cref{fig:condensation_criterion_fid}, we plot the minimum value of the $z$ shell-averaged value of $t_\mathrm{ti}/t_\mathrm{ff}$ versus the amplitude of logarithmic density fluctuations for gas with $T>10^6~\mathrm{K}$. %The grey symbols show \change{$\min(t_\mathrm{ti})/t_\mathrm{ff}$} at $t=0$, the dashed lines show their evolution till $t=\min(t_\mathrm{mp},t_\mathrm{end})$ and the colored symbols show their values at $t=\min(t_\mathrm{mp},t_\mathrm{end})$. 
The dash-dotted black line shows the condensation curve proposed by \citetalias{Mohapatra2023MNRAS}, scaled down\footnote{There is some ambiguity in defining $t_\mathrm{ti}$ since the value of $\mathrm{d}\ln\Lambda/\mathrm{d}T$ changes for $10^6~\mathrm{K}<T<10^7\mathrm{K}$. Further, it is non-trivial to define the density-dependence of the heating by the SNIa ($\alpha_\mathrm{heat}$ parameter in eq.~\ref{eq:t_ti}). We find that scaling down the \citetalias{Mohapatra2023MNRAS} condensation curve by a factor of $0.6$ separates the single and multiphase simulations of our study well.} by a factor of $0.6$.

We observe a clear impact of the role played by the SNIa-driving on the occurrence of multiphase condensation. In the $f_\mathrm{SN}0.01$ and $f_\mathrm{SN}0.1$ runs, the seed density fluctuations at $t=0$ are quickly damped by viscous forces and the weak SNIa driving does not generate large density fluctuations. As a result, the simulations no longer satisfy the condensation criterion and remain in a single-phase till $t=t_\mathrm{end}$. The stronger SNIa driving in the $f_\mathrm{SN}0.5$ and $f_\mathrm{SN}0.99$ runs increases the amplitude of density fluctuations. However, the larger number of injected SNIa also drive an outflow at initial times (see Figs \ref{fig:mass_fluxes_fid} and \ref{fig:ener_evol_fid}) and overheat small regions of the gas where the remnants deposit their energy. These processes decrease the gas density and as a result increase \change{$\min(t_\mathrm{ti})/t_\mathrm{ff}$}. After accounting for the combined effect of the two, the system still satisfies the condensation criterion and forms multiphase gas.

\section{Effect of varying other parameters and summary}\label{sec:other-params-summary}

In the previous section, we discussed the effects of varying the level of SNIa heating while keeping all other parameters constant. In this section, we vary different simulation parameters such as the mean gas density, the strength of gravity, inclusion of heating due to AGB winds and assess the impact of switching SNIa heating on/off in these systems. These help us understand the local effects of SNIa heating on the ISM in different regions of early-type galaxies.

\subsection{Scaling of fluctuations with the rms Mach number}\label{subsec:fluc_scaling_mach}
\begin{figure*}
		\centering
	\includegraphics[width=2.0\columnwidth]{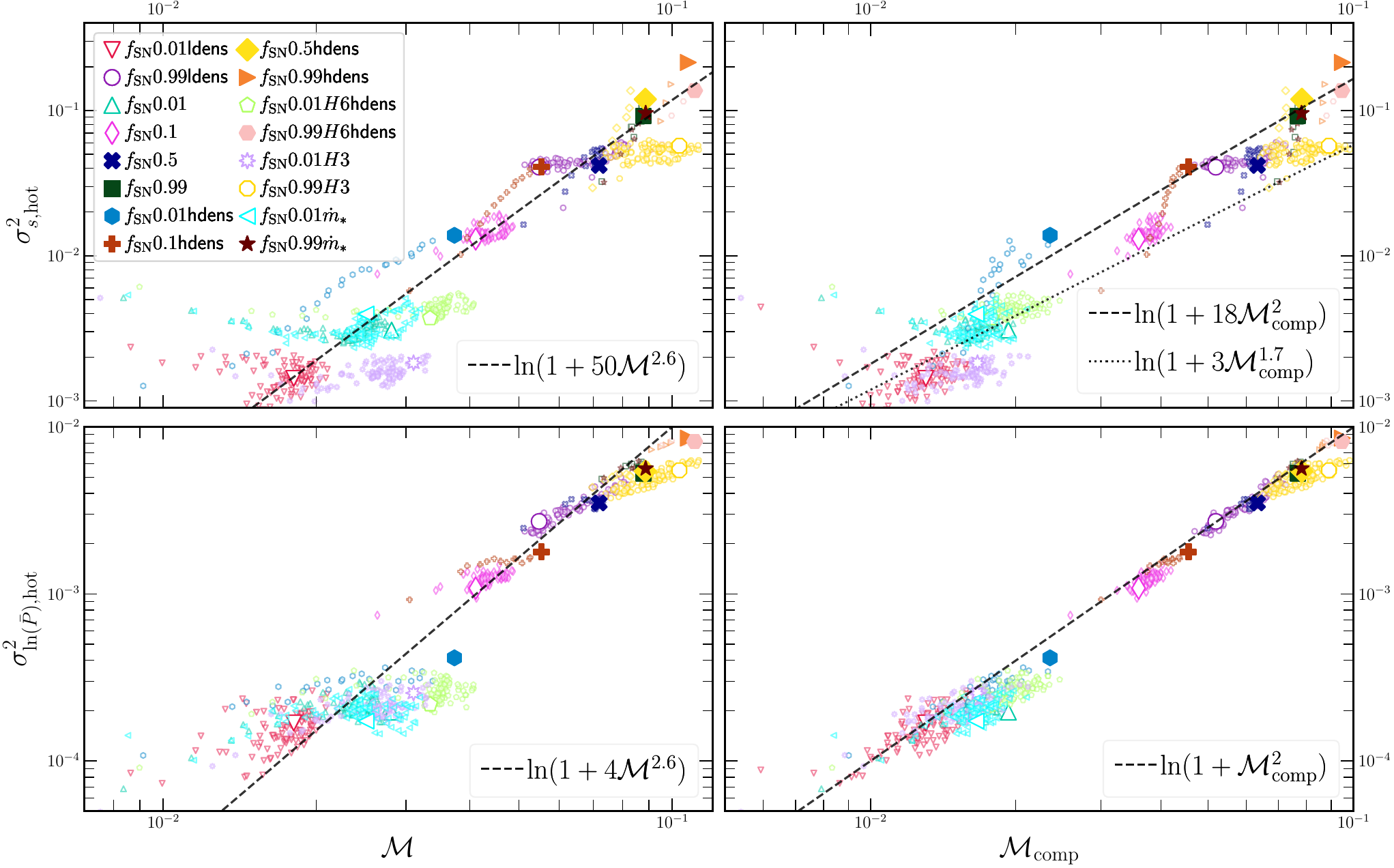}	
	\caption{\emph{First row}: The square of the amplitude of logarithmic density fluctuations ($\sigma_s^2$) plotted against the Mach number ($\mathcal{M}$) (\emph{left column}) and its compressive component $\mathcal{M}_\mathrm{comp}$. The larger symbols represent the data-points at $t=\min(t_\mathrm{mp},t_\mathrm{end})$ and the smaller symbols show their values are at earlier times. The filled symbols denote simulations that form multiphase gas. (\emph{right column}). The dotted line on the right panel shows the predicted scaling relation for subsonic turbulence with compressive forcing from \citealt{konstandin2012}. \emph{Second row}: Similar to the first row, but for the square of logarithmic pressure fluctuations $\sigma_{\ln(P/\mean{P})}^2$ instead. The dashed lines in all four panels show our empirical fits to the data. The pressure fluctuations show a tight scaling with $\mathcal{M}_\mathrm{comp}$. }
	\label{fig:sig_mach}
\end{figure*}

Observational studies of the hot gas around giant elliptical galaxies or galaxy clusters often rely on indirect techniques such as using the amplitude of surface brightness fluctuations \citep[which are dependent on the gas density, e.g.][]{zhuravleva2014turbulent,Zhuravleva2015MNRAS,zhuravleva2018} or fluctuations in the thermal Sunyaev-Zeldovich effect to infer the amplitude of the turbulent gas velocities \citep[which depend on the thermal pressure of the gas, e.g.][]{khatri2016,Romero2023ApJ}. On the other hand, direct measurements of the turbulent velocities using the high spectral-resolution telescope XRISM\footnote{\url{https://heasarc.gsfc.nasa.gov/docs/xrism/about/}} can be used to infer the state of perturbations in the hot ISM. 
In this subsection, we present the scaling relation between the gas density/pressure fluctuations and the turbulent Mach number for SNIa-driven turbulence. 

In \Cref{fig:sig_mach}, we show the joint evolution of $\sigma_{s,\mathrm{hot}}^2$ (row 1) and $\sigma_{\ln(\bar{P}),\mathrm{hot}}^2$ (row 2) versus $\mathcal{M}$ (column 1) and its compressive component $\mathcal{M}_\mathrm{comp}$ (column 2). The dashed lines in each panel show empirical fits to the data. The dotted line in the top right panel shows the $\sigma_s$--$\mathcal{M}_\mathrm{comp}$ scaling relation for homogeneous turbulence forced on large scales from \cite{konstandin2012}. 

In general, we find that both $\sigma_{s,\mathrm{hot}}$ and $\sigma_{\ln(\bar{P}),\mathrm{hot}}$ increase with increasing $\mathcal{M}$ or $\mathcal{M}_\mathrm{comp}$. The amplitude of pressure fluctuations is much smaller than that of density fluctuations. The fluctuations are larger for runs with higher $f_\mathrm{SN}$, in line with our expectations. The multiphase runs (represented by the filled markers) are typically associated with larger $f_\mathrm{SN}$ and thus are found in the top right of each panel, whereas the single-phase regions are typically found in the bottom left, with a few exceptions. Among the runs with the same $f_\mathrm{SN}$ but different densities, the `ldens' (`hdens') runs have a lower (higher) cooling rate, hence a weaker (stronger) SNIa driving and smaller (larger) $\sigma_{s,\mathrm{hot}}$, $\sigma_{\ln(\bar{P}),\mathrm{hot}}$ compared to the fiducial set. Variations in the other parameters such as `$H$' or the inclusion of stellar heating do not have any significant effect on these results.  

Both $\sigma_{s,\mathrm{hot}}$ and $\sigma_{\ln(\bar{P}),\mathrm{hot}}$ follow power-law scaling with $\mathcal{M}$ and $\mathcal{M}_\mathrm{comp}$ with some scatter around the empirical fits. We find a tight relation between $\sigma_{\ln(\bar{P}),\mathrm{hot}}$ and $\mathcal{M}_\mathrm{comp}$. 
Similar to \citetalias{MLi2020ApJb}, we find both fluctuations to be much larger than predicted by scaling relations in literature, such as \cite{konstandin2012,Mohapatra2021MNRAS,Mohapatra2022MNRASc}. Unlike the proposed scaling relations, which were based on idealized turbulence driven by exciting particular modes in $k$-space, the supernovae overheat small regions of the gas which expand, rise buoyantly and drive turbulence in the ISM. Heating by the SNIa and radiative cooling of the ISM are also associated with isobaric density perturbations, as seen in \cref{fig:temp_dens_phase_diagram}. We expect the differences between such idealized driving and driving by the SNIa to be due to the heating and cooling of the gas, as well as the differences in driving scale, modes, etc. 

Compared to \citetalias{MLi2020ApJb}, who study SNIa driven turbulence in an unstratified box (i.e.~without external gravity), we observe some key differences. 
The amplitude of perturbations that we observe are smaller by a factor of $4$ or more compared to that of \citetalias{MLi2020ApJb} for the same $\mathcal{M}$. This is likely due to the smaller ratio between the power in compressive and solenoidal modes in our simulations, even though the contribution from compressive modes still dominate the total power. In our simulations, we find that the bubbles inflated due to the energy deposited by the SNIa rise buoyantly and form mushroom-shaped clouds, which are a typical characteristic of the Rayleigh-Taylor (RT) instability (see fig.~\ref{fig:2dslice_bubble}). The motion of the hot bubble gas with respect to the ambient medium also generates the Kelvin-Helmholtz (KH) instability. Both the RT and KH instabilities can drive solenoidal turbulence in the ISM and increase its contribution to the total power. 

Although the amplitude of density and pressure fluctuations generated by SNIa-driven turbulence is large, they may be difficult difficult to detect since they occur on $\lesssim 100~\mathrm{pc}$ scales. First of all, small-scale fluctuations would be canceled out due to averaging along the line of sight. Secondly, most current X-ray and microwave telescopes lack the resolution to resolve small scales even for nearby massive elliptical galaxies ($R_\mathrm{fade}\sim46~\mathrm{pc}\sim0.5~$arc-second for the M87 galaxy, which is close to the resolution limit of the Chandra telescope). Future X-ray telescopes such as AXIS\footnote{\url{https://axis.astro.umd.edu/}} would be useful to measure these ISM properties with their higher angular resolutions and better sensitivities. \change{The synchrotron emission from the electrons accelerated in the SNIa-shocks could be detected in radio wavelengths. However, it would be challenging to distinguish them from the radio emission from other sources, such as shocks due to AGN activity and mergers. }

\subsection{Condition for multiphase gas formation–all runs}\label{subsec:conditions_multi_all_runs}
\begin{figure}
		\centering
	\includegraphics[width=\columnwidth]{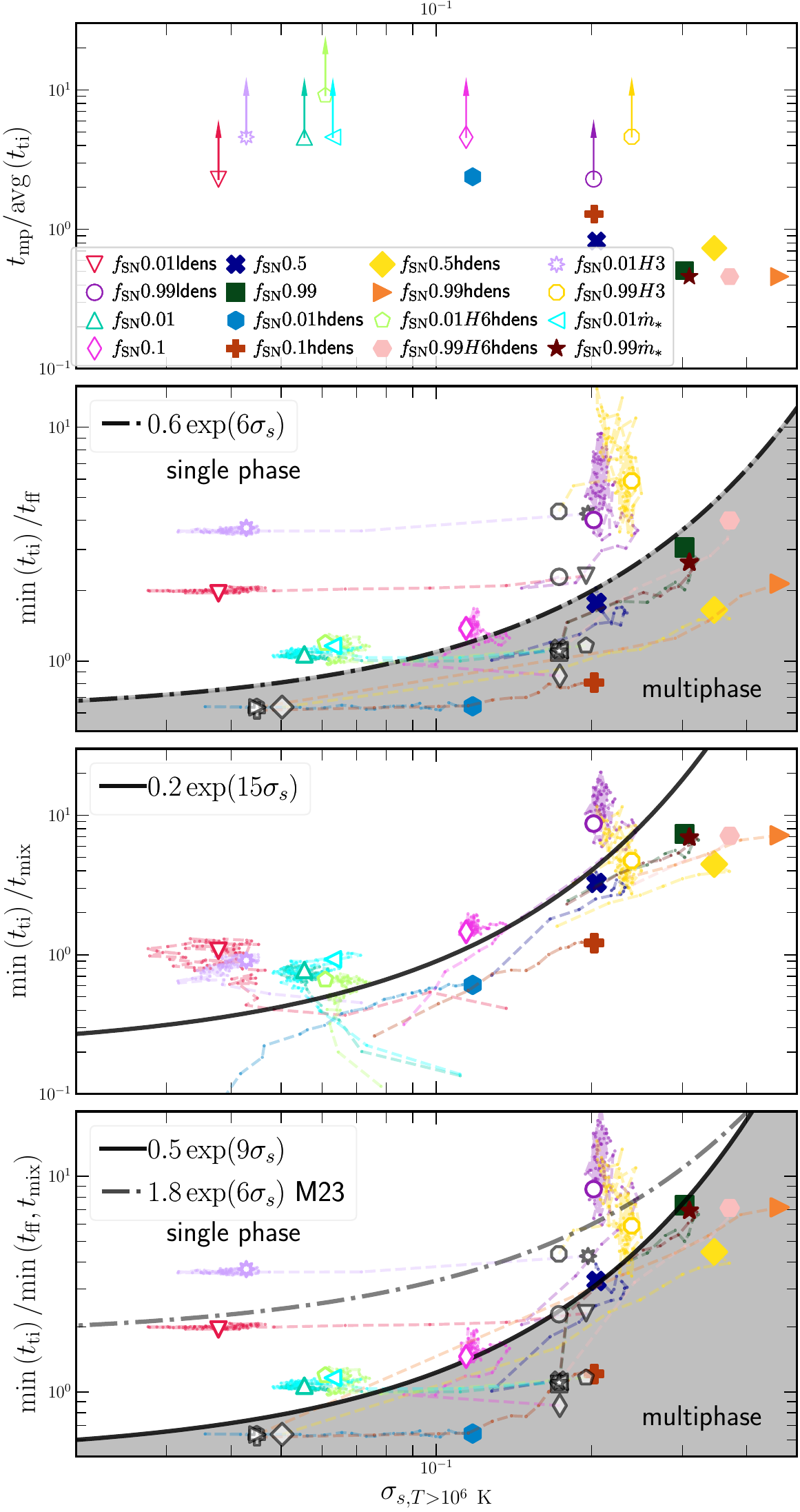}	
	\caption{\emph{First row:} The time taken by a simulation to form multiphase gas normalized by the initial thermal instability time-scale $t_\mathrm{mp}/\mathrm{avg}(t_\mathrm{ti})_{t=0}$ plotted against the amplitude of logarithmic density fluctuations in the hot phase $\sigma_{s,\mathrm{hot}}$. When the simulation forms multiphase gas, we show the data points as filled markers. When they remain single phase till $t=t_\mathrm{end}$, we show the lower-limit of this ratio and mark them as unfilled data points and an upward arrow. \emph{Second row:} The minimum value of the ratio $t_\mathrm{ti}/t_\mathrm{ff}$ plotted against $\sigma_s$. The dark grey points show the value of this ratio at $t=0$ and the dashed lines show the evolution of this ratio till $t=\min(t_\mathrm{mp},t_\mathrm{end})$.  The dash-dotted curve shows the $t_\mathrm{ti}/t_\mathrm{ff}-\sigma_s$ condensation criterion, which separates the plot into single and multi-phase regions. \emph{Third row:} Similar to the second row, but showing the ratio $t_\mathrm{ti}/t_\mathrm{mix}$ instead. \change{\emph{Fourth row:} Here we show the larger of the two ratios shown in rows 2 and 3. The dot-dashed line shows the condensation criterion proposed by \citetalias{Mohapatra2023MNRAS}. The solid line is a re-scaled version of this criterion which fits our results better.} 
    }
	\label{fig:condensation_criterion_summary}
\end{figure}

In this subsection, we revisit the criteria for multiphase gas condensation for all our simulations. In addition to \change{$\min(t_\mathrm{ti})/t_\mathrm{ff}$} we discussed earlier, we also check the importance of turbulent mixing in suppressing multiphase condensation, as proposed by \cite{Gaspari2018ApJ}. 

In the first row of \Cref{fig:condensation_criterion_summary}, we show the ratio $t_\mathrm{mp}/\mean{t_{\mathrm{ti},t=0}}$ (averaged over the entire volume) versus $\sigma_{s,\mathrm{hot}}$. Seven out of our fourteen simulations form multiphase gas, denoted using filled markers. We show $t_\mathrm{end}/\mean{t_{\mathrm{ti},t=0}}$ as the lower limit of $t_\mathrm{mp}/\mean{t_{\mathrm{ti},t=0}}$ for the single-phase runs, which are denoted by unfilled data points. For most of the multiphase runs,  $t_\mathrm{mp}/\mean{t_{\mathrm{ti},t=0}}<1$. Since the SNIa driving generates large density fluctuations, locally dense regions cool in less than the $\mean{t_{\mathrm{ti},t=0}}$. The $f_\mathrm{SN}0.01$hdens and the $f_\mathrm{SN}0.1$hdens runs are the two exceptions since the SNIa driving is weak for these runs.

We show the ratio \change{$\min(t_\mathrm{ti})/t_\mathrm{ff}$} versus $\sigma_{s,\mathrm{hot}}$ for all runs in the second row. The values of these two quantities at $t=0$ are denoted by the unfilled dark-gray markers and the dashed lines show their evolution with time till $\min(t_\mathrm{mp},t_\mathrm{end})$. We start with $\sigma_{s,\mathrm{hot}}\approx0.2$ for all runs except the `hdens' set, for which we set $\sigma_{s,\mathrm{hot}}\approx0.05$. In general, we find that $\sigma_{s,\mathrm{hot}}$ decreases with time for the runs with $f_\mathrm{SN}=0.01$ that remain single phase and increases with time for the runs with $f_\mathrm{SN}=0.99$ that form multiphase gas. The value of $\sigma_{s,\mathrm{hot}}$ stays roughly constant for $f_\mathrm{SN}=0.99$ runs that remain in a single phase. For most of the `hdens' runs, $\sigma_{s,\mathrm{hot}}$ increases with time, which we associate with the onset of thermal instability. 

Now we shift our focus to the values of this ratio at $t=\min(t_\mathrm{ti},t_\mathrm{end})$, represented by the colored markers.  We refer the reader to ~Table \ref{tab:sim_params} for important simulation parameters. We find that the runs with stronger stratification ($f_\mathrm{SN}0.01H3$ and $f_\mathrm{SN}0.99H3$) do not form multiphase gas due to the shorter $t_\mathrm{ff}$, which increases the value of \change{$\min(t_\mathrm{ti})/t_\mathrm{ff}$}. Comparing the high-density `hdens' set and the fiducial set, we find that both the $f_\mathrm{SN}0.01$hdens and $f_\mathrm{SN}0.1$hdens runs form multiphase gas (in addition to the $f_\mathrm{SN}0.5$hdens and $f_\mathrm{SN}0.99$hdens runs) due to the shorter $t_\mathrm{ti}$. However, once we double the strength of stratification, the $f_\mathrm{SN}0.01H6$hdens run no longer forms multiphase gas due to a smaller $t_\mathrm{ff}$, whereas the $f_\mathrm{SN}0.99H6$hdens run still does. In the `ldens' runs, due to the longer $t_\mathrm{ti}$ neither $f_\mathrm{SN}0.01$ldens nor $f_\mathrm{SN}0.99$ldens runs form multiphase gas.  The runs with AGB winds ($f_\mathrm{SN}0.01\dot{m}_*$ and $f_\mathrm{SN}0.99\dot{m}_*$) show similar trends as their fiducial counterparts. The exponential condensation curve that we introduced in Section \ref{subsec:cond_multiphase_fid} separates all our single and multiphase simulations well, i.e. the final value of \change{$\min(t_\mathrm{ti})/t_\mathrm{ff}$} is smaller (larger) than the condensation curve if the simulation forms (does not form) multiphase gas. 

In the third row, we show the ratio  \change{$\min(t_\mathrm{ti})/t_\mathrm{mix}$} as a function of $\sigma_{s,\mathrm{hot}}$. We also draw an exponential curve in an attempt to separate the simulations that form multiphase gas from the ones that do not. Except the `$f_\mathrm{SN}0.99H3$' run which has strong stratification, this curve also separates between the single and multi-phase simulations. %In contrast with the results of \citetalias{Mohapatra2023MNRAS}, we find that multiphase condensation takes place even when $t_\mathrm{ti}/t_\mathrm{mix}\gg1$ (the condensation curve parameters $c_1$ and $c_2$ in $c_1\times\exp(c_2\sigma_s)$ in our empirical fit are much larger). 
% \footnote{\citetalias{Mohapatra2023MNRAS} show that condensation occurs when both \change{$\min(t_\mathrm{ti})/t_\mathrm{ff}$} and \change{$\min(t_\mathrm{ti})/t_\mathrm{mix}$} are smaller than a $\sigma_{s,\mathrm{hot}}$ dependent threshold. Since $t_\mathrm{ti}/t_\mathrm{mix}$ is larger than $t_\mathrm{ti}/t_\mathrm{ff}$ for most of our simulations, the joint ratio $\min(t_\mathrm{ti}/\min(t_\mathrm{ff},t_\mathrm{mix}))$ simply reduces to \change{$\min(t_\mathrm{ti})/t_\mathrm{mix}$} for our simulations.} 
% \change{It is steeper than the condensation curve in \citetalias{Mohapatra2023MNRAS}, who studied multiphase condensation due to turbulence driven on large scales.} 

\change{In the fourth row, we show the joint ratio $\min(t_\mathrm{ti})/\min(t_\mathrm{ff},t_\mathrm{mix})$, which is used to define the multiphase condensation criterion in \citetalias{Mohapatra2023MNRAS}. The dot-dashed line shows the \citetalias{Mohapatra2023MNRAS} condensation curve whereas the solid line shows a re-scaled version, which is a better empirical fit to the data in this work. Thus, both short $t_\mathrm{ff}$ and short $t_\mathrm{mix}$ can prevent multiphase condensation. Among our simulations with the same values of $t_\mathrm{ff}$ and initial $t_\mathrm{ti}$,  increasing $f_\mathrm{SN}$  leads to larger density perturbations. The increase in the solenoidal component of turbulence (due to the larger SNIa rate) is generally insufficient to mix these perturbations with the ambient gas and prevent multiphase condensation.}

% ==================================
\section{Caveats and Future Work}\label{sec:caveats-future}
In this section we discuss some of the short-comings of this study. We also outline some interesting aspects to consider in future works.

\subsection*{Boundary conditions}
We have modeled a small $1.5~\mathrm{kpc}^3$ patch of the ISM of an elliptical galaxy using a cuboidal box oriented along the direction of stratification of the ISM.
%We have kept the density and pressure fixed at the upper and lower $z$ boundaries. We set the $z$-direction velocities to be diode, except for gas with $T\geq T_\mathrm{cutoff}$ at the lower boundary. 
Our boundary conditions are outlined in detail in Section \ref{subsubsec:boundary_conditions}. We have also tested the effect of the  outflow for $v_z$ at the $z$ boundaries, hydrostatic equilibrium for $P$ and $\rho$ or constant ratio boundary conditions. We find that the evolution of our multiphase simulations for $t>t_\mathrm{mp}$ is sensitive to the choice of boundary conditions. If we do not fix the density and pressure at the upper and lower boundaries, once any cold gas forms it triggers a large-scale cooling flow and the simulation domain loses all its mass. Most of the other choices that we tested face this issue. The choice of boundary conditions does not have a significant effect on our simulations that remain in a single phase at all times.

\subsection*{Geometry}
The hot ISM in elliptical galaxies is expected to follow an elliptical/spherical distribution whereas we model it as a plane-parallel atmosphere. This difference in geometry can affect the energy and mass outflow rates since the hot gas is expected to expand and cool at larger radii. We plan to conduct global galaxy-scale simulations with the appropriate geometry in the future, which will let us address the fate of the SNe heated gas.  This will also resolve the challenge of boundary condition sensitivity highlighted in the previous paragraph.   

\subsection*{Heating model}
We include an additional heating term that replenishes a $(1-f_\mathrm{SN})$ fraction of the radiative losses in each $z$-shell. This modification also improved the stability of the atmosphere against a large-scale cooling flow. This overly idealized heating model is motivated by the observational evidence for global thermal stability in massive galaxies but in detail it cannot be correct and the formation of multiphase gas and the long-term evolution of the hot ISM may be sensitive to the details of the correct heating function.   

Many massive elliptical galaxies have AGN and the AGN jets can drive large-scale motions in the ISM. The in-fall of satellite galaxies can also add mass and drive turbulence in the ISM. We have not considered their effects in this study.

\subsection*{Resolution}
We have performed all simulations listed in Table \ref{tab:sim_params} at two resolutions--$512^2\times768$ and $256^2\times384$. We find the results of our single-phase runs are convergent with increasing the resolution. Among the multiphase runs, we find that the results diverge for $t\geq t_\mathrm{mp}$. The lower resolution runs form more cold gas compared to their high-resolution counterparts. This may be due to the effects of excessive averaging at the boundary layers between the hot and cold regions which forms more gas at intermediate temperatures ($10^5~\mathrm{K}<T<10^6~\mathrm{K}$). The intermediate temperature gas cools fast and forms more cold gas in the lower resolution simulations. 
%The isochoric cooling of gas that we find in \Cref{fig:temp_dens_phase_diagram} could also be sensitive to the resolution.

\subsection*{Physics}
We have ignored the effect of important physics such as magnetic fields \citep{Hopkins2020MNRAS,Wang2021MNRAS,Buie2022ApJ}, cosmic rays \citep{Kempski2020MNRAS,Butsky2020ApJ,Beckmann2022A&A} and conduction \citep{Parrish2009ApJ,El-Badry2019MNRAS}, see \cite{Faucher-Giguere2023ARAA} for a review. Magnetic fields and cosmic rays are expected to be energetically important 
%comparable to the kinetic energy in galaxies 
and can affect the properties of the thermal energy-driven outflow. Conduction can affect the energy exchange between the SNIa inflated bubbles and the ISM. All three can affect the formation of cold gas and its kinematics. In future studies, we plan to include some of these physical properties and study their impact.

\subsection*{Chemical Evolution}
We have ignored the evolution of the chemical composition of the ISM due to metals injected into the ISM by the SNIa.  SNIa are one of the main sources of elements such as \texttt{Fe}, \texttt{Co}, \texttt{Ni}, etc. The gas cooling rate at $T\sim10^6~\mathrm{K}$ is sensitive to the chemical composition of the ISM. We plan to include heavy element injection and transport in future works and study their properties such as their radial extent, spatial variations, etc.
% =============================
\section{Conclusion}\label{sec:Conclusion}
In this paper we model a $1.5~\mathrm{kpc}^3$ local stratified patch of the hot ISM of a massive elliptical galaxy. We study the effect of different strengths of heating due to type Ia supernovae (SNIa), motivated by the observational fact that the SNIa heating rate is of order the radiative cooling rate in many massive galaxies (see, e.g., Fig. \ref{fig:rad_profiles_analytic}).  We fix the heating rate due to the randomly injected SNIa to a fraction $f_\mathrm{SN}$ of the net cooling rate of the ISM and compare the hot ISM properties against a uniform shell heating model typically used in idealized simulations.
%the latter is often used in cosmological simulations to model the SNIa. 
We have conducted a total of 16 simulations, where we vary ISM properties such as gas density and the strength of stratification. 
Here we summarize some of our key findings:
% We first summarize the findings of our fiducial runs where we keep all simulation parameters constant and increase the ratio between the SNIa heating rate and the net cooling rate (section~\ref{sec:results-fid})
\begin{itemize}
    \item The SNIa deposit their energy in small $\sim20~\mathrm{pc}$ size regions of the ISM. These regions expand and rise buoyantly, driving turbulence in the ISM. The turbulence is associated with large density fluctuations (Figs.~\ref{fig:2dproj_fid} and \ref{fig:2dslice_bubble}). 
    \item The high-density regions have a short cooling time. Since the ISM at these temperatures is thermally unstable, if the ratio of the thermal instability growth time to the free fall time ($t_\mathrm{ti}/t_\mathrm{ff}$) is small enough, the dense regions cool down to the cooling cutoff temperature at $10^{4.2}~\mathrm{K}$, they are out of hydrostatic equilibrium and rain down through the bottom of the box. Since the SNIa seed these density fluctuations \change{\citep[also seen in][]{MLi2020ApJa}}, a larger SNIa injection rate is more likely to trigger multiphase condensation (Figs.~\ref{fig:condensation_criterion_fid}).   Much of the literature has focused on the formation of multiphase gas by thermal instability \citep[for eg.][]{sharma2012thermal,mccourt2012,Choudhury2019} or AGN driven-uplift of gas \citep[eg.][]{Pulido2018ApJ,Huvsko2023MNRAS}.   Our results show that SNIa are also efficient sources of multiphase gas production in the hot ISM/ICM of massive galaxies, groups, and clusters.  A spatially and temporally resolved treatment of SNIa is likely critical for understanding the formation of multiphase gas in massive galaxies and its role in fueling star formation and black hole growth.
    \item As the SNIa rate is fixed, once the gas density drops due to multiphase condensation, the SNIa overheat the ISM and drive an outflow. The net mass remaining in the simulation decreases with increasing SNIa rate (figs.~\ref{fig:mass_etot_evol_fid}, \ref{fig:mass_fluxes_fid}).  \change{Even in the simulations that  do not form multiphase gas, whenever the net heating exceeds the net cooling, the SNIa drive an outflow which further decreases the net mass and the cooling rate.}  This evolution reflects the fact that heating by SNIa is inevitably unstable:   unlike core-collapse SNe or (plausibly) AGN there is no connection between the SNIa rate and the radiative cooling of the hot gas. Global simulations will be required to understand the ultimate outcome of this instability.
    %In comparison to the mass lost to outflows at the boundaries, the mass added to the ISM by SNIa is negligible. 
    \item In the initial phases of our simulations (prior to multiphase gas condensation), the total energy flux is set by the sum of convective and thermal wind fluxes, whereas sound waves and the wind kinetic energy carry negligible amounts of energy. At late times, after multiphase gas condensation, the convective energy flux also drops and the wind becomes the dominant energy transport mechanism (Fig.~\ref{fig:vert_prof_fsn05}). Multiphase condensation and outflows also strongly alter the $z$-profile of gas density (Fig.~\ref{fig:vert_prof_dens}).
    % time evolution?
    \item The SNIa drive subsonic turbulence in the ISM which causes isobaric perturbations in it (figs.~\ref{fig:dens_temp_mach_pdfs_fid}, ~\ref{fig:temp_dens_phase_diagram}). The amplitudes of the density and pressure fluctuations are proportional to the compressive component of the rms Mach number. However, the fluctuations generated by SNIa  are much larger than predicted by the density fluctuations--rms Mach number scaling relation in homogeneous turbulence  \citep{konstandin2012}, \change{in agreement with \cite{MLi2020ApJb}}. We expect that these differences are due to the additional thermodynamics in our simulations, i.e., perturbations associated with the direct heating of the ISM by the SNIa and the radiative cooling of the ISM.  
    %Differences in the driving scale and modes between idealized turbulence simulations of \cite{konstandin2012} and turbulence driven by the SNIa can also contribute to the mismatch.
    \item For all 16 simulations, we find that multiphase condensation occurs if the criterion \change{$\min(t_\mathrm{ti})/t_\mathrm{ff}\leq c_1\exp(c_2\sigma_s)$} is satisfied. We obtain $c_1=0.6$ and $c_2=6$ using an empirical fit (Fig.~\ref{fig:condensation_criterion_summary}). In contrast with the simulations of idealized driven turbulence in \cite{Mohapatra2023MNRAS}, we find turbulent mixing does not suppress multiphase condensation even when $t_\mathrm{ti}/t_\mathrm{mix}\gg10$. This is because the SNIa drive compressive turbulence which is poor at mixing perturbations in the gas. Thus one needs to be careful in interpreting the importance of $t_\mathrm{ti}/t_\mathrm{mix}$ in observations, where it is difficult to determine the source/driving mode of turbulence.
    
\end{itemize}

\section{Acknowledgements} 
% \begin{acknowledgments}

% The authors would like to thank the anonymous referee for a constructive report, which helped to improve this work.
RM would like to thank Minghao Guo, Patrick Mullen and Jim Stone for help with setting up the simulations using Athenak and post-processing the results. We thank Chang-Goo Kim, Prateek Sharma, Romain Teyssier and Eve Ostriker for useful discussions. \change{We thank the anonymous referee for their comments, which helped improve this paper.}
This work was supported in part by a Simons Investigator award from the Simons Foundation (EQ) and by NSF grant AST-2107872.
The analysis presented in this article was performed in part on computational resources managed and supported by Princeton Research Computing, a consortium of groups including the Princeton Institute for Computational Science and Engineering (PICSciE) and the Office of Information Technology’s High Performance Computing Center and Visualization Laboratory at Princeton University.
We also used the Delta GPU machine at National Center for Supercomputing Applications, Illinois, United States through allocations PHY230106 and PHY230045 from the Advanced Cyberinfrastructure Coordination Ecosystem: Services and Support (ACCESS) program, which is supported by National Science Foundation grants 2138259, 2138286, 2138307, 2137603, and 2138296. The Delta research computing project supported by the National Science Foundation (award OCI 2005572), and the State of Illinois. Delta is a joint effort of the University of Illinois at Urbana-Champaign and its National Center for Supercomputing Applications.

% \end{acknowledgments}

% This work used the following software/packages:
\software{\texttt{Athena++} \citep{Stone2020ApJS}, \texttt{matplotlib} \citep{Hunter4160265}, \texttt{cmasher} \citep{Ellert2020JOSS}, \texttt{scipy} \citep{Virtanen2020}, \texttt{NumPy} \citep{Harris2020}, \texttt{CuPy} \citep{Okuta2017CuPyA}, \texttt{h5py} \citep{collette_python_hdf5_2014} and \texttt{astropy} \citep{astropy2018}}.

%%%%%%%%%%%%%%%%%%%%%%%%%%%%%%%%%%%%%%%%%%%%%%%%%%

\section{Data Availability}
All relevant data associated with this article is available upon reasonable request to the corresponding author.

\section{Additional Links}
Movies of density, temperature, density fluctuations and pressure fluctuations along slices at $x=0$ are available at the following links:
\begin{itemize}
    \item `Fiducial' runs: \url{https://youtu.be/EhY205ezfsU} 
    \item `hdens' runs: \url{https://youtu.be/xXRj2j7zrYE} 
    \item `ldens' runs: \url{https://youtu.be/LHbx7Cp2mkE} 
    \item `H6hdens' runs: \url{https://youtu.be/_Ti4wZ0F_x8} 
    \item `H3' runs: \url{https://youtu.be/nlyCZBC_Krs}
    \item `$\dot{m}_*$' runs: \url{https://youtu.be/xNnO2jJZHmM}.
\end{itemize}

%%%%%%%%%%%%%%%%%%%%%%%%%%%%%%%%%%%%%%%%%%%%%%%%%%

%%%%%%%%%%%%%%%%% APPENDICES %%%%%%%%%%%%%%%%%%%%%

\appendix
% \begin{multicols}[2]
\renewcommand\thefigure{\thesection \arabic{figure}} 
\setcounter{figure}{0}   
\setcounter{table}{0}   
\section{Effect of including AGB winds}\label{app:effect_AGB_wind}
\begin{figure}%[h]
		\centering
	\includegraphics[width=1.0\columnwidth]{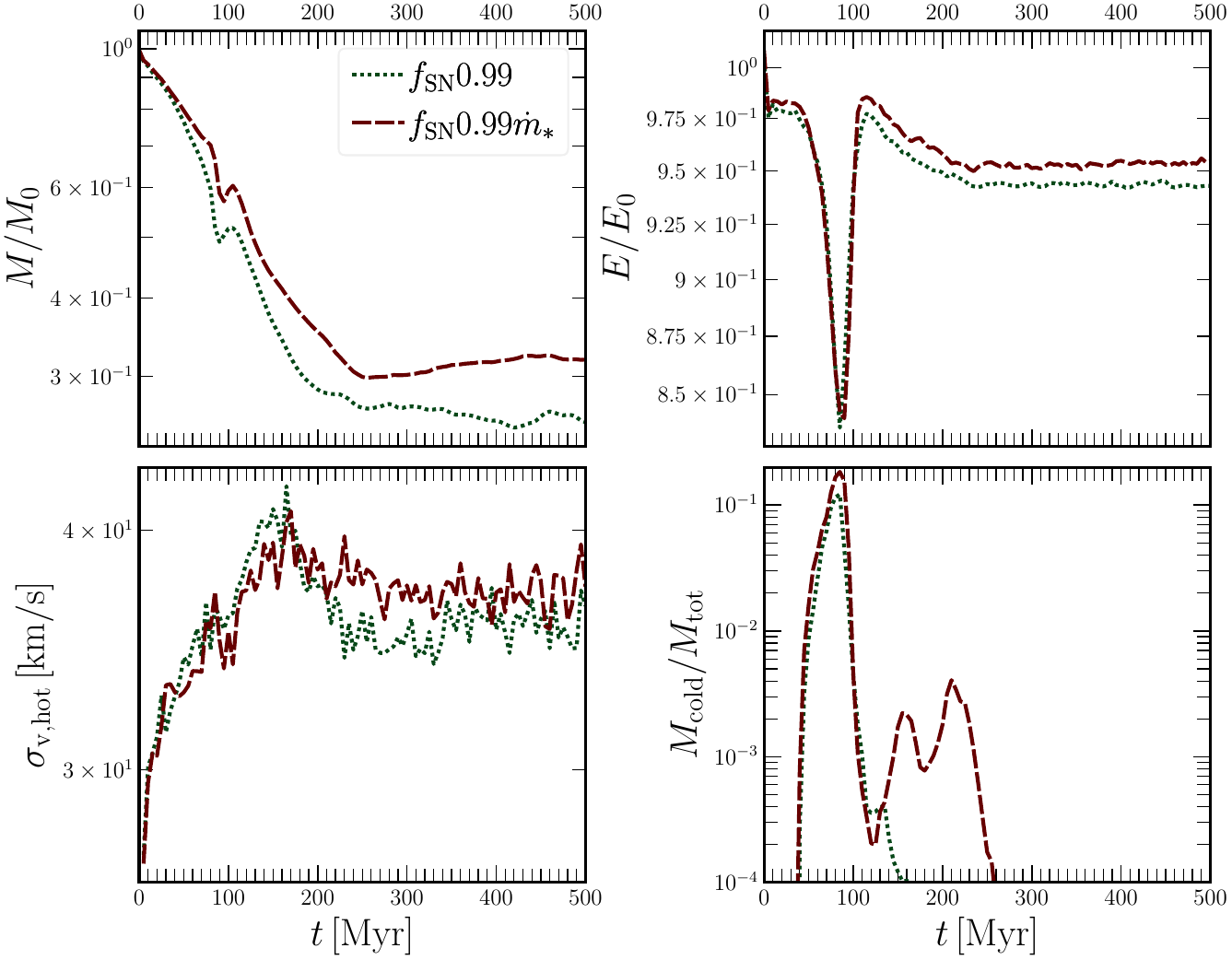}	
	\caption{Time-evolution of net mass (row 1, col 1), energy (row 1, col 2), velocity dispersion of hot gas (row 2, col 1) and mass fraction of cold gas (row 2, col 2) for our fiducial $f_\mathrm{SN}0.99$ run and the $f_\mathrm{SN}0.99\dot{m}_*$ run implementing mass and energy injection from AGB winds. We find no major differences between the two, except a larger fraction of cold gas formation for the $f_\mathrm{SN}0.99\dot{m}_*$ run.}
	\label{fig:mass_etot_evol_mdot_stellar}
\end{figure}

In Section \ref{subsubsec:heating_AGB_wind_rate}, we discussed that the energy injected into the ISM due to the thermalisation of the material ejected by AGB stars can contribute to its heating. The mass injected by these stars can also change the density of the ISM and its cooling rate. We have described our implementation of mass and energy input due to AGB wind ejecta in Section \ref{subsubsec:AGB_wind_injection}. Here, we study their effect on the time-evolution of important parameters in our simulations. 

In \Cref{fig:mass_etot_evol_mdot_stellar}, we show the evolution of the net mass, the net energy, the hot gas dispersion velocity and the mass fraction of cold gas for our fiducial $f_\mathrm{SN}0.99$ run and the $f_\mathrm{SN}0.99\dot{m}_*$ run. 
Comparing the two, we do not find any major differences. We do observe that the net mass in the $f_\mathrm{SN}0.99\dot{m}_*$ run increases slowly at late times, possibly due to mass injection from the AGB winds. The additional mass in the system raises the gas density and is also responsible for the formation and existence of cold gas in the simulation domain till $\sim250~\mathrm{Myr}$, compared to $150~\mathrm{Myr}$ for the $f_\mathrm{SN}0.99$ run.

% \begin{figure}%[h]
%     \vspace{0.5em}
% 		\centering
% 	\includegraphics[width=1.0\columnwidth]{NGC_5044_profs.pdf}	
% 	\caption{\change{Radial profiles of gas density (row 1), temperature (row 2), entropy (row 3) and heating due to SNIa, AGB stars and radiative cooling (row 4). To construct the stellar mass distribution, we used a Hernquist distribution with $M_*=6\times10^{11}M_\sun$ and $R_\mathrm{eff}=6.07~\mathrm{kpc}$. These values are consistent with the best-fit profiles for NGC 5044, shown in \cite{Voit2015ApJ803L21V}. The ratio of SNIa heating to the net cooling varies with $r$.}}
% \end{figure}
% \end{multicols}


\bibliographystyle{aasjournal}
\bibliography{refs.bib} % if your bibtex file is called example.bib

\begin{thebibliography}{}
\expandafter\ifx\csname natexlab\endcsname\relax\def\natexlab#1{#1}\fi
\providecommand{\url}[1]{\href{#1}{#1}}
\providecommand{\dodoi}[1]{doi:~\href{http://doi.org/#1}{\nolinkurl{#1}}}
\providecommand{\doeprint}[1]{\href{http://ascl.net/#1}{\nolinkurl{http://ascl.net/#1}}}
\providecommand{\doarXiv}[1]{\href{https://arxiv.org/abs/#1}{\nolinkurl{https://arxiv.org/abs/#1}}}

\bibitem[{{Allen} {et~al.}(2006){Allen}, {Dunn}, {Fabian}, {Taylor}, \& {Reynolds}}]{Allen2006MNRAS}
{Allen}, S.~W., {Dunn}, R.~J.~H., {Fabian}, A.~C., {Taylor}, G.~B., \& {Reynolds}, C.~S. 2006, \mnras, 372, 21, \dodoi{10.1111/j.1365-2966.2006.10778.x}

\bibitem[{{Anderson} \& {Sunyaev}(2018)}]{Anderson2018A&A}
{Anderson}, M.~E., \& {Sunyaev}, R. 2018, \aap, 617, A123, \dodoi{10.1051/0004-6361/201732510}

\bibitem[{{Ar{\'e}valo} {et~al.}(2016){Ar{\'e}valo}, {Churazov}, {Zhuravleva}, {Forman}, \& {Jones}}]{Arevalo2016ApJ}
{Ar{\'e}valo}, P., {Churazov}, E., {Zhuravleva}, I., {Forman}, W.~R., \& {Jones}, C. 2016, \apj, 818, 14, \dodoi{10.3847/0004-637X/818/1/14}

\bibitem[{{Astropy Collaboration} {et~al.}(2018){Astropy Collaboration}, {Price-Whelan}, {Sip{\H{o}}cz}, {G{\"u}nther}, {Lim}, {Crawford}, {Conseil}, {Shupe}, {Craig}, {Dencheva}, {Ginsburg}, {Vand erPlas}, {Bradley}, {P{\'e}rez-Su{\'a}rez}, {de Val-Borro}, {Aldcroft}, {Cruz}, {Robitaille}, {Tollerud}, {Ardelean}, {Babej}, {Bach}, {Bachetti}, {Bakanov}, {Bamford}, {Barentsen}, {Barmby}, {Baumbach}, {Berry}, {Biscani}, {Boquien}, {Bostroem}, {Bouma}, {Brammer}, {Bray}, {Breytenbach}, {Buddelmeijer}, {Burke}, {Calderone}, {Cano Rodr{\'\i}guez}, {Cara}, {Cardoso}, {Cheedella}, {Copin}, {Corrales}, {Crichton}, {D'Avella}, {Deil}, {Depagne}, {Dietrich}, {Donath}, {Droettboom}, {Earl}, {Erben}, {Fabbro}, {Ferreira}, {Finethy}, {Fox}, {Garrison}, {Gibbons}, {Goldstein}, {Gommers}, {Greco}, {Greenfield}, {Groener}, {Grollier}, {Hagen}, {Hirst}, {Homeier}, {Horton}, {Hosseinzadeh}, {Hu}, {Hunkeler}, {Ivezi{\'c}}, {Jain}, {Jenness}, {Kanarek}, {Kendrew}, {Kern}, {Kerzendorf}, {Khvalko}, {King}, {Kirkby}, {Kulkarni},
  {Kumar}, {Lee}, {Lenz}, {Littlefair}, {Ma}, {Macleod}, {Mastropietro}, {McCully}, {Montagnac}, {Morris}, {Mueller}, {Mumford}, {Muna}, {Murphy}, {Nelson}, {Nguyen}, {Ninan}, {N{\"o}the}, {Ogaz}, {Oh}, {Parejko}, {Parley}, {Pascual}, {Patil}, {Patil}, {Plunkett}, {Prochaska}, {Rastogi}, {Reddy Janga}, {Sabater}, {Sakurikar}, {Seifert}, {Sherbert}, {Sherwood-Taylor}, {Shih}, {Sick}, {Silbiger}, {Singanamalla}, {Singer}, {Sladen}, {Sooley}, {Sornarajah}, {Streicher}, {Teuben}, {Thomas}, {Tremblay}, {Turner}, {Terr{\'o}n}, {van Kerkwijk}, {de la Vega}, {Watkins}, {Weaver}, {Whitmore}, {Woillez}, {Zabalza}, \& {Astropy Contributors}}]{astropy2018}
{Astropy Collaboration}, {Price-Whelan}, A.~M., {Sip{\H{o}}cz}, B.~M., {et~al.} 2018, \aj, 156, 123, \dodoi{10.3847/1538-3881/aabc4f}

\bibitem[{{Bambic} \& {Reynolds}(2019)}]{Bambic2019ApJ}
{Bambic}, C.~J., \& {Reynolds}, C.~S. 2019, \apj, 886, 78, \dodoi{10.3847/1538-4357/ab4daf}

\bibitem[{{Barkhudaryan} {et~al.}(2019){Barkhudaryan}, {Hakobyan}, {Karapetyan}, {Mamon}, {Kunth}, {Adibekyan}, \& {Turatto}}]{Barkhudaryan2019MNRAS}
{Barkhudaryan}, L.~V., {Hakobyan}, A.~A., {Karapetyan}, A.~G., {et~al.} 2019, \mnras, 490, 718, \dodoi{10.1093/mnras/stz2585}

\bibitem[{{Beckmann} {et~al.}(2022){Beckmann}, {Dubois}, {Pellissier}, {Olivares}, {Polles}, {Hahn}, {Guillard}, \& {Lehnert}}]{Beckmann2022A&A}
{Beckmann}, R.~S., {Dubois}, Y., {Pellissier}, A., {et~al.} 2022, \aap, 665, A129, \dodoi{10.1051/0004-6361/202142527}

\bibitem[{{B{\^\i}rzan} {et~al.}(2004){B{\^\i}rzan}, {Rafferty}, {McNamara}, {Wise}, \& {Nulsen}}]{Birzan2004ApJ}
{B{\^\i}rzan}, L., {Rafferty}, D.~A., {McNamara}, B.~R., {Wise}, M.~W., \& {Nulsen}, P.~E.~J. 2004, \apj, 607, 800, \dodoi{10.1086/383519}

\bibitem[{{Buie} {et~al.}(2022){Buie}, {Scannapieco}, \& {Mark Voit}}]{Buie2022ApJ}
{Buie}, E., {Scannapieco}, E., \& {Mark Voit}, G. 2022, \apj, 927, 30, \dodoi{10.3847/1538-4357/ac4bc2}

\bibitem[{{Butsky} {et~al.}(2020){Butsky}, {Fielding}, {Hayward}, {Hummels}, {Quinn}, \& {Werk}}]{Butsky2020ApJ}
{Butsky}, I.~S., {Fielding}, D.~B., {Hayward}, C.~C., {et~al.} 2020, \apj, 903, 77, \dodoi{10.3847/1538-4357/abbad2}

\bibitem[{{Calzadilla} {et~al.}(2022){Calzadilla}, {McDonald}, {Donahue}, {McNamara}, {Fogarty}, {Gaspari}, {Gitti}, {Russell}, {Tremblay}, {Voit}, \& {Ubertosi}}]{Calzadilla2022ApJ}
{Calzadilla}, M.~S., {McDonald}, M., {Donahue}, M., {et~al.} 2022, \apj, 940, 140, \dodoi{10.3847/1538-4357/ac9790}

\bibitem[{{Choudhury} \& {Reynolds}(2022)}]{Choudhury2022MNRAS}
{Choudhury}, P.~P., \& {Reynolds}, C.~S. 2022, \mnras, 514, 3765, \dodoi{10.1093/mnras/stac1457}

\bibitem[{{Choudhury} \& {Sharma}(2016)}]{choudhury2016}
{Choudhury}, P.~P., \& {Sharma}, P. 2016, \mnras, 457, 2554, \dodoi{10.1093/mnras/stw152}

\bibitem[{{Choudhury} {et~al.}(2019){Choudhury}, {Sharma}, \& {Quataert}}]{Choudhury2019}
{Choudhury}, P.~P., {Sharma}, P., \& {Quataert}, E. 2019, \mnras, 488, 3195, \dodoi{10.1093/mnras/stz1857}

\bibitem[{{Ciotti} {et~al.}(1991){Ciotti}, {D'Ercole}, {Pellegrini}, \& {Renzini}}]{Ciotti1991ApJ}
{Ciotti}, L., {D'Ercole}, A., {Pellegrini}, S., \& {Renzini}, A. 1991, \apj, 376, 380, \dodoi{10.1086/170289}

\bibitem[{Collette(2013)}]{collette_python_hdf5_2014}
Collette, A. 2013, Python and HDF5 (O'Reilly)

\bibitem[{{Conroy} {et~al.}(2009){Conroy}, {Gunn}, \& {White}}]{Conroy2009ApJ}
{Conroy}, C., {Gunn}, J.~E., \& {White}, M. 2009, \apj, 699, 486, \dodoi{10.1088/0004-637X/699/1/486}

\bibitem[{{Conroy} {et~al.}(2015){Conroy}, {van Dokkum}, \& {Kravtsov}}]{Conroy2015ApJ}
{Conroy}, C., {van Dokkum}, P.~G., \& {Kravtsov}, A. 2015, \apj, 803, 77, \dodoi{10.1088/0004-637X/803/2/77}

\bibitem[{{Crain} {et~al.}(2015){Crain}, {Schaye}, {Bower}, {Furlong}, {Schaller}, {Theuns}, {Dalla Vecchia}, {Frenk}, {McCarthy}, {Helly}, {Jenkins}, {Rosas-Guevara}, {White}, \& {Trayford}}]{Crain2015MNRAS}
{Crain}, R.~A., {Schaye}, J., {Bower}, R.~G., {et~al.} 2015, \mnras, 450, 1937, \dodoi{10.1093/mnras/stv725}

\bibitem[{{Das} {et~al.}(2021){Das}, {Choudhury}, \& {Sharma}}]{Das2021MNRAS}
{Das}, H.~K., {Choudhury}, P.~P., \& {Sharma}, P. 2021, \mnras, 502, 4935, \dodoi{10.1093/mnras/stab382}

\bibitem[{{Donahue} \& {Voit}(2022)}]{Donahue2022PhR}
{Donahue}, M., \& {Voit}, G.~M. 2022, \physrep, 973, 1, \dodoi{10.1016/j.physrep.2022.04.005}

\bibitem[{{Draine}(2011)}]{Draine2011piimbook}
{Draine}, B.~T. 2011, {Physics of the Interstellar and Intergalactic Medium} (Princeton University Press)

\bibitem[{{El-Badry} {et~al.}(2019){El-Badry}, {Ostriker}, {Kim}, {Quataert}, \& {Weisz}}]{El-Badry2019MNRAS}
{El-Badry}, K., {Ostriker}, E.~C., {Kim}, C.-G., {Quataert}, E., \& {Weisz}, D.~R. 2019, \mnras, 490, 1961, \dodoi{10.1093/mnras/stz2773}

\bibitem[{{Fabian}(2012)}]{Fabian2012ARA&A}
{Fabian}, A.~C. 2012, \araa, 50, 455, \dodoi{10.1146/annurev-astro-081811-125521}

\bibitem[{{Faucher-Gigu{\`e}re} \& {Oh}(2023)}]{Faucher-Giguere2023ARAA}
{Faucher-Gigu{\`e}re}, C.-A., \& {Oh}, S.~P. 2023, \araa, 61, 131, \dodoi{10.1146/annurev-astro-052920-125203}

\bibitem[{{Federrath} {et~al.}(2008){Federrath}, {Klessen}, \& {Schmidt}}]{Federrath2008}
{Federrath}, C., {Klessen}, R.~S., \& {Schmidt}, W. 2008, \apjl, 688, L79, \dodoi{10.1086/595280}

\bibitem[{{Federrath} {et~al.}(2010){Federrath}, {Roman-Duval}, {Klessen}, {Schmidt}, \& {Mac Low}}]{federrath2010}
{Federrath}, C., {Roman-Duval}, J., {Klessen}, R.~S., {Schmidt}, W., \& {Mac Low}, M.~M. 2010, \aap, 512, A81, \dodoi{10.1051/0004-6361/200912437}

\bibitem[{{Fielding} {et~al.}(2017){Fielding}, {Quataert}, {Martizzi}, \& {Faucher-Gigu{\`e}re}}]{Fielding2017MNRAS}
{Fielding}, D., {Quataert}, E., {Martizzi}, D., \& {Faucher-Gigu{\`e}re}, C.-A. 2017, \mnras, 470, L39, \dodoi{10.1093/mnrasl/slx072}

\bibitem[{{Fielding} {et~al.}(2020){Fielding}, {Ostriker}, {Bryan}, \& {Jermyn}}]{Fielding2020ApJ}
{Fielding}, D.~B., {Ostriker}, E.~C., {Bryan}, G.~L., \& {Jermyn}, A.~S. 2020, \apjl, 894, L24, \dodoi{10.3847/2041-8213/ab8d2c}

\bibitem[{{Gaspari} {et~al.}(2012){Gaspari}, {Ruszkowski}, \& {Sharma}}]{Gaspari2012}
{Gaspari}, M., {Ruszkowski}, M., \& {Sharma}, P. 2012, \apj, 746, 94, \dodoi{10.1088/0004-637X/746/1/94}

\bibitem[{{Gaspari} {et~al.}(2017){Gaspari}, {Temi}, \& {Brighenti}}]{Gaspari2017MNRAS}
{Gaspari}, M., {Temi}, P., \& {Brighenti}, F. 2017, \mnras, 466, 677, \dodoi{10.1093/mnras/stw3108}

\bibitem[{{Gaspari} {et~al.}(2018){Gaspari}, {McDonald}, {Hamer}, {Brighenti}, {Temi}, {Gendron-Marsolais}, {Hlavacek-Larrondo}, {Edge}, {Werner}, {Tozzi}, {Sun}, {Stone}, {Tremblay}, {Hogan}, {Eckert}, {Ettori}, {Yu}, {Biffi}, \& {Planelles}}]{Gaspari2018ApJ}
{Gaspari}, M., {McDonald}, M., {Hamer}, S.~L., {et~al.} 2018, \apj, 854, 167, \dodoi{10.3847/1538-4357/aaaa1b}

\bibitem[{{Gatuzz} {et~al.}(2023{\natexlab{a}}){Gatuzz}, {Sanders}, {Dennerl}, {Liu}, {Fabian}, {Pinto}, {Eckert}, {Russell}, {Tamura}, {Walker}, \& {ZuHone}}]{Gatuzz2023MNRASa}
{Gatuzz}, E., {Sanders}, J.~S., {Dennerl}, K., {et~al.} 2023{\natexlab{a}}, \mnras, 520, 4793, \dodoi{10.1093/mnras/stad447}

\bibitem[{{Gatuzz} {et~al.}(2023{\natexlab{b}}){Gatuzz}, {Sanders}, {Dennerl}, {Liu}, {Fabian}, {Pinto}, {Eckert}, {Walker}, \& {ZuHone}}]{Gatuzz2023MNRASb}
---. 2023{\natexlab{b}}, \mnras, 525, 6394, \dodoi{10.1093/mnras/stad2716}

\bibitem[{{Guo} {et~al.}(2023){Guo}, {Stone}, {Kim}, \& {Quataert}}]{MGuo2023ApJ}
{Guo}, M., {Stone}, J.~M., {Kim}, C.-G., \& {Quataert}, E. 2023, \apj, 946, 26, \dodoi{10.3847/1538-4357/acb81e}

\bibitem[{Harris {et~al.}(2020)Harris, Millman, van~der Walt, Gommers, Virtanen, Cournapeau, Wieser, Taylor, Berg, Smith, Kern, Picus, Hoyer, van Kerkwijk, Brett, Haldane, del R{\'{i}}o, Wiebe, Peterson, G{\'{e}}rard-Marchant, Sheppard, Reddy, Weckesser, Abbasi, Gohlke, \& Oliphant}]{Harris2020}
Harris, C.~R., Millman, K.~J., van~der Walt, S.~J., {et~al.} 2020, {Array programming with NumPy},  Nature Research, \dodoi{10.1038/s41586-020-2649-2}

\bibitem[{{Hernquist}(1990)}]{Hernquist1990ApJ}
{Hernquist}, L. 1990, \apj, 356, 359, \dodoi{10.1086/168845}

\bibitem[{{Hopkins} {et~al.}(2020){Hopkins}, {Chan}, {Garrison-Kimmel}, {Ji}, {Su}, {Hummels}, {Kere{\v{s}}}, {Quataert}, \& {Faucher-Gigu{\`e}re}}]{Hopkins2020MNRAS}
{Hopkins}, P.~F., {Chan}, T.~K., {Garrison-Kimmel}, S., {et~al.} 2020, \mnras, 492, 3465, \dodoi{10.1093/mnras/stz3321}

\bibitem[{Hunter(2007)}]{Hunter4160265}
Hunter, J.~D. 2007, Computing in Science Engineering, 9, 90, \dodoi{10.1109/MCSE.2007.55}

\bibitem[{{Hu{\v{s}}ko} \& {Lacey}(2023)}]{Huvsko2023MNRAS}
{Hu{\v{s}}ko}, F., \& {Lacey}, C.~G. 2023, \mnras, 521, 4375, \dodoi{10.1093/mnras/stad793}

\bibitem[{{Kempski} \& {Quataert}(2020)}]{Kempski2020MNRAS}
{Kempski}, P., \& {Quataert}, E. 2020, \mnras, 493, 1801, \dodoi{10.1093/mnras/staa385}

\bibitem[{{Khatri} \& {Gaspari}(2016)}]{khatri2016}
{Khatri}, R., \& {Gaspari}, M. 2016, \mnras, 463, 655, \dodoi{10.1093/mnras/stw2027}

\bibitem[{{Kim} \& {Ostriker}(2015)}]{CGKim2015ApJ}
{Kim}, C.-G., \& {Ostriker}, E.~C. 2015, \apj, 802, 99, \dodoi{10.1088/0004-637X/802/2/99}

\bibitem[{{Konstandin} {et~al.}(2012){Konstandin}, {Girichidis}, {Federrath}, \& {Klessen}}]{konstandin2012}
{Konstandin}, L., {Girichidis}, P., {Federrath}, C., \& {Klessen}, R.~S. 2012, \apj, 761, 149, \dodoi{10.1088/0004-637X/761/2/149}

\bibitem[{{Lemaster} \& {Stone}(2009)}]{Lemaster2009ApJ}
{Lemaster}, M.~N., \& {Stone}, J.~M. 2009, \apj, 691, 1092, \dodoi{10.1088/0004-637X/691/2/1092}

\bibitem[{{Li} {et~al.}(2020{\natexlab{a}}){Li}, {Li}, {Bryan}, {Ostriker}, \& {Quataert}}]{MLi2020ApJa}
{Li}, M., {Li}, Y., {Bryan}, G.~L., {Ostriker}, E.~C., \& {Quataert}, E. 2020{\natexlab{a}}, \apj, 894, 44, \dodoi{10.3847/1538-4357/ab86b4}

\bibitem[{{Li} {et~al.}(2020{\natexlab{b}}){Li}, {Li}, {Bryan}, {Ostriker}, \& {Quataert}}]{MLi2020ApJb}
---. 2020{\natexlab{b}}, \apj, 898, 23, \dodoi{10.3847/1538-4357/ab9c22}

\bibitem[{{Li} {et~al.}(2019){Li}, {Bryan}, \& {Quataert}}]{YLi2019ApJ}
{Li}, Y., {Bryan}, G.~L., \& {Quataert}, E. 2019, \apj, 887, 41, \dodoi{10.3847/1538-4357/ab4bca}

\bibitem[{{Li} {et~al.}(2020{\natexlab{c}}){Li}, {Gendron-Marsolais}, {Zhuravleva}, {Xu}, {Simionescu}, {Tremblay}, {Lochhaas}, {Bryan}, {Quataert}, {Murray}, {Boselli}, {Hlavacek-Larrondo}, {Zheng}, {Fossati}, {Li}, {Emsellem}, {Sarzi}, {Arzamasskiy}, \& {Vishniac}}]{Li2020ApJ}
{Li}, Y., {Gendron-Marsolais}, M.-L., {Zhuravleva}, I., {et~al.} 2020{\natexlab{c}}, \apjl, 889, L1, \dodoi{10.3847/2041-8213/ab65c7}

\bibitem[{{Li} {et~al.}(2018){Li}, {Yuan}, {Mo}, {Yoon}, {Gan}, {Ho}, {Wang}, {Ostriker}, \& {Ciotti}}]{YPLi2018ApJ}
{Li}, Y.-P., {Yuan}, F., {Mo}, H., {et~al.} 2018, \apj, 866, 70, \dodoi{10.3847/1538-4357/aade8b}

\bibitem[{{Main} {et~al.}(2017){Main}, {McNamara}, {Nulsen}, {Russell}, \& {Vantyghem}}]{Main2017MNRAS}
{Main}, R.~A., {McNamara}, B.~R., {Nulsen}, P.~E.~J., {Russell}, H.~R., \& {Vantyghem}, A.~N. 2017, \mnras, 464, 4360, \dodoi{10.1093/mnras/stw2644}

\bibitem[{{Maoz} \& {Graur}(2017)}]{Maoz2017ApJ}
{Maoz}, D., \& {Graur}, O. 2017, \apj, 848, 25, \dodoi{10.3847/1538-4357/aa8b6e}

\bibitem[{{Martizzi} {et~al.}(2015){Martizzi}, {Faucher-Gigu{\`e}re}, \& {Quataert}}]{Martizzi2015MNRAS}
{Martizzi}, D., {Faucher-Gigu{\`e}re}, C.-A., \& {Quataert}, E. 2015, \mnras, 450, 504, \dodoi{10.1093/mnras/stv562}

\bibitem[{{McCourt} {et~al.}(2012){McCourt}, {Sharma}, {Quataert}, \& {Parrish}}]{mccourt2012}
{McCourt}, M., {Sharma}, P., {Quataert}, E., \& {Parrish}, I.~J. 2012, \mnras, 419, 3319, \dodoi{10.1111/j.1365-2966.2011.19972.x}

\bibitem[{{McNamara} \& {Nulsen}(2007)}]{McNamara2007ARA&A}
{McNamara}, B.~R., \& {Nulsen}, P.~E.~J. 2007, \araa, 45, 117, \dodoi{10.1146/annurev.astro.45.051806.110625}

\bibitem[{{Merritt} \& {Ferrarese}(2001)}]{Merritt2001MNRAS}
{Merritt}, D., \& {Ferrarese}, L. 2001, \mnras, 320, L30, \dodoi{10.1046/j.1365-8711.2001.04165.x}

\bibitem[{{Mohapatra} {et~al.}(2020){Mohapatra}, {Federrath}, \& {Sharma}}]{Mohapatra2020}
{Mohapatra}, R., {Federrath}, C., \& {Sharma}, P. 2020, \mnras, 493, 5838, \dodoi{10.1093/mnras/staa711}

\bibitem[{{Mohapatra} {et~al.}(2021){Mohapatra}, {Federrath}, \& {Sharma}}]{Mohapatra2021MNRAS}
---. 2021, \mnras, 500, 5072, \dodoi{10.1093/mnras/staa3564}

\bibitem[{{Mohapatra} {et~al.}(2022){Mohapatra}, {Federrath}, \& {Sharma}}]{Mohapatra2022MNRASc}
---. 2022, \mnras, 514, 3139, \dodoi{10.1093/mnras/stac1610}

\bibitem[{{Mohapatra} \& {Sharma}(2019)}]{Mohapatra2019}
{Mohapatra}, R., \& {Sharma}, P. 2019, \mnras, 484, 4881, \dodoi{10.1093/mnras/stz328}

\bibitem[{{Mohapatra} {et~al.}(2023){Mohapatra}, {Sharma}, {Federrath}, \& {Quataert}}]{Mohapatra2023MNRAS}
{Mohapatra}, R., {Sharma}, P., {Federrath}, C., \& {Quataert}, E. 2023, \mnras, 525, 3831, \dodoi{10.1093/mnras/stad2574}

\bibitem[{{Molero} {et~al.}(2023){Molero}, {Matteucci}, \& {Ciotti}}]{Molero2023MNRAS}
{Molero}, M., {Matteucci}, F., \& {Ciotti}, L. 2023, \mnras, 518, 987, \dodoi{10.1093/mnras/stac3066}

\bibitem[{Okuta {et~al.}(2017)Okuta, Unno, Nishino, Hido, \& Loomis}]{Okuta2017CuPyA}
Okuta, R., Unno, Y., Nishino, D., Hido, S., \& Loomis, C. 2017, in Proceedings of Workshop on Machine Learning Systems (LearningSys) in The Thirty-first Annual Conference on Neural Information Processing Systems (NIPS).
\newblock \url{http://learningsys.org/nips17/assets/papers/paper_16.pdf}

\bibitem[{{Olivares} {et~al.}(2019){Olivares}, {Salome}, {Combes}, {Hamer}, {Guillard}, {Lehnert}, {Polles}, {Beckmann}, {Dubois}, {Donahue}, {Edge}, {Fabian}, {McNamara}, {Rose}, {Russell}, {Tremblay}, {Vantyghem}, {Canning}, {Ferland }, {Godard}, {Peirani}, \& {Pineau des Forets}}]{Olivares2019A&A}
{Olivares}, V., {Salome}, P., {Combes}, F., {et~al.} 2019, \aap, 631, A22, \dodoi{10.1051/0004-6361/201935350}

\bibitem[{{Olivares} {et~al.}(2023){Olivares}, {Su}, {Forman}, {Gaspari}, {Andrade-Santos}, {Salome}, {Nulsen}, {Edge}, {Combes}, \& {Jones}}]{Olivares2023ApJ}
{Olivares}, V., {Su}, Y., {Forman}, W., {et~al.} 2023, \apj, 954, 56, \dodoi{10.3847/1538-4357/ace359}

\bibitem[{{Parrish} {et~al.}(2009){Parrish}, {Quataert}, \& {Sharma}}]{Parrish2009ApJ}
{Parrish}, I.~J., {Quataert}, E., \& {Sharma}, P. 2009, \apj, 703, 96, \dodoi{10.1088/0004-637X/703/1/96}

\bibitem[{{Parrish} {et~al.}(2010){Parrish}, {Quataert}, \& {Sharma}}]{Parrish2010ApJ}
---. 2010, \apjl, 712, L194, \dodoi{10.1088/2041-8205/712/2/L194}

\bibitem[{{Peterson} {et~al.}(2003){Peterson}, {Kahn}, {Paerels}, {Kaastra}, {Tamura}, {Bleeker}, {Ferrigno}, \& {Jernigan}}]{Peterson2003ApJ}
{Peterson}, J.~R., {Kahn}, S.~M., {Paerels}, F.~B.~S., {et~al.} 2003, \apj, 590, 207, \dodoi{10.1086/374830}

\bibitem[{{Pillepich} {et~al.}(2018){Pillepich}, {Springel}, {Nelson}, {Genel}, {Naiman}, {Pakmor}, {Hernquist}, {Torrey}, {Vogelsberger}, {Weinberger}, \& {Marinacci}}]{Pillepich2018MNRAS}
{Pillepich}, A., {Springel}, V., {Nelson}, D., {et~al.} 2018, \mnras, 473, 4077, \dodoi{10.1093/mnras/stx2656}

\bibitem[{{Prasad} {et~al.}(2015){Prasad}, {Sharma}, \& {Babul}}]{prasad2015}
{Prasad}, D., {Sharma}, P., \& {Babul}, A. 2015, \apj, 811, 108, \dodoi{10.1088/0004-637X/811/2/108}

\bibitem[{{Prasad} {et~al.}(2018){Prasad}, {Sharma}, \& {Babul}}]{prasad2018}
---. 2018, \apj, 863, 62, \dodoi{10.3847/1538-4357/aacce8}

\bibitem[{{Pulido} {et~al.}(2018){Pulido}, {McNamara}, {Edge}, {Hogan}, {Vantyghem}, {Russell}, {Nulsen}, {Babyk}, \& {Salom{\'e}}}]{Pulido2018ApJ}
{Pulido}, F.~A., {McNamara}, B.~R., {Edge}, A.~C., {et~al.} 2018, \apj, 853, 177, \dodoi{10.3847/1538-4357/aaa54b}

\bibitem[{{Rafferty} {et~al.}(2008){Rafferty}, {McNamara}, \& {Nulsen}}]{Rafferty2008ApJ}
{Rafferty}, D.~A., {McNamara}, B.~R., \& {Nulsen}, P.~E.~J. 2008, \apj, 687, 899, \dodoi{10.1086/591240}

\bibitem[{{Romero} {et~al.}(2023){Romero}, {Gaspari}, {Schellenberger}, {Bhandarkar}, {Devlin}, {Dicker}, {Forman}, {Khatri}, {Kraft}, {Di Mascolo}, {Mason}, {Moravec}, {Mroczkowski}, {Nulsen}, {Orlowski-Scherer}, {Perez Sarmiento}, {Sarazin}, {Sievers}, \& {Su}}]{Romero2023ApJ}
{Romero}, C.~E., {Gaspari}, M., {Schellenberger}, G., {et~al.} 2023, \apj, 951, 41, \dodoi{10.3847/1538-4357/acd3f0}

\bibitem[{{Scannapieco} \& {Bildsten}(2005)}]{Scannapieco2005ApJ}
{Scannapieco}, E., \& {Bildsten}, L. 2005, \apjl, 629, L85, \dodoi{10.1086/452632}

\bibitem[{{Sharma} {et~al.}(2012){Sharma}, {McCourt}, {Quataert}, \& {Parrish}}]{sharma2012thermal}
{Sharma}, P., {McCourt}, M., {Quataert}, E., \& {Parrish}, I.~J. 2012, \mnras, 420, 3174, \dodoi{10.1111/j.1365-2966.2011.20246.x}

\bibitem[{{Smith} {et~al.}(2017){Smith}, {Bryan}, {Glover}, {Goldbaum}, {Turk}, {Regan}, {Wise}, {Schive}, {Abel}, {Emerick}, {O'Shea}, {Anninos}, {Hummels}, \& {Khochfar}}]{Smith2017MNRAS}
{Smith}, B.~D., {Bryan}, G.~L., {Glover}, S. C.~O., {et~al.} 2017, \mnras, 466, 2217, \dodoi{10.1093/mnras/stw3291}

\bibitem[{{Stone} {et~al.}(2020){Stone}, {Tomida}, {White}, \& {Felker}}]{Stone2020ApJS}
{Stone}, J.~M., {Tomida}, K., {White}, C.~J., \& {Felker}, K.~G. 2020, \apjs, 249, 4, \dodoi{10.3847/1538-4365/ab929b}

\bibitem[{{Su} {et~al.}(2019){Su}, {Hopkins}, {Hayward}, {Ma}, {Faucher-Gigu{\`e}re}, {Kere{\v{s}}}, {Orr}, {Chan}, \& {Robles}}]{KySu2019MNRAS}
{Su}, K.-Y., {Hopkins}, P.~F., {Hayward}, C.~C., {et~al.} 2019, \mnras, 487, 4393, \dodoi{10.1093/mnras/stz1494}

\bibitem[{{Tang} {et~al.}(2009{\natexlab{a}}){Tang}, {Wang}, {Lu}, \& {Mo}}]{Tang2009MNRASa}
{Tang}, S., {Wang}, Q.~D., {Lu}, Y., \& {Mo}, H.~J. 2009{\natexlab{a}}, \mnras, 392, 77, \dodoi{10.1111/j.1365-2966.2008.14057.x}

\bibitem[{{Tang} {et~al.}(2009{\natexlab{b}}){Tang}, {Wang}, {Mac Low}, \& {Joung}}]{Tang2009MNRASb}
{Tang}, S., {Wang}, Q.~D., {Mac Low}, M.-M., \& {Joung}, M.~R. 2009{\natexlab{b}}, \mnras, 398, 1468, \dodoi{10.1111/j.1365-2966.2009.15206.x}

\bibitem[{{Tang} \& {Churazov}(2017)}]{Tang2017MNRAS}
{Tang}, X., \& {Churazov}, E. 2017, \mnras, 468, 3516, \dodoi{10.1093/mnras/stx590}

\bibitem[{{Tremblay} {et~al.}(2018){Tremblay}, {Combes}, {Oonk}, {Russell}, {McDonald}, {Gaspari}, {Husemann}, {Nulsen}, {McNamara}, {Hamer}, {O'Dea}, {Baum}, {Davis}, {Donahue}, {Voit}, {Edge}, {Blanton}, {Bremer}, {Bulbul}, {Clarke}, {David}, {Edwards}, {Eggerman}, {Fabian}, {Forman}, {Jones}, {Kerman}, {Kraft}, {Li}, {Powell}, {Randall}, {Salom{\'e}}, {Simionescu}, {Su}, {Sun}, {Urry}, {Vantyghem}, {Wilkes}, \& {ZuHone}}]{Tremblay2018ApJ}
{Tremblay}, G.~R., {Combes}, F., {Oonk}, J.~B.~R., {et~al.} 2018, \apj, 865, 13, \dodoi{10.3847/1538-4357/aad6dd}

\bibitem[{{Trott} {et~al.}(2021){Trott}, {Berger-Vergiat}, {Poliakoff}, {Rajamanickam}, {Lebrun-Grandie}, {Madsen}, {Al Awar}, {Gligoric}, {Shipman}, \& {Womeldorff}}]{Trott2021CSE}
{Trott}, C., {Berger-Vergiat}, L., {Poliakoff}, D., {et~al.} 2021, Computing in Science and Engineering, 23, 10, \dodoi{10.1109/MCSE.2021.3098509}

\bibitem[{{Trujillo} {et~al.}(2011){Trujillo}, {Ferreras}, \& {de La Rosa}}]{Trujillo2011MNRAS}
{Trujillo}, I., {Ferreras}, I., \& {de La Rosa}, I.~G. 2011, \mnras, 415, 3903, \dodoi{10.1111/j.1365-2966.2011.19017.x}

\bibitem[{{van der Velden}(2020)}]{Ellert2020JOSS}
{van der Velden}, E. 2020, The Journal of Open Source Software, 5, 2004, \dodoi{10.21105/joss.02004}

\bibitem[{Virtanen {et~al.}(2020)Virtanen, Gommers, Oliphant, Haberland, Reddy, Cournapeau, Burovski, Peterson, Weckesser, Bright, van~der Walt, Brett, Wilson, Millman, Mayorov, Nelson, Jones, Kern, Larson, Carey, Polat, Feng, Moore, VanderPlas, Laxalde, Perktold, Cimrman, Henriksen, Quintero, Harris, Archibald, Ribeiro, Pedregosa, van Mulbregt, Vijaykumar, Bardelli, Rothberg, Hilboll, Kloeckner, Scopatz, Lee, Rokem, Woods, Fulton, Masson, H{\"a}ggstr{\"o}m, Fitzgerald, Nicholson, Hagen, Pasechnik, Olivetti, Martin, Wieser, Silva, Lenders, Wilhelm, Young, Price, Ingold, Allen, Lee, Audren, Probst, Dietrich, Silterra, Webber, Slavi{\v c}, Nothman, Buchner, Kulick, Sch{\"o}nberger, de~Miranda~Cardoso, Reimer, Harrington, Rodr{\'\i}guez, Nunez-Iglesias, Kuczynski, Tritz, Thoma, Newville, K{\"u}mmerer, Bolingbroke, Tartre, Pak, Smith, Nowaczyk, Shebanov, Pavlyk, Brodtkorb, Lee, McGibbon, Feldbauer, Lewis, Tygier, Sievert, Vigna, Peterson, More, Pudlik, Oshima, Pingel, Robitaille, Spura, Jones, Cera, Leslie, Zito,
  Krauss, Upadhyay, Halchenko, V{\'a}zquez-Baeza, \& Contributors}]{Virtanen2020}
Virtanen, P., Gommers, R., Oliphant, T.~E., {et~al.} 2020, Nature Methods, 17, 261, \dodoi{10.1038/s41592-019-0686-2}

\bibitem[{{Voit}(2018)}]{Voit2018ApJ}
{Voit}, G.~M. 2018, \apj, 868, 102, \dodoi{10.3847/1538-4357/aae8e2}

\bibitem[{{Voit}(2021)}]{Voit2021ApJ}
---. 2021, \apjl, 908, L16, \dodoi{10.3847/2041-8213/abe11f}

\bibitem[{{Voit} {et~al.}(2015){Voit}, {Donahue}, {O'Shea}, {Bryan}, {Sun}, \& {Werner}}]{Voit2015ApJ803L21V}
{Voit}, G.~M., {Donahue}, M., {O'Shea}, B.~W., {et~al.} 2015, \apjl, 803, L21, \dodoi{10.1088/2041-8205/803/2/L21}

\bibitem[{{Voit} {et~al.}(2020){Voit}, {Bryan}, {Prasad}, {Frisbie}, {Li}, {Donahue}, {O'Shea}, {Sun}, \& {Werner}}]{Voit2020ApJ}
{Voit}, G.~M., {Bryan}, G.~L., {Prasad}, D., {et~al.} 2020, \apj, 899, 70, \dodoi{10.3847/1538-4357/aba42e}

\bibitem[{{Wang} {et~al.}(2021){Wang}, {Ruszkowski}, {Pfrommer}, {Oh}, \& {Yang}}]{Wang2021MNRAS}
{Wang}, C., {Ruszkowski}, M., {Pfrommer}, C., {Oh}, S.~P., \& {Yang}, H. Y.~K. 2021, \mnras, 504, 898, \dodoi{10.1093/mnras/stab966}

\bibitem[{{Wang} \& {Yang}(2022)}]{Wang2022MNRAS}
{Wang}, S.-C., \& {Yang}, H. Y.~K. 2022, \mnras, 512, 5100, \dodoi{10.1093/mnras/stac788}

\bibitem[{{Werner} {et~al.}(2012){Werner}, {Allen}, \& {Simionescu}}]{Werner2012MNRAS}
{Werner}, N., {Allen}, S.~W., \& {Simionescu}, A. 2012, \mnras, 425, 2731, \dodoi{10.1111/j.1365-2966.2012.21245.x}

\bibitem[{{Werner} {et~al.}(2013){Werner}, {Oonk}, {Canning}, {Allen}, {Simionescu}, {Kos}, {van Weeren}, {Edge}, {Fabian}, {von der Linden}, {Nulsen}, {Reynolds}, \& {Ruszkowski}}]{Werner2013ApJ}
{Werner}, N., {Oonk}, J.~B.~R., {Canning}, R.~E.~A., {et~al.} 2013, \apj, 767, 153, \dodoi{10.1088/0004-637X/767/2/153}

\bibitem[{{Werner} {et~al.}(2014){Werner}, {Oonk}, {Sun}, {Nulsen}, {Allen}, {Canning}, {Simionescu}, {Hoffer}, {Connor}, {Donahue}, {Edge}, {Fabian}, {von der Linden}, {Reynolds}, \& {Ruszkowski}}]{Werner2014MNRAS}
{Werner}, N., {Oonk}, J.~B.~R., {Sun}, M., {et~al.} 2014, \mnras, 439, 2291, \dodoi{10.1093/mnras/stu006}

\bibitem[{{Zhuravleva} {et~al.}(2018){Zhuravleva}, {Allen}, {Mantz}, \& {Werner}}]{zhuravleva2018}
{Zhuravleva}, I., {Allen}, S.~W., {Mantz}, A., \& {Werner}, N. 2018, \apj, 865, 53, \dodoi{10.3847/1538-4357/aadae3}

\bibitem[{{Zhuravleva} {et~al.}(2014){Zhuravleva}, {Churazov}, {Schekochihin}, {Allen}, {Ar{\'e}valo}, {Fabian}, {Forman}, {Sanders}, {Simionescu}, {Sunyaev}, {Vikhlinin}, \& {Werner}}]{zhuravleva2014turbulent}
{Zhuravleva}, I., {Churazov}, E., {Schekochihin}, A.~A., {et~al.} 2014, \nat, 515, 85, \dodoi{10.1038/nature13830}

\bibitem[{{Zhuravleva} {et~al.}(2015){Zhuravleva}, {Churazov}, {Ar{\'e}valo}, {Schekochihin}, {Allen}, {Fabian}, {Forman}, {Sanders}, {Simionescu}, {Sunyaev}, {Vikhlinin}, \& {Werner}}]{Zhuravleva2015MNRAS}
{Zhuravleva}, I., {Churazov}, E., {Ar{\'e}valo}, P., {et~al.} 2015, \mnras, 450, 4184, \dodoi{10.1093/mnras/stv900}

\end{thebibliography}
\end{document}